\documentclass[aps,PRB,twocolumn,longbibliography,superscriptaddress]{revtex4-2}
\usepackage{amsmath,amssymb}
\usepackage{subcaption}
\usepackage{float}
\usepackage{ragged2e}
\usepackage{graphicx}
\graphicspath{ {figure} }
\usepackage{dcolumn}
\usepackage{bbold}
\usepackage[hidelinks]{hyperref}
\hypersetup{
  colorlinks,
  citecolor=magenta,
  linkcolor=blue,
  urlcolor=blue}
\usepackage{verbatim,color,xcolor}
\usepackage{soul}

\newcommand{\ua}{{\uparrow}}
\newcommand{\da}{{\downarrow}}
\newcommand{\Wt}{{\Tilde{W}}}
\newcommand{\Bp}{{B_\perp}}
\newcommand{\Kua}{{|K\uparrow \rangle}}

\newcommand{\Kpua}{{|K'\uparrow \rangle}}

\begin{document}

\title{Magnon transmission across $\nu=1|-1|1$ mono-layer graphene junction as a probe of electronic structure}

\author{Suman Jyoti De}
\affiliation{Department of Physics, McGill University, Montreal, Quebec, Canada H3A 2T8}
\affiliation{Harish-Chandra Research Institute, A CI of Homi Bhabha National Institute, Chhatnag  Road, Jhunsi, Prayagraj 211019, India}
\author{Sumathi Rao}
\affiliation{International Centre for Theoretical Sciences (ICTS-TIFR),
Shivakote, Hesaraghatta Hobli, Bangalore 560089, India}
\author{Ganpathy Murthy}
\affiliation{Department of Physics \& Astronomy, University of Kentucky, Lexington, KY 40506, USA}

\begin{abstract}
   We study magnon transmission across gate-controlled junctions in the $n=0$ manifold of Landau levels in monolayer graphene, in the presence of both spin and valley Zeeman fields. Specifically, we consider the $1|-1|1$ sandwich geometry. The nature of the interfaces between regions of different filling turns out to be crucial for magnon transmission. Using the Hartree-Fock approximation, we find that either the spin or the valley degrees of freedom of the occupied one-body states rotate across the interfaces. If the interfaces exhibit spin rotation, magnon transmission is suppressed at high energies, while if the interfaces have valley rotation, magnon transmission becomes perfect at high energies. The valley Zeeman coupling, which arises from partial alignment with the encapsulating Boron Nitride, is independent of perpendicular magnetic field $B$, while the spin Zeeman and other anisotropic couplings scale linearly with $B$. This allows the tuning of the relative strength of the valley Zeeman coupling in situ by varying $B$, which can drive phase transitions of the interfaces between spin-rotated and valley-rotated phases, leading to magnon transmission being either vanishing or perfect at high energies. Our analysis, along with the experimental measurements, can be used to determine the anisotropic couplings in the sample.
\end{abstract}

\maketitle

\section{Introduction}
\label{sec:intro}

The quantum Hall effect (QHE), initially discovered in semiconductor heterostructures \cite{IQHE_Discovery_1980,Thouless_1982}, is the simplest manifestation of band topology \cite{Kane_Mele_2DTI_PhysRevLett.95.146802,TIs_RevModPhys.82.3045}. The kinetic energy in the strong orbital magnetic field is quenched into a discrete set of highly degenerate Landau levels. When an integer number of Landau levels are filled, the system is insulating in the bulk. The topological nontriviality of the bulk leads to protected chiral edge modes \cite{Halperin_Edge_1982} which carry the charge and heat currents detected in transport. The absence of kinetic energy within each Landau level means that the ground states of electrons in partially filled Landau levels are controlled by interactions, leading to the fractional quantum Hall effect (FQHE) \cite{FQHE_Discovery_1982, Laughlin_1983}. 

The dominance of interactions in the quantum Hall regime is not limited to fractional fillings. If internal degeneracies such as spin and valley are present, each orbital Landau level can be filled by multiple flavors. At integer fillings, the many-body ground state is selected by the interactions, a phenomenon known as quantum Hall ferromagnetism \cite{Shivaji_Skyrmion, QHFM_Fertig_1989, QHFM_Yang_etal_1994, QHFM_Moon_etal_1995}.  

The QHE has found a new and remarkable manifestation in monolayer graphene (MLG) \cite{Berger_etal_2004, Novoselov_etal_2004, Zhang_etal_2005, neto2009:rmp}, an atomically thin two-dimensional(2D) material with a honeycomb lattice. MLG shows well-quantized Hall plateaus\cite{neto2009:rmp, zhang2006:nu0} in the presence of a strong orbital $\Bp$ field at low temperatures. 

The electronic band structure of MLG has two inequivalent points in the Brillouin zone (BZ), the $K$ and $K'$ valleys, where the conduction and valence bands touch each other linearly in Dirac crossings. The low energy, long wavelength features of MLG are thus dictated by the linear Dirac spectrum close to the two valleys \cite{neto2009:rmp}. At charge neutrality, the chemical potential lies at the Dirac points. 
When an orbital $\Bp$ field is turned on, the kinetic energy near each Dirac point becomes quantized into a set of particle-hole symmetric Landau levels, with $E_n\propto sgn(n)\sqrt{\Bp|n|}$ \cite{neto2009:rmp}. The internal (near) degeneracies are now spin and valley, leading to fourfold, nearly degenerate Landau levels. Coming to the edge structure, all the $n>0$ Landau levels produce chiral modes with a particle-like dispersion, and all the $n<0$ Landau levels lead to chiral modes with a hole-like dispersion.  The $n=0$ manifold of Landau levels is special, in that the wave functions are composed of an equal superposition of particle-like and hole-like momentum states. Near an edge, because of intervalley scattering, the orbital $K$ and $K'$ $n=0$ Landau levels combine to produce one particle-like and one hole-like edge mode \cite{brey2006:nu0}. Another special feature of the $n=0$ LL is that the electronic wave function in a particular valley is completely localized on a particular sublattice of the honeycomb lattice (valley-sublattice locking). Thus, in the zero-energy $n=0$ manifold of Landau levels (called the zero Landau levels, or ZLLs), one needs a four-component spinor to describe the internal spin/valley degrees of freedom. When none of the ZLLs are filled, the filling factor $\nu$ is defined to be $\nu=-2$, which is the same as its Hall conductance in dimensionless units. When a single ZLL is filled $\nu=-1$, and if a single one is empty $\nu=1$. The charge neutral state with two filled ZLLs and two empty ones has $\nu=0$.

Let us first consider the one-body terms in the Hamiltonian. The Zeeman coupling $E_Z$ is always present. In addition, in MLG samples encapsulated with hexagonal Boron Nitride (HBN), partial alignment of the graphene lattice with that of the HBN produces a sublattice potential that favors one sublattice over the other \cite{KV_DGG_2013,KV_expt_2013,KV_HBN_Jung2015,KV_HBN_Jung2017}. In the ZLLs, this means favoring one valley over the other. For this reason, we will call this coupling the valley Zeeman coupling $E_V$. 

The case of $\nu=0$ (charge neutrality), though not the main topic of this work, is the most complex and has led to the development of many experimental techniques \cite{jiang2007:nu0,young2012:nu0, Maher_Kim_etal_2013,young2014:nu0, Magnon_transport_Yacoby_2018, Spin_transport_Lau2018, Magnon_Transport_Assouline_2021, Magnon_transport_Zhou_2022,li2019:stm, STM_Yazdani2021visualizing, STM_Coissard_2022} aimed at elucidating its nature. Two of the four ZLLs are filled. When $E_Z>0$ and only long-range Coulomb interactions are present, the ground state is a quantum spin Hall insulator with maximal spin polarization \cite{abanin2006:nu0,Brey_Fertig_2006}. The two counter-propagating chiral modes have opposite spins, and ought to produce a two-terminal edge conductance of $2e^2/h$. However, experimentally, at a purely perpendicular field, MLG is a vanilla insulator with no edge modes \cite{jiang2007:nu0,young2012:nu0, Maher_Kim_etal_2013,young2014:nu0}. In tilted field experiments, when a $B_{\parallel}$ as well as a $B_{\perp}$ are applied and $E_Z$ is increased to a large value, the state does approach the fully polarized spin ferromagnet, and the expected edge conductance is asymptotically recovered \cite{young2014:nu0}. 

In addition to the long-range Coulomb interaction, there are lattice scale interactions that break the $SU(4)$ symmetry and compete with the Zeeman and valley Zeeman couplings in selecting the ground state \cite{alicea2006:gqhe, KYang_SU4_Skyrmion_2006, Herbut1, Herbut2,abanin2006:nu0,brey2006:nu0}. It has long been known that at the four-fermion level, the low-energy effective theory must conserve the number of electrons in each valley separately, leading to a $U(1)_v$ symmetry \cite{alicea2006:gqhe}. In a seminal work, Kharitonov proposed an ultra-short-range (USR) model \cite{kharitonov2012:nu0,Kharitonov_Edge_2012} for the residual interactions which has two couplings, a valley Ising coupling $u_z$, and a valley $XY$ coupling $u_{xy}$, both expected to scale linearly with $B_{\perp}$. He then solved this model in the Hartree-Fock (HF) approximation to obtain a phase diagram in the $u_z-u_{xy}$ parameter space. At charge neutrality, the following quantum Hall phases are seen \cite{kharitonov2012:nu0}: A fully spin-polarized quantum spin Hall insulator (FM), a canted antiferromagnet (CAF), a bond-ordered (BO) state, presumably with a Kekul\'e distortion (and thus often called KD, though we will use the notation BO), and a charge density wave (CDW). It is believed that at a purely perpendicular field, the system is either in a BO state or a CAF state, and upon increasing $E_Z$ it undergoes a transition to the fully spin-polarized state. More recently, even more complex phase diagrams have been proposed both  for $\nu=0$ \cite{Das_Kaul_Murthy_2022, Phase_diagram_nu0_Suman2023,Stefanides_Sodemann2022,Stefanidis_Sodemann2023} and $\nu=\pm 1$~\cite{Lian_Rosch_Goerbig_2016, Atteia_Goerbig_2021} systems by
relaxing the ultra-short-range assumption for the anisotropic residual couplings. 

Many experimental techniques have been used to probe the $\nu=0$ state. Apart from transport \cite{jiang2007:nu0,young2012:nu0, Maher_Kim_etal_2013,young2014:nu0}, the state has been probed by scanning tunneling microscopy/spectroscopy  (STM) \cite{li2019:stm,STM_Yazdani2021visualizing, STM_Coissard_2022} and by the transmission of magnons \cite{Magnon_transport_Yacoby_2018, Spin_transport_Lau2018, Magnon_Transport_Assouline_2021, Magnon_transport_Zhou_2022}. Magnons are collective excitations which carry spin, and are always present in systems which have a nonzero spin polarization. These two techniques probe different order parameters. While magnon transmission probes whether the state in question has any spin polarization, current STM experiments are not spin-resolved. They detect charge density at the atomic scale, and can thus detect the presence of charge and/or bond order. STM experiments ubiquitously show bond order and CDW order \cite{li2019:stm,STM_Yazdani2021visualizing,STM_Coissard_2022}. As an aside, we note that while bond order and CAF order do not coexist in Kharitonov's phase diagram \cite{kharitonov2012:nu0}, removing the restriction of ultra-short range interactions allows them to coexist \cite{Das_Kaul_Murthy_2022,Phase_diagram_nu0_Suman2023,Stefanides_Sodemann2022,Stefanidis_Sodemann2023} at $\nu=0$. Based on a combination of STM and magnon transmission experiments, the current consensus is that at low $B_{\perp}$, the system is a spin singlet and has bond order. As the  field $B_{\perp}$ increases, the couplings $u_z$ and $u_{xy}$ increase, while $E_V$ remains the same. As detected by magnon transmission \cite{Magnon_transport_Zhou_2022}, the system makes a transition into a magnetic state, presumably the CAF phase, at some critical value of $B_{\perp}$. Based on Kharitonov's phase diagram this puts physical systems in the region of the parameter space where $u_z>0$ while $u_{xy}<0$. 

Our goal in this work is to thoroughly examine a much simpler system, which is a ``sandwich" of $\nu=1$, $\nu=-1$ and $\nu=1$, denoted as $1|-1|1$.  Such a system has been examined before in the limiting case when $E_V=0$~\cite{Wei_Huang_MacDonald2021}. We study it in full generality in the neighborhood of the physical region of the $u_z,\ u_{xy}$ parameter space with nonzero $E_Z,\ E_V$. Real samples of graphene are believed to have $u_z>0,\ u_{xy}<0$. We will examine both $u_z>|u_{xy}|$, which puts the system at $E_Z=E_V=0$ in the AF phase at $\nu=0$, and $u_z<|u_{xy}|$, for which the system is in the bond-ordered phase at $E_Z=E_V=0$ at $\nu=0$. Since $E_Z,u_z,u_{xy}$ all scale linearly with $B_{\perp}$ while $E_V$ remains constant, one can access many different regimes simply by varying $B_{\perp}$. The first step is to examine the Hartree-Fock (HF) ground states for each of $\nu=\pm1$. It is important to note that for $\nu=\pm1$ ultra-short-range interactions are unable to pick out a unique bulk ground state if $E_Z=E_V=0$. We will always have both types of Zeeman fields nonzero in what follows, because in this situation, the bulk ground states at $\nu=\pm1$ are unique. As we will show, the ordering of the HF energies in the $\nu=\pm1$ states, which depends on the coupling constants $u_z,\ u_{xy}$ as well as $E_Z,\ E_V$,  plays a key role in the nature of the interface between $\nu=1$ and $\nu=-1$. The middle layer of the sandwich can undergo transitions between different ground states as $B_{\perp}$ is varied. Note that magnon scattering in a skyrmion crystal with a $1|1\pm\delta\nu|1$ sandwich also has been studied earlier \cite{Magnon_scattering_Nilotpal_Roderich_Doucot2023}.

Experimentally, magnons can be generated at the the contacts in a $\nu=1$ system \cite{Magnon_transport_Yacoby_2018, Young_Skyrmion_Solid_Graphene_2019, Magnon_Transport_Assouline_2021, Spin_transport_Lau2018, Magnon_transport_Zhou_2022, Magnons_Yacoby2022} by creating a potential difference between the co-propagating edge channels of opposite spins at an edge between $\nu=2$ and $\nu=1$. When the bias voltage between the two channels exceeds the spin flip energy $2E_Z$, a magnon is created by a spin-flip process, which can then be transmitted through the $1|\nu_m|1$ junction, where $\nu_m$ is the filling fraction in the middle region.
The magnon transmission probability through the sandwich is studied as a function of the incident magnon energy, either by local conductance measurements or non-local voltage measurements.
In an earlier work \cite{Magnon_transport_Yacoby_2018} (believed to be for  $E_V\approx0$) it was found that for both configurations $1|1|1$ and $1|-1|1$, the magnons were largely transmitted, while
for $1|0|1$, they were largely reflected for energies near threshold ($2E_Z$). Although this can be understood through kinematic constraints~ \cite{Wei_Huang_MacDonald2021}, later experiments~\cite{Magnon_Transport_Assouline_2021} show  magnons
are largely reflected for $1|-1|1$, perhaps because of oblique incidence at the interface. More recent experiments \cite{Magnon_transport_Zhou_2022} performed in the range $0<\nu<1$ of the middle region show that the transmission across the junction can be changed by tuning the external perpendicular
magnetic field $B_{\perp}$. Thus, even for the simpler system, the experimental situation is far from clear.  This is a strong motivation for our detailed study 
including all allowed parameters.  Our main result in this paper is to show that magnon transmission through the $1|-1|1$ sandwich can detect various phases in the intermediate region as a function of  $B_{\perp}$.
 
Another important motivation for us is the possibility of determining all the coupling parameters using magnon transmission across the sandwich. The idea is as follows: The value of $E_V$ can be determined by zero-$B_\perp$ measurements of the gap~\cite{KV_DGG_2013}. Using this and the known value of $E_Z$, one is left with the free parameters $u_z,\ u_{xy}$. The three parameters $E_Z,u_z,u_{xy}$ all scale linearly with $B_\perp$. The Coulomb interaction scales with $\sqrt{B_\perp}$, whereas $E_V$ is independent of $B_\perp$. Thus, by varying $B_\perp$ one is able to vary the ratios of these parameters over a wide range. The ratios determine the ordering of the one-body levels, and consequently the structure of the junction, and the order parameters in the middle region of the sandwich. Of particular interest are specific ratios of the coupling constants where phase transitions in the order parameters of the middle region occur, which are reflected in the magnon transmission probability. Thus, magnon transmission over a wide range of $B_\perp$ would ideally allow us to determine $u_z/B_\perp$ and $u_{xy}/B_\perp$ for a given sample.

The rest of the paper is organized as follows. In Section II, we present the Hamiltonian and understand the bulk HF spectrum for $\nu=\pm1$. It will turn out that the ordering of the levels in energy plays a key role in the nature of the interface between $\nu=1$ and $\nu=-1$, which in turn determines the transmission/reflection of collective excitations through it.   We will examine how the various parameters of the model ($E_Z,\ E_V$ and the anisotropic couplings) enter in determining this ordering of levels.  Since all the parameters except $E_V$ scale linearly with $B_\perp$, we can alter the ordering of levels simply by varying $B_\perp$.  In Sec. III we will present the full Hamiltonian of the system, including the interfaces, and apply the HF approximation. Here we will explicitly see how the bulk ordering of levels is the deciding factor in the structure of the interface. In Sec. IV we study the collective excitations via the time-dependent Hartree-Fock (TDHF) approximation, and introduce the bulk collective excitations which are the scattering states for the magnon transmission problem. Following earlier work \cite{Wei_Huang_MacDonald2021} we also set up the formalism to study magnon transmission and reflection through the $1|-1|1$ junction using the TDHF equations. In Section V we present our results followed by a discussion and conclusion in Sec VI.

\section{The bulk Hamiltonian and Hartree Fock ground state}
\label{sec:bulk_HF}
The Hamiltonian of the $N=0$ LLs in MLG, because of its sub-lattice and valley locking,  can be written using four levels denoted by their spin($\ua, \da$) and valley indices ($K, K'$). We work in the Landau gauge $\vec{A}=(0,\Bp x,0)$, where $\Bp$ is the external magnetic field perpendicular to the sample. Note that in the following, the magnetic length $\ell$ is defined as $\ell=\sqrt{h/eB_\perp}$.

The bulk Hamiltonian for the ZLLs in MLG for generic symmetry-allowed interactions (first proposed by Kharitonov \cite{kharitonov2012:nu0}) is
\begin{align}
    \mathcal{H}^{bulk} = \frac{\pi \ell^2}{A} \sum_{k_1,k_2,\vec{q}} e^{i(\phi(k_1,\vec{q})+\phi(k_2,-\vec{q}))} e^{-\frac{(q^2 \ell^2)}{2}} \hspace{0.1cm} \times \hspace{0.2cm}
    \nonumber \\
    \big[ \sum_{\alpha=x,y,z} v_\alpha(q) : (\vec{c}_{k_1-q_y}^{\hspace{0.1cm} \dagger} \tau_\alpha \vec{c}_{k_1}) (\vec{c}_{k_2+q_y}^{\hspace{0.1cm} \dagger} \tau_\alpha \vec{c}_{k_2}) : \hspace{0.5cm}
    \nonumber \\ 
    + V(q)~ :(\vec{c}_{k_1-q_y}^{\hspace{0.1cm} \dagger} \vec{c}_{k_1})(\vec{c}_{k_2+q_y}^{\hspace{0.1cm} \dagger} \vec{c}_{k_2}):  \big] \hspace{0.5cm}
    \nonumber \\
    - \sum_{k} E_Z \vec{c}_k^{\hspace{0.1cm} \dagger} \sigma_z \vec{c}_k 
    - \sum_{k} E_V \vec{c}_k^{\hspace{0.1cm} \dagger} \tau_z \vec{c}_k \hspace{1.8cm} 
    \label{eq:Ha_bulk}
\end{align}
where $\phi(k,\vec{q})=\ell^2(-q_x k + \frac{1}{2} q_x q_y)$ and $\vec{c}_k=(c_{k K\ua},c_{k K\da},c_{k K'\ua},c_{k K'\da})^T$. The matrices $\tau_\alpha$ and $\sigma_\alpha$, $\alpha=0,x,y,z$, with $0$ denoting the identity matrix, are  Pauli matrices acting in the valley and spin spaces respectively. More explicitly, $\tau_\alpha\equiv \tau_\alpha\otimes\sigma_0$.
Here, $V(q_x,q_y)=\frac{E_c}{\sqrt{q^2 \ell^2 + q_0^2}}$ is the screened Coulomb interaction. We have used the ultra-short-range (USR) assumption for the residual anisotropic interactions \cite{kharitonov2012:nu0}, implying that $v_\alpha(q)$ are independent of momentum; $v_\alpha(q)=2\pi\ell^2 u_\alpha$. The valley $XY$ coupling is given by $u_x=u_y=u_{xy}$.
As seen in earlier work~\cite{Das_Kaul_Murthy_2022, Phase_diagram_nu0_Suman2023,Stefanides_Sodemann2022,Stefanidis_Sodemann2023}, relaxing the USR assumption does lead to the appearance of new phases at $\nu=0$. However, as long as $E_Z,E_V>0$, which we will assume throughout this work, the USR assumption does not seem to have a strong effect on the phases of $\nu=\pm1$ in the physical region of the parameter space~\cite{Lian_Rosch_Goerbig_2016,Lian_Goerbig_2017,Atteia_Goerbig_2021}, which is why we continue to use the USR assumption here. We will keep the Coulomb screening wavevector $q_0$ small in our analysis. $E_Z$ is the spin Zeeman term, which denotes the coupling of the electron spin with the external magnetic field, $E_Z \propto \mu_b \Bp$, where $\mu_b$ is the Bohr magneton. $E_V$ is the valley Zeeman/sublattice potential term, which breaks the valley degeneracy in MLG and favors the $K$ valley over the $K'$ valley in the noninteracting limit. 
As explained in the introduction, this term is usually generated 
from the partial misalignment of the substrate layer, (such as HBN layer) with the graphene layer~\cite{KV_DGG_2013,KV_expt_2013,KV_HBN_Jung2015,KV_HBN_Jung2017}. 

At this point, it is useful to look at the symmetries of the Hamiltonian. For $E_Z=E_V=0$ the Hamiltonian is invariant under $SU(2)_s\otimes U(1)_v\otimes Z_{2v}$, where the subscripts $s,v$ stand for spin and valley respectively. Once one allows for nonzero $E_Z,E_V$, the symmetry reduces to $U(1)_s\otimes U(1)_v$. It is also worth noting that while the $U(1)_s$ symmetry holds very generally, the $U(1)_v$ symmetry is valid only for interactions at the four-fermion level~\cite{alicea2006:gqhe}. Once one includes six-fermion interactions, the conservation of momentum up to a reciprocal lattice vector will reduce the $U(1)_v$ symmetry to $Z_{3v}$. This has the important consequence that when the $U(1)_v$ symmetry is spontaneously broken, the would-be Goldstone modes will be gapped by the reduction of the continuous symmetry $U(1)_v$ to the discrete $Z_{3v}$.

Using the HF approximation and restricting to translational invariant ground states up to an intervalley coherence, the bulk HF Hamiltonian from Eq.~\ref{eq:Ha_bulk} can be expressed as
\begin{align}
    \mathcal{H}_{HF}^{bulk} = \sum_{i,j} \big[\sum_{m,n} (V_{ijmn}-V_{inmj})\Delta_{mn} -V_{ex}\Delta_{ji} \big]c^\dagger_i c_j \nonumber \\
    -\vec{c}^{\hspace{0.1cm}\dagger}(E_Z \sigma_z + E_V \tau_z) \vec{c} 
    \hspace{2.5cm}
    \label{eq:HFHa_bulk}
\end{align}
where $V_{ijmn}=\sum_{\alpha} u_\alpha (\tau_\alpha)_{ij} (\tau_\alpha)_{mn}$ and $V_{ex}=\int \frac{E_c}{2\pi} ~\ell^2 d^2q \frac{e^{-[\frac{q^2\ell^2}{2}]}}{\sqrt{q^2\ell^2 + q_0^2}}$ is the exchange contribution of the screened Coulomb interaction. The Hamiltonian is written in the basis $\vec{c}^{\hspace{0.1cm}\dagger}=(c_{K\ua}^\dagger,c_{K\da}^\dagger,c_{K'\ua}^\dagger,c_{K'\da}^\dagger)$. $\Delta$ is completely known in terms of the one-body expectation values $\Delta_{ij}=\langle \text{HF}| c^\dagger_i c_j |\text{HF} \rangle$ for the Slater determinant ground state $|\text{HF} \rangle$.

Now, let us look at the single-particle HF energies of the various levels at $\nu=\pm 1$. For $\nu=\pm 1$ with $E_Z, E_V > 0$, the electron spin and valley indices are good quantum numbers so the projector $\Delta$ for the occupied state is diagonal in the basis $({K\uparrow},{K\downarrow},{K'\uparrow},{K'\downarrow})^T$.
For  bulk $\nu=-1$ the occupied level is ${K\uparrow}$, implying $\Delta=diag(1,0,0,0)$, leading to the single-particle HF energies
\begin{align}
    E_{K\uparrow}=-V_{ex}-E_V-E_Z, \nonumber \\
    E_{K\downarrow}=-E_V+E_Z+u_z, \nonumber \\
    E_{K'\uparrow}=E_V-E_Z-u_z-2u_{xy}, \nonumber \\
    E_{K'\downarrow}=E_V+E_Z-u_z
\end{align}
For  bulk $\nu=1$ the unoccupied state is ${K'\downarrow}$, implying  $\Delta=diag(1,1,1,0)$, leading to the single-particle HF energies
\begin{align}
    E_{K\uparrow}=-V_{ex}-E_V-E_Z-2u_{xy}, \nonumber \\
    E_{K\downarrow}=-V_{ex}-E_V+E_Z, \nonumber \\
    E_{K'\uparrow}=-V_{ex}+E_V-E_Z-2u_z-2u_{xy}, \nonumber \\
    E_{K'\downarrow}=E_V+E_Z-u_z-2u_{xy}
\end{align}

One of the central thrusts of this paper is to examine magnon transmission as  $B_\perp$ changes, while maintaining the filling fractions across the junction. This is very reasonable experimentally \cite{Magnon_transport_Zhou_2022}. All the parameters of the Hamiltonian(\ref{eq:Ha_bulk}) except $E_V$ change with $\Bp$. The couplings in the Hamiltonian depend on $\Bp$ as follows~\cite{kharitonov2012:nu0}, 
\begin{align}
    E_c \propto \frac{1}{\ell} = \sqrt{\Bp}~ E_c^0,
    \nonumber \\
    u_\alpha \propto \frac{1}{\ell^2} = \Bp~ u_\alpha^0,
    \nonumber \\
    E_Z = \Bp~ E_Z^0
    \label{eq:Bp_dependence}
 \end{align}
where $B_\perp$ is in Tesla,  and $E_c^0$,  
$u_\alpha^0$, and $E_Z^0$ are the strengths of the parameters at a reference perpendicular field of $B_\perp^0=1$ $Tesla$. We keep $B_\perp^0$ general in what follows to maintain flexibility. As we will show in the section on results (Section \ref{sec:results}), the magnon transmission amplitude depends strongly on the structure of the junctions and the intermediate region, which in turn is determined by the ordering of single-particle levels in the three regions. Since this is an important finding of our paper we will illustrate it here in some detail with examples. 

Let us focus on the energy differences between the single-particle energies of the HF levels. For $\nu=1$, the energy differences between the occupied levels are 
\begin{eqnarray}
    E_{K\uparrow}-E_{K'\uparrow}&=&2(\Bp u_z^0 -E_V)
    \nonumber \\
    E_{K\downarrow}-E_{K'\uparrow}&=&2(\Bp[E_Z^0+(u_z^0-|u_{xy}^0|)]-E_V)
    \nonumber \\
    E_{K\uparrow}-E_{K\downarrow}&=&2\Bp(|u_{xy}^0|-E_Z^0)
\end{eqnarray}

\begin{figure}[h]
    \centering
    \includegraphics[width=0.45\textwidth,height=5cm]{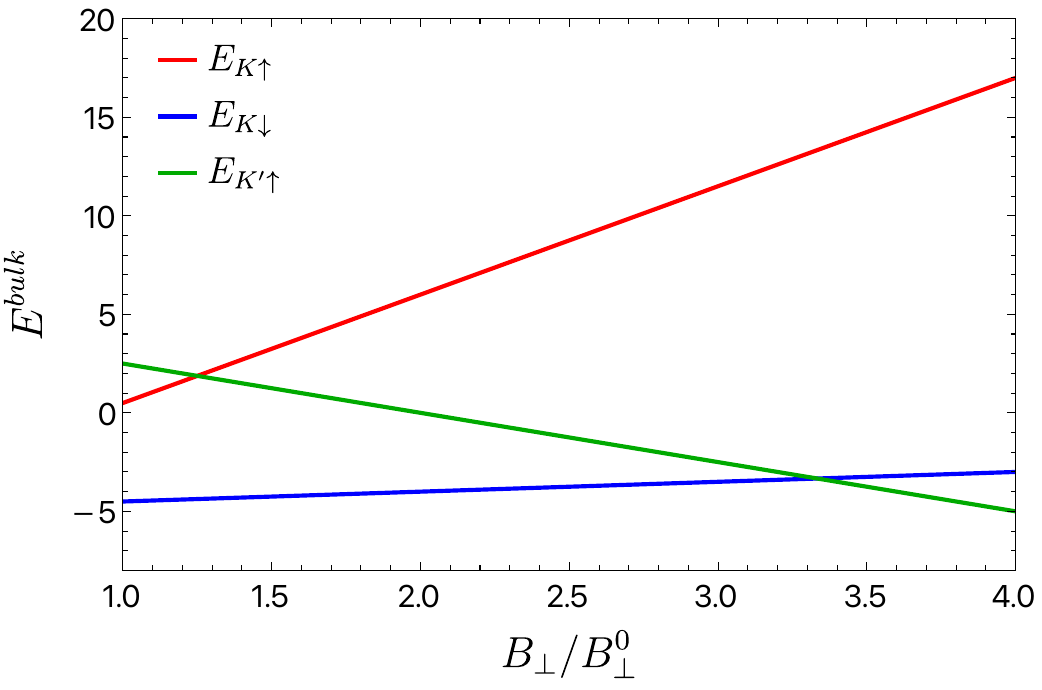}
    \caption{HF energies of the occupied levels for  bulk $\nu=1$ as a function of $\Bp/B_\perp^0$ with  $V_{ex}=0$,~$E_Z^0=0.5$,~$E_V=5$,~$u_z^0=4$ and $u_{xy}^0=-3$ i.e. the case when $E_V >~(u_z^0 - |u_{xy}^0|) >~E_Z^0$. For $\Bp/B_\perp^0 \simeq 1$,  the ordering of the occupied levels is given by $E_{K\downarrow} < E_{K\uparrow} <E_{K'\uparrow}$. As we increase $\Bp$, the energy of the state ${K'\uparrow}$ decreases linearly while the other energies increase, thus changing the order of the occupied levels.}
    \label{fig:bulk_ordering_uz_g_uxy_Ev_5}
\end{figure}

Since there are multiple energy scales involved, we will consider two illustrative cases, leaving the detailed investigation of all the possibilities to later sections. In both cases we will look only at what is believed to be the physical region of anisotropic couplings, given by $u_z>0,\ u_{xy}<0$. In Fig.~ (\ref{fig:bulk_ordering_uz_g_uxy_Ev_5})  we focus on the case $u_z>|u_{xy}|$. Assuming that the field is perpendicular, and that $u_z^0-|u_{xy}^0|>E_Z^0$. Finally, assuming a partially aligned HBN substrate, we take $E_V>u_z^0-|u_{xy}^0|$. As seen in Fig.~ (\ref{fig:bulk_ordering_uz_g_uxy_Ev_5}), there are three different orderings as a function of $B_\perp$. For  $B_\perp<E_V/u_z^0$ (roughly $1.3T$ for the parameters chosen) the ordering of the levels is  $E_{{K\downarrow}} < E_{{K\uparrow}} <E_{{K'\uparrow}}$. For intermediate values $\frac{E_V}{E_Z^0+(u_z^0-|u_{xy}^0|)}>B_\perp>\frac{E_V}{u_z^0}$, the ordering becomes $E_{{K\downarrow}}<E_{{K'\uparrow}}<E_{{K\uparrow}}$. Finally, for large values of $B_\perp$,  beyond roughly  $3.4T$ for the parameters chosen, the ordering is $E_{{K'\uparrow}}<E_{{K\downarrow}}<E_{{K\uparrow}}$.

\begin{figure}[h]
    \centering
    \includegraphics[width=0.45\textwidth,height=5cm]{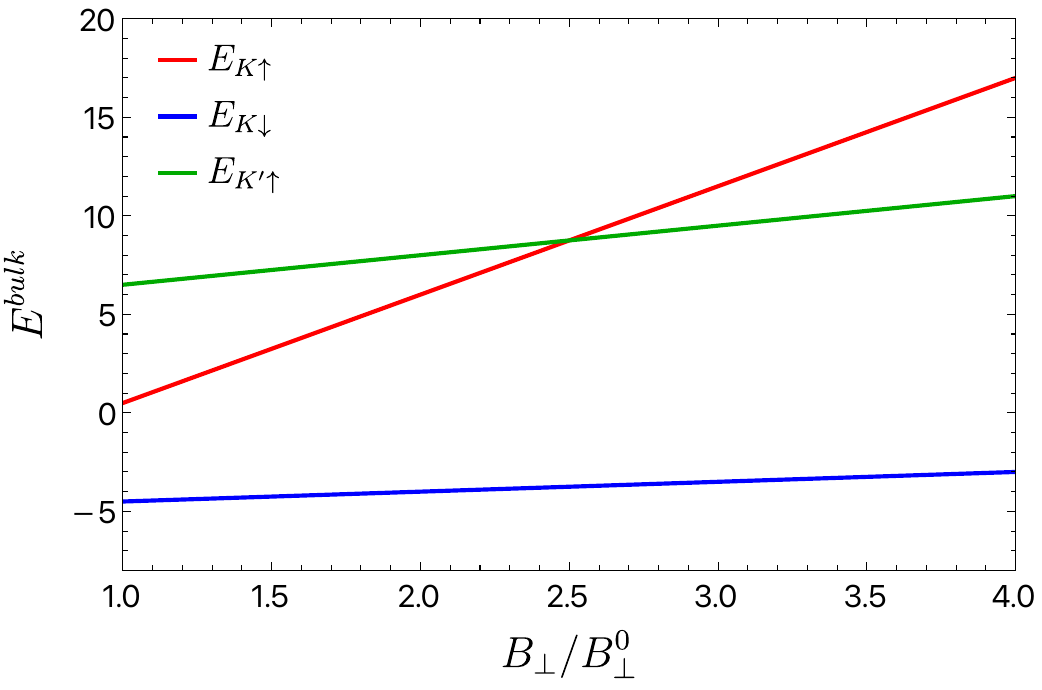}
    \caption{HF energies of the occupied levels for bulk $\nu=1$ as a function of $\Bp/B_\perp^0$ with  $V_{ex}=0$,~$E_Z^0=0.5$,~ $E_V=5$,~$u_z^0=2$ and $u_{xy}^0=-3$ i.e. the case when $E_V >(|u_{xy}^0|-u_z^0) >E_Z^0$. For $\Bp/B_\perp^0 \simeq 1$,  the ordering of the occupied levels are again $E_{K\downarrow} < E_{K\uparrow} <E_{K'\uparrow}$, but as we increase $\Bp$, the energies of all the occupied levels increase. We find that  the ordering of the occupied levels changes only once, in contrast to Fig.~(\ref{fig:bulk_ordering_uz_g_uxy_Ev_5}) where it changes twice.}
    \label{fig:bulk_ordering_uxy_g_uz_Ev_5}
\end{figure}

As a second illustrative example, let us consider a case where, at $E_Z=E_V=0$, the system would have been in the BO state in the physical region of anisotropic couplings, that is, $u_z<|u_{xy}|$. We will consider a sceniario similar to the one in the previous paragraph, with $E_V > (|u_{xy}^0|-u_z^0) > E_Z^0$.  Now there are only two different types of orderings of single-particle energy levels, as seen in Fig.~ (\ref{fig:bulk_ordering_uxy_g_uz_Ev_5}). For $B_\perp<E_V/u_z^0$ (roughly $2.5T$ for the parameters chosen) the ordering is $E_{K\downarrow}<E_{K\uparrow}<E_{K'\uparrow}$. For larger $B_\perp$ the ordering switches to $E_{K\downarrow}<E_{K'\uparrow}<E_{K\uparrow}$. 

From these examples it is clear that for fixed  $E_V$ and purely perpendicular field, the ratio of  $u_z$ and $u_{xy}$ (that is, whether $\frac{u_z}{|u_{xy}|} > 1$ or $\frac{u_z}{|u_{xy}|} < 1$) as well as the magnitude of $B_\perp$ control the energetic ordering of filled levels at $\nu=1$. While any static ground state average will not be affected by the ordering of the filled levels, we will show that this ordering has drastic effects on the junctions, and thence on the magnon transmission.

Thus far we have focused on the bulk. Now we will describe the full inhomogeneous Hamiltonian of our system with the $\nu=1|-1|1$ sandwich and explicitly show how the bulk energetic ordering of occupied levels in $\nu=1$ and the strength of $B_\perp$ affect the self-consistent HF results.

\section{Hamiltonian and HF approximations with junctions}
\label{sec:Hamiltonian_junction_HF}
 The  Hamiltonian of the system in the presence of $\nu=1|-1|1$ sandwich is almost identical to Eq.~\ref{eq:Ha_bulk}, the only difference being the positive background charges enforcing the different fillings in the different regions of the sandwich. 
\begin{align}
    \mathcal{H} = \frac{\pi \ell^2}{A} \big[ \sum_{k_1,k_2,\vec{q}} \hspace{0.2cm} \sum_{\alpha=x,y,z} e^{i(\phi(k_1,\vec{q})+\phi(k_2,-\vec{q}))} e^{-\frac{(q^2 \ell^2)}{2}} \hspace{0.5cm}
    \nonumber \\
    \times \hspace{0.1cm} v_\alpha(q) : (\vec{c}_{k_1-q_y}^{\hspace{0.1cm} \dagger} \tau_\alpha \vec{c}_{k_1}) (\vec{c}_{k_2+q_y}^{\hspace{0.1cm} \dagger} \tau_\alpha \vec{c}_{k_2}) : + \hspace{0.6cm}
    \nonumber \\ 
    \sum_{\vec{q}} V(q) : (\rho(\vec{q}) - \rho_b(\vec{q})) (\rho(-\vec{q}) - \rho_b(-\vec{q})) : \big] \hspace{0.6cm}
    \nonumber \\
    - \sum_{k} E_Z \vec{c}_k^{\hspace{0.1cm} \dagger} \sigma_z \vec{c}_k 
    - \sum_{k} E_V \vec{c}_k^{\hspace{0.1cm} \dagger} \tau_z \vec{c}_k \hspace{2cm} 
    \label{eq:Ha_junc}
\end{align}
where $\phi(k,\vec{q})=\ell^2(-q_x k + \frac{1}{2} q_x q_y)$, and $\rho(\vec{q})$ is the Fourier transform of the electron density $\rho_0(x,y)=\psi_0(x,y)^\dagger \psi_0(x,y)$ for the ZLL. $\rho_b(\vec{q})$ is the Fourier transform of the positive background charge density, which we choose to be 
\begin{align}
    \rho_{b}(x,y) = \frac{1}{2 \pi \ell^2}
    \begin{cases}
    3 ,& x < -W/2 \\
    1, & -W/2 \leq x \leq W/2 \\
    3, & x > W/2
    \end{cases}
    \label{eq:bg}
\end{align}
Note that $\rho_b$  is independent of $y$. As can be inferred from Eq.~\ref{eq:bg}, the positive background ``tries" to maintain a filling of $\nu=1$ (three of four ZLLs filled) for $|x|>W/2$, and a filling of $\nu=-1$ (one of four ZLLs filled) for the region $|x|<W/2$. The edges are sharp, that is, there is an abrupt change in the background charge density at $\pm W/2$. As we know from previous work, smooth edge potentials can induce edge reconstructions~\cite{EdgeR_MacD_1990,EdgeR_First_1992,EdgeR_Dempsey_etal_1993,EdgeR_Chamon_Wen_1994,EdgeR_Meir_1994,MSF2014,EdgeR_Mode_Switch_2017,EdgeR_Amartya2021}, a complication that we do not want here. The width of the middle region is fixed by the device geometry, and does not change with $B_{\perp}$. At the reference value  $B_\perp^0=1~Tesla$, the dimensionless width of the middle region is given by ${\tilde W}^0=W/\ell^0$. As $B_\perp$ increases the dimensionless width ${\tilde W}={\tilde W}^0\sqrt{B_\perp}$ increases. This affects the nature of the interfaces and consequently the amplitude of magnon transmission through the system. 

As with the Hamiltonian of Eq.~\ref{eq:Ha_bulk}, this Hamiltonian also has the symmetry group $U(1)_s\otimes U(1)_v$, with the usual caveat about would-be valley Goldstone modes becoming gapped when six-fermion interactions are included. 

In the HF approximation, one reduces the two-body interaction terms in the Hamiltonian(\ref{eq:Ha_junc}) to one-body terms generated by taking averages assuming a single Slater determinantal (SSD) state. Each SSD can be uniquely characterized by all possible one-body averages. In our problem, assuming that translation invariance in the $y$-direction is not broken spontaneously, these averages are
\begin{align}
    \langle c_{k_1 i}^\dagger c_{k_2 j} \rangle = \delta_{k_1,k_2} \Delta_{ij}(k_1)
    \label{eq:Delta}
\end{align}
$i,j$ runs from $1$ to $4$ and denotes the four possible nearly degenerate ZLLs. The inhomogeneity in the problem manifests itself as a nontrivial dependence of the $\Delta$ matrices on the guiding center index $k$. We will make use of the $U(1)_s\otimes U(1)_v$ symmetry to rotate the state in the spin and valley spaces so as to make the $\Delta$ matrices real. We get the following HF Hamiltonian in the basis $\vec{c}_k=(c_{k K\ua},c_{k K\da},c_{k K'\ua},c_{k K'\da})^T$ as
\begin{align}
    \mathcal{H}_{HF}= \frac{\ell}{L_y} \sum_{k_1,k_2} \sum_{i,j} \big[ \sum_{m,n} \big( V_{ijmn}(k_1-k_2) ~- \nonumber \\ 
    V_{inmj}(k_1-k_2) \big) \Delta_{mn}(k_2) c_{k_1 i}^\dagger c_{k_1 j} ~+
    \nonumber \\
    V_H(k_1-k_2) \Delta_{jj}(k_2) c_{k_1 i}^\dagger c_{k_1 i} ~-
    \nonumber \\
    V_F(k_1-k_2) \Delta_{ji}(k_2) c_{k_1 i}^\dagger c_{k_1 j} \big] ~- 
    \nonumber \\
    \sum_k V_{bg}(k) \vec{c}_{k}^{\hspace{0.1cm} \dagger} \vec{c}_{k}
    - \sum_{k} E_Z \vec{c}_k^{\hspace{0.1cm} \dagger} \sigma_z \vec{c}_k 
    - \sum_{k} E_V \vec{c}_k^{\hspace{0.1cm} \dagger} \tau_z \vec{c}_k 
    \label{eq:HFHa_junc}
\end{align}
where,
\begin{align}
    V_{ijmn}(k)= \sqrt{2 \pi} ~ e^{-[\frac{k^2 \ell^2}{2}]} \sum_\alpha u_\alpha (\tau_\alpha)_{ij} (\tau_\alpha)_{mn} , 
    \nonumber \\ 
    V_H(k)=\int_{-\infty}^\infty \ell dq ~ V(q,0) e^{-[\frac{q^2 \ell^2}{2}]} \cos(q k \ell^2) , 
    \nonumber \\ 
    V_F(k)=\int_{-\infty}^\infty \ell dq ~ V(q,k) e^{-[\frac{(q^2 + k^2) \ell^2}{2}]}  
\end{align} 
and 
\begin{align}
    V_{bg}(k)= 2\ell^2 \int_0^\infty dq V(q,0) e^{-[\frac{q^2 \ell^2}{4}]} {\text {Re}}(e^{-iq k \ell^2} \rho_b(-q)) .
\end{align}
As usual in applications of HF~\cite{EdgeR_Mode_Switch_2017,EdgeR_Amartya2021,Wei_Huang_MacDonald2021}, one starts with a ``seed" configuration of the $\Delta$ matrices. The HF Hamiltonian is diagonalized, states below the chemical potential are occupied, and the resulting ground state is used to find an improved set of $\Delta$ matrices. The process is repeated until self-consistency is achieved, in the sense that the $\Delta$ matrices on the next step match the $\Delta$ matrices on the previous step to some desired level of precision. Once self-consistency has been achieved, the eigenvalues of the $\Delta$ matrix at every $k$ can only be $0$ or $1$, representing the occupations of the energy levels at that $k$. During the iterative process, the chemical potential is maintained such that the system is charge neutral overall.

\begin{figure}[h]
    \centering
    \includegraphics[width=0.45\textwidth,height=7cm]{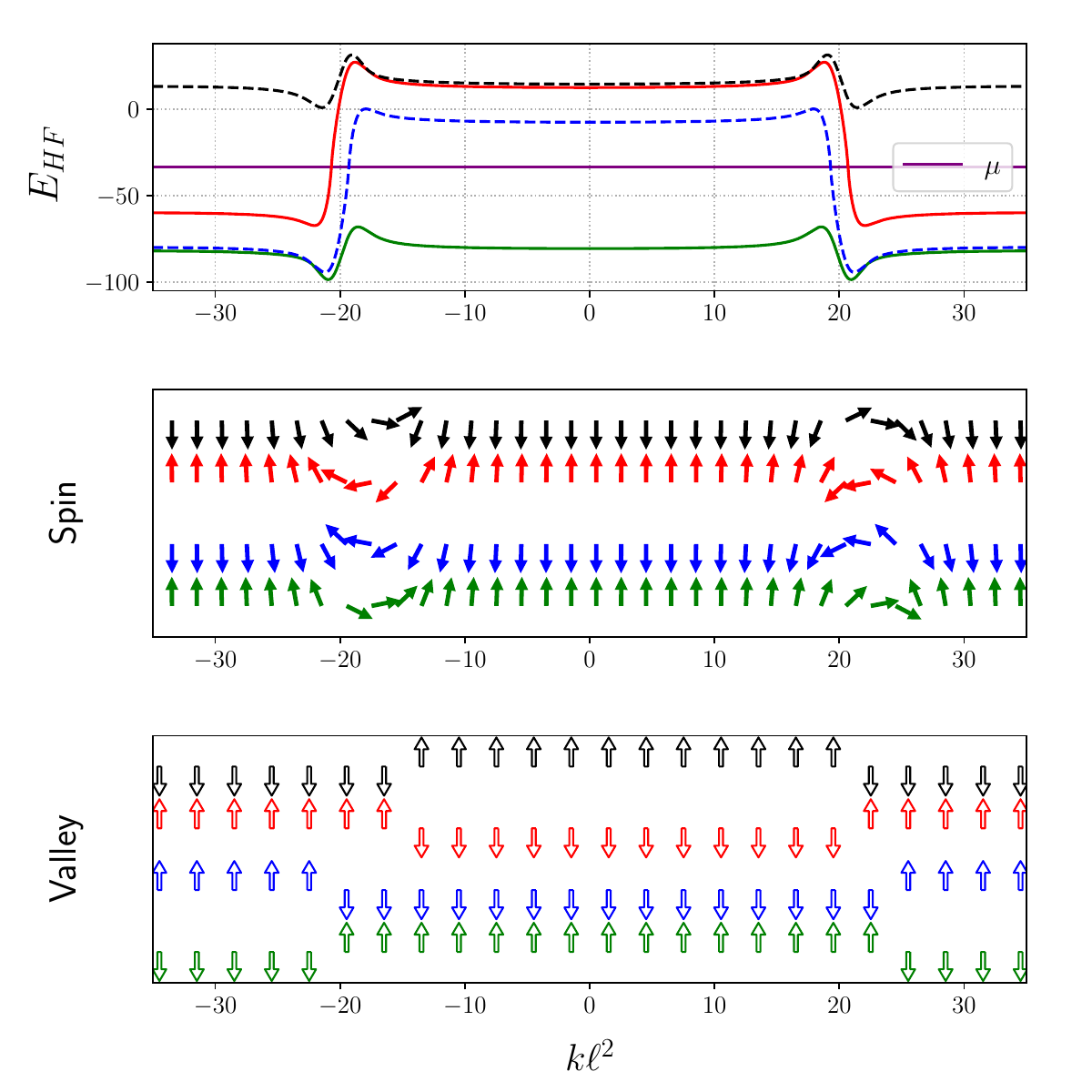}
    \caption{HF energies (top panel), spin structure (middle panel), and valley structure (bottom panel) of each self-consistent HF level for $\Bp=4B_\perp^0$. The Hamiltonian parameters  are $\Wt^0=20,~ E_C^0=30,~E_Z^0=0.5,~ E_V=5.0,~u_z^0=4,~u_{xy}^0=-3$. The spin and valley structure of each HF level is shown by plotting the vectors $\vec{\sigma}=(\langle\sigma_x\rangle, \langle\sigma_z\rangle)$(thin arrows for spin) and  $\vec{\tau}=(\langle\tau_x\rangle,\langle\tau_z\rangle)$(thick arrows for valley). $\langle\sigma_z\rangle=+1,-1$ is identified with $\uparrow$,$\downarrow$ spin and $\langle\tau_z\rangle =+1,-1$ with the valleys  $K$,$K'$ respectively. 
     The direction of the arrows of spin and valley represent the orientation of the averages of $\vec{\tau},~\vec{\sigma}$ in the $xz$ plane of each internal space.  The colors of the arrows are the same as that of the corresponding energy levels.  For the parameters chosen,  the bulk $\nu=1$ ordering is $E_{K'\uparrow} < E_{K\downarrow} < E_{K\uparrow} < E_{K'\downarrow}$ and the self-consistent ground state  prefers many-body spin rotation, {whereas} the valley indices of each level flip discontinuously  across the interfaces.}
    \label{fig:HF_uz_g_uxy}
\end{figure}

We will present some HF results here to explicitly show how the bulk ordering in $\nu=1$, which is dictated by the ratio of $u_z^0$ and $u_{xy}^0$ and $\Bp$ (for fixed $E_V, E_Z^0$), determines the self-consistent HF state of the junction. In Landau gauge, the momentum $k$ along the periodic direction $y$ is related to the guiding center position $X_k=k \ell^2$ along the $x$-direction. The figures we present in what follows show the HF single-particle energies and the spin-valley directions of each HF state as functions of the guiding center position $X_k$. Since we have used the $U(1)_s\otimes U(1)_v$ symmetries to make $\Delta$ real, the averages of $\vec{\tau},~\vec{\sigma}$ lie in the $xz$ plane in each internal space. We will simply present the directions as arrows with an $\uparrow$ representing $K$ in the case of valley, and $\uparrow$ representing $\langle \sigma_z \rangle = 1$ in the case of spin.

\begin{figure}[h]
    \centering
    \includegraphics[width=0.45\textwidth,height=7cm]{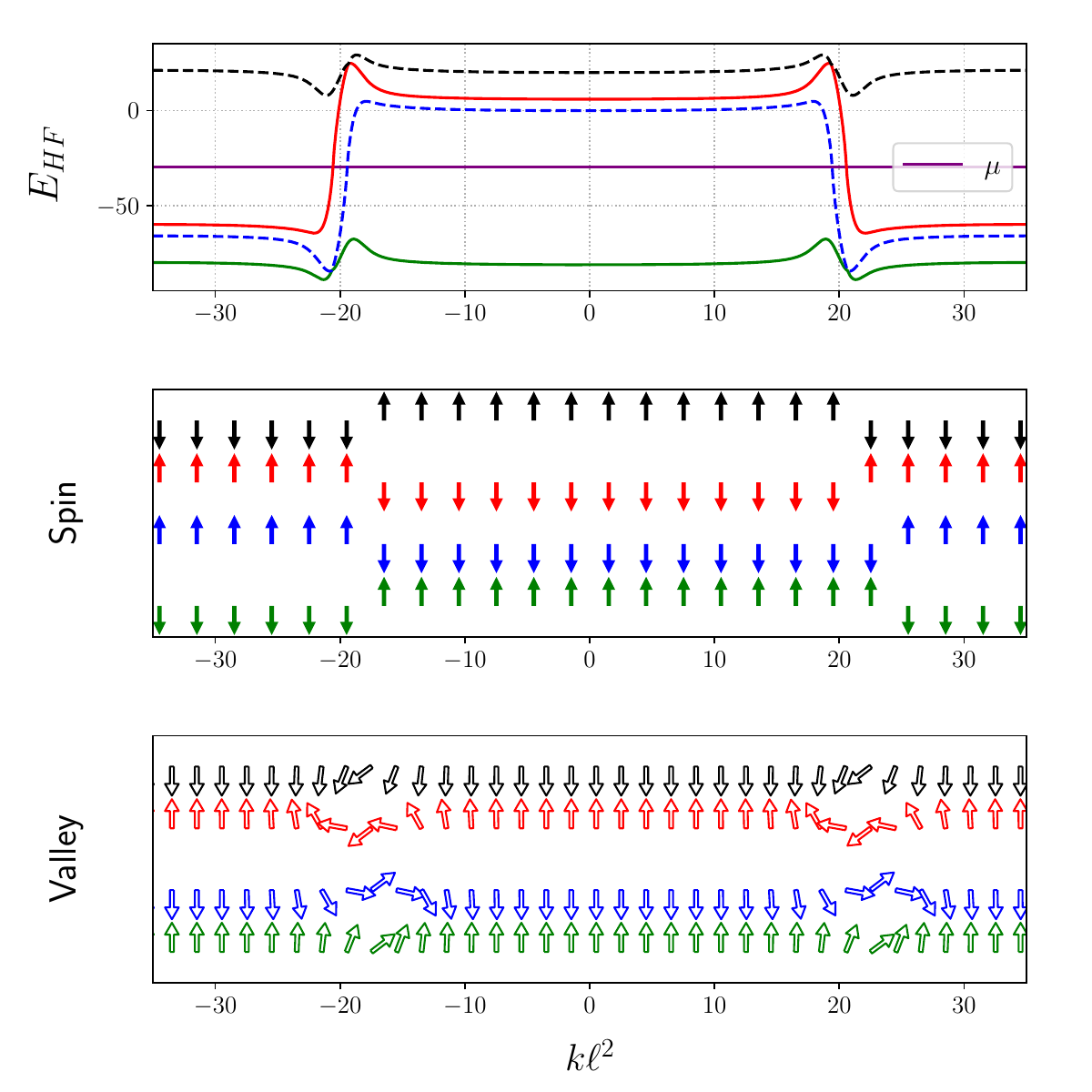}
    \caption{HF energies (top panel), spin structure (middle panel), and valley structure (bottom panel) of each self-consistent HF level for $\Bp=4B_\perp^0$. The Hamiltonian parameters are $\Wt^0=20,~ E_C^0=30,~E_Z^0=0.5,~E_V=5.0,~u_z^0=2,~u_{xy}^0=-3$. We have used the same convention to plot the spin and valley structure here as in Fig.~(\ref{fig:HF_uz_g_uxy}), with thin arrows for spin and thick arrows for valley. In this case,  the bulk ordering of the energy levels $E_{K\downarrow} < E_{K'\uparrow} < E_{K\uparrow} < E_{K'\downarrow}$, is different from the case in Fig.~(\ref{fig:HF_uz_g_uxy}). Here, we find self-consistently that the ground state prefers valley rotation and the spin flips discontinuously across the interfaces.}
    \label{fig:HF_uxy_g_uz}
\end{figure}

First, consider the case where $u_z^0>|u_{xy}^0|$, shown in Fig.~(\ref{fig:HF_uz_g_uxy}). We have chosen the parameters $\Bp/B_\perp^0=4$ with the other parameters identical to those of   Fig.~(\ref{fig:bulk_ordering_uz_g_uxy_Ev_5}).  The ordering of the HF levels deep in the bulk of $\nu=1$ is $E_{K'\uparrow} < E_{K\downarrow} < E_{K\uparrow} < E_{K'\downarrow}$, whereas the filled state for bulk  $\nu=-1$ is ${K\uparrow}$. As can be seen from the directions of the spin and valley for each single-particle state, the system prefers to spontaneously break the  $U(1)_s$ symmetry at each interface, rotating the spins continuously. However, the valley degree of freedom remains polarized either at $K$ or $K'$, keeping the $U(1)_v$ symmetry intact. A level crossing occurs at the interface between the two lowest levels, discontinuously exchanging their valley polarizations.

Next, let us consider the case $u_z^0<|u_{xy}^0|$, shown in Fig.~(\ref{fig:HF_uxy_g_uz}). For $\Bp/B_\perp^0=4$ with the other parameters being chosen as in Fig.~(\ref{fig:bulk_ordering_uxy_g_uz_Ev_5}),  the ordering of the HF levels in the $\nu=1$ bulk is  $E_{K\downarrow} < E_{K'\uparrow} < E_{K\uparrow} < E_{K'\downarrow}$, whereas the filled state for bulk $\nu=-1$ is ${K\uparrow}$. Now a spontaneous breaking of the $U(1)_v$ symmetry occurs, and there is a continuous rotation of the valley polarization of each single-particle state. However, the spin directions remain frozen at either $\uparrow$ or $\downarrow$, with an abrupt change occuring at the interfaces via a level crossing between the two lowest levels. 

The above two examples show the importance of the ratio $u_z^0/|u_{xy}^0|$ to the structure of the interface. We will present many more examples in Section \ref{sec:results} and show how the structure of the interfaces in turn affects magnon transmission across the junction. However, we will first examine the collective excitations of the system and review the magnon scattering formalism~\cite{Wei_Huang_MacDonald2021} using the time-dependent Hartree-Fock (TDHF) method.

\begin{figure*}
    \centering
    \begin{subfigure}{0.33\textwidth}
        \caption{$\Bp=B_\perp^0$}
        \includegraphics[width=\textwidth,height=7cm]{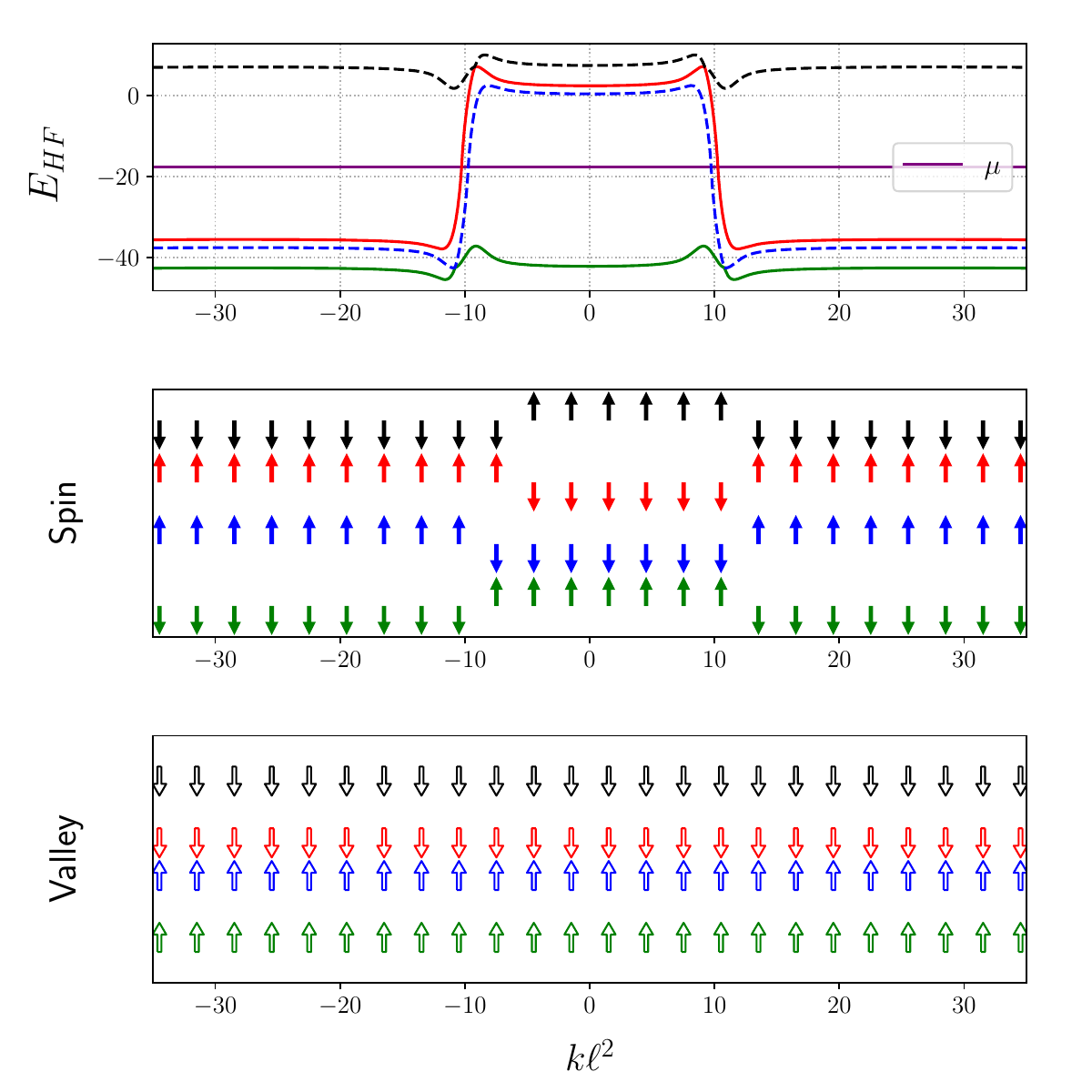}
    \end{subfigure}
    \hspace{-0.2cm}
    \begin{subfigure}{0.33\textwidth}
        \caption{$\Bp=2B_\perp^0$}
        \includegraphics[width=\textwidth,height=7cm]{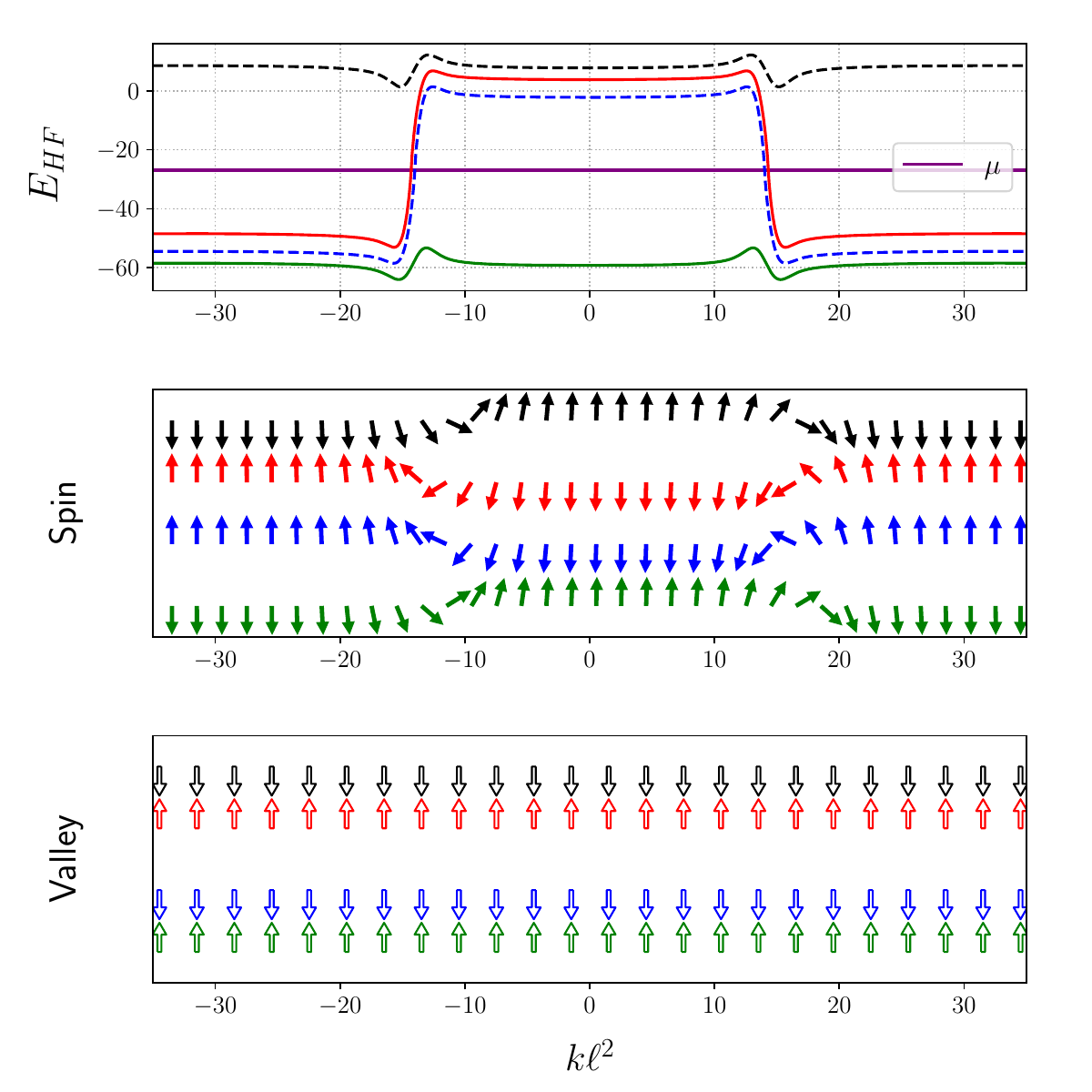}
    \end{subfigure}
    \hspace{-0.2cm}
    \begin{subfigure}{0.33\textwidth}
        \caption{$\Bp=4B_\perp^0$}
        \includegraphics[width=\textwidth,height=7cm]{HF_B_perb_4.0_gxy_-12.0_gz_16.0_Ez_2.0_Ev_5.0.pdf}
    \end{subfigure}
    \caption{Hartree-Fock energies (first row), spin structure (second row), and valley structure (third row) of each energy level for three different values of the perpendicular magnetic field $B_\perp^0,2B_\perp^0,4B_\perp^0$. The colors of the arrows for the spin and valley are the same as that of the corresponding energy levels. At $B_\perp^0$ the Hamiltonian parameters are $\Wt^0=20, E_C^0=30, u_z^0=4.0, u_{xy}^0=-3.0, E_Z^0=0.5, E_V=5.0$, identical to Fig.~(\ref{fig:bulk_ordering_uz_g_uxy_Ev_5}). We have chosen the values of $\Bp$ such that we sample all three different regions of bulk ordering shown in Fig.~(\ref{fig:bulk_ordering_uz_g_uxy_Ev_5}).  For $\Bp=B_\perp^0$ (first column) both the spin and valley indices of each HF level remain a good quantum number but with increasing $\Bp$ (second and third columns) the system prefers a spin rotated ground state close to the interfaces while the valley remains a good quantum number.}
    \label{fig:HF_uz_g_uxy_Ev_5}
\end{figure*}

\begin{figure*}
    \centering
    \begin{subfigure}{0.33\textwidth}
        \caption{$\Bp=1B_\perp^0$}
        \includegraphics[width=\textwidth,height=4cm]{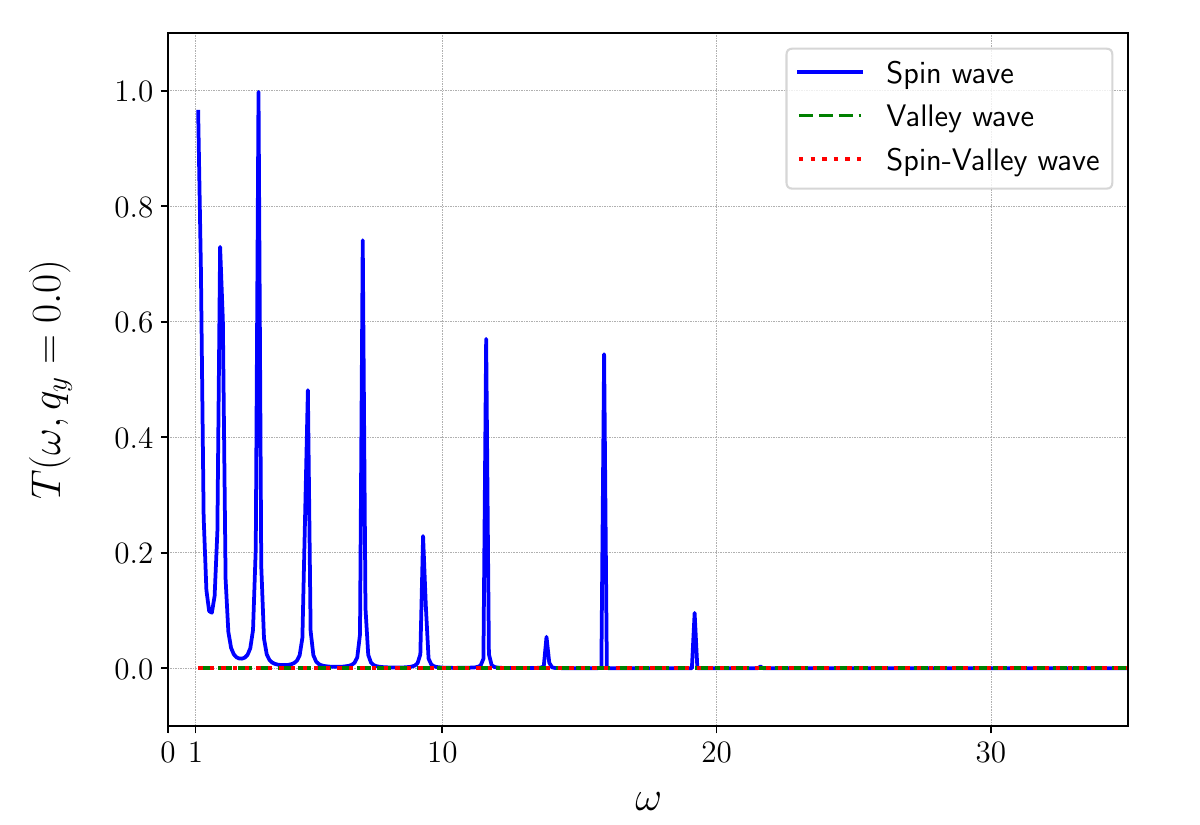}
    \end{subfigure}
    \hspace{-0.2cm}
    \begin{subfigure}{0.33\textwidth}
        \caption{$\Bp=2B_\perp^0$}
        \includegraphics[width=\textwidth,height=4cm]{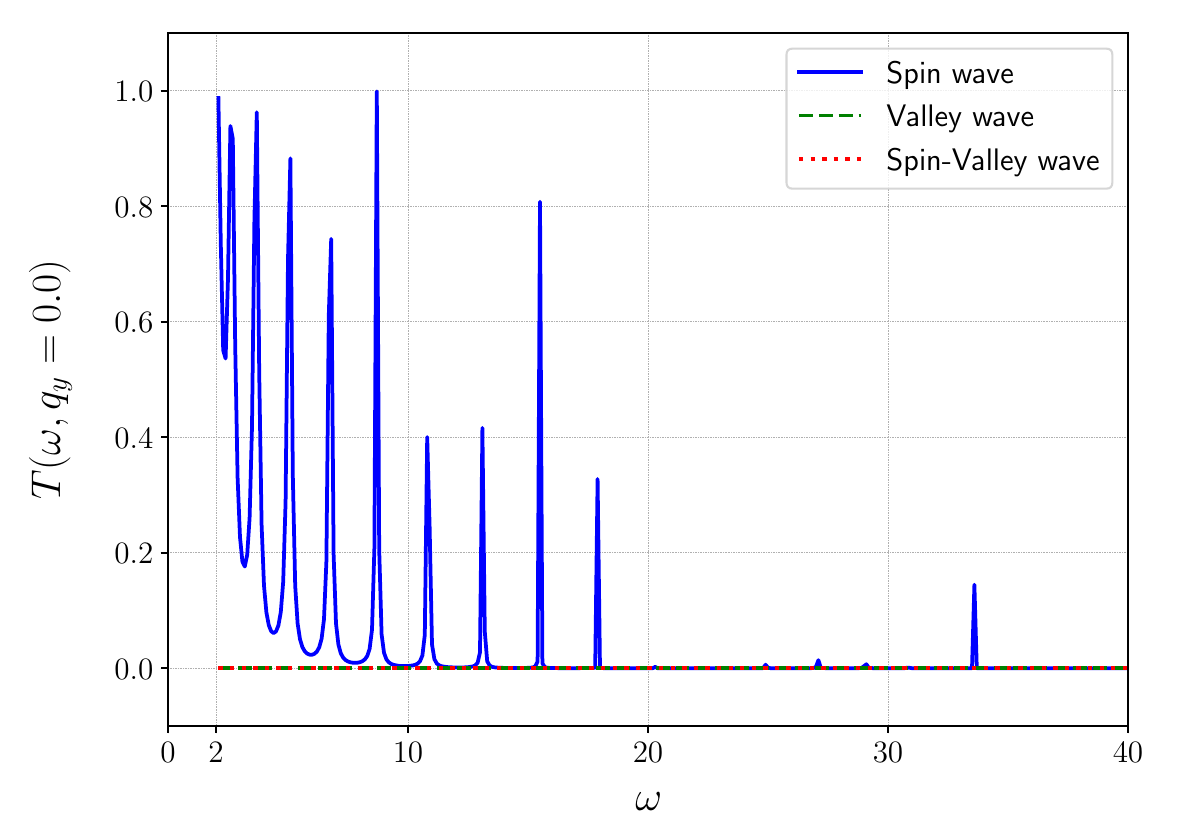}
    \end{subfigure}
    \hspace{-0.2cm}
    \begin{subfigure}{0.33\textwidth}
        \caption{$\Bp=4B_\perp^0$}
        \includegraphics[width=\textwidth,height=4cm]{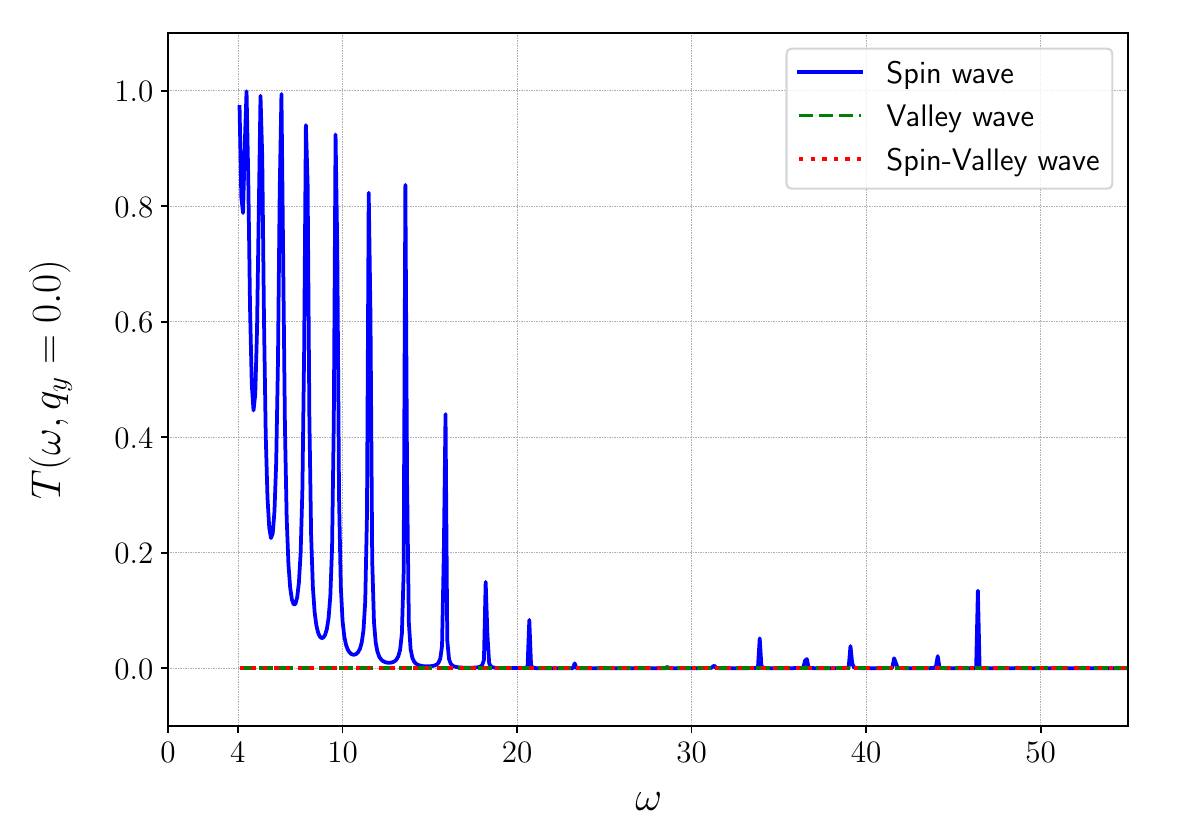}
    \end{subfigure}
    \caption{Transmission amplitudes as a function of incoming magnon energy at $q_y=0$(normal incidence) corresponding to the HF state in Fig.~(\ref{fig:HF_uz_g_uxy_Ev_5}). At $B_\perp^0$ the Hamiltonian parameters are $\Wt^0=20, E_C^0=30, u_z^0=4.0, u_{xy}^0=-3.0, E_Z^0=0.5, E_V=5.0$. For this set of parameters, the transmission amplitude of the spin magnon mode is strongly suppressed at higher energies and consists of some low energy peaks. The transmission behaviors remain qualitatively the same as $\Bp$ increases. Both the valley-wave and the spin-valley wave modes remain inactive throughout.}
    \label{fig:TDHF_uz_g_uxy_Ev_5}
\end{figure*}

\begin{figure*}
    \centering
    \begin{subfigure}{0.33\textwidth}
        \caption{$\Bp=1B_\perp^0$}
        \includegraphics[width=\textwidth,height=7cm]{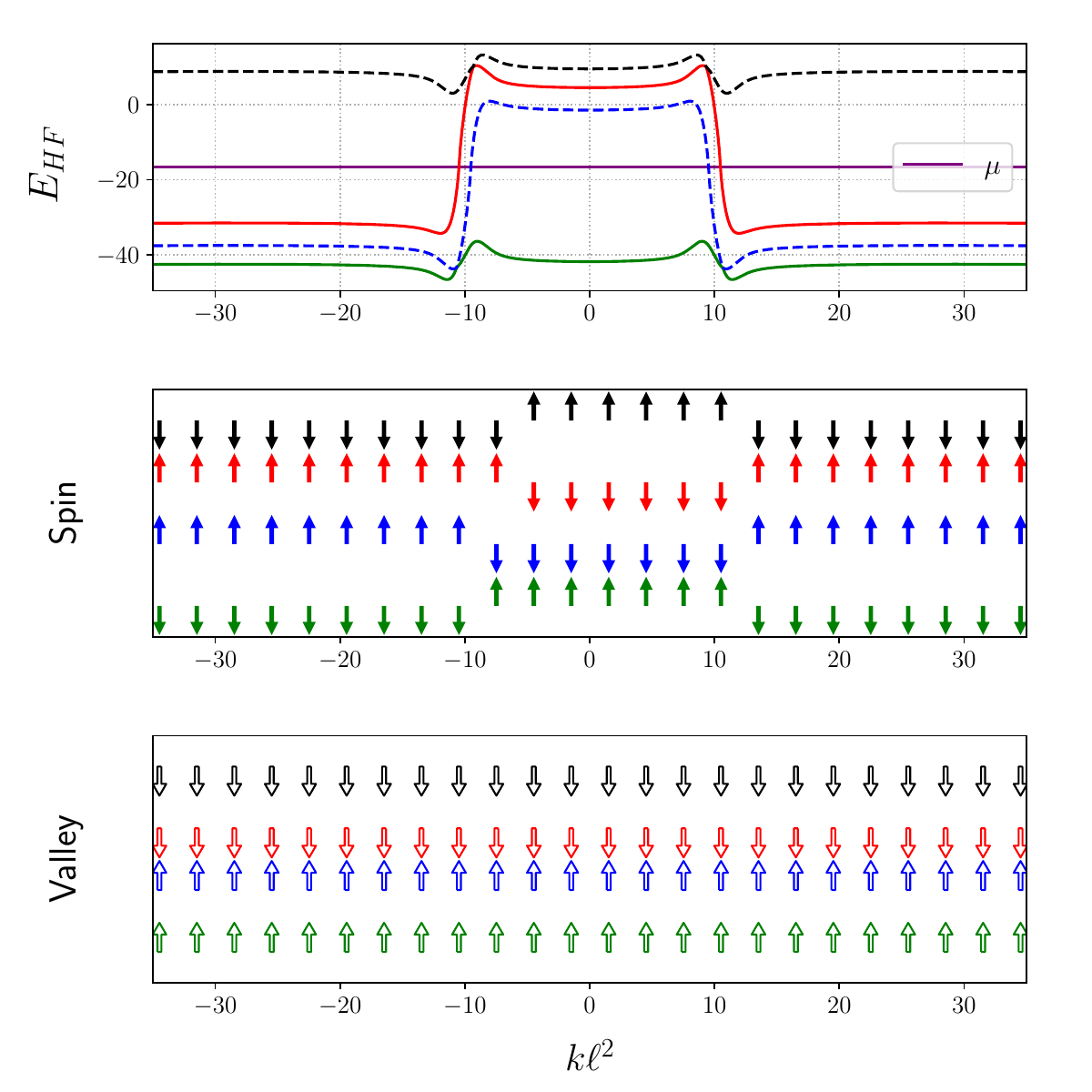}
    \end{subfigure}
    \hspace{-0.2cm}
    \begin{subfigure}{0.33\textwidth}
        \caption{$\Bp=2B_\perp^0$}
        \includegraphics[width=\textwidth,height=7cm]{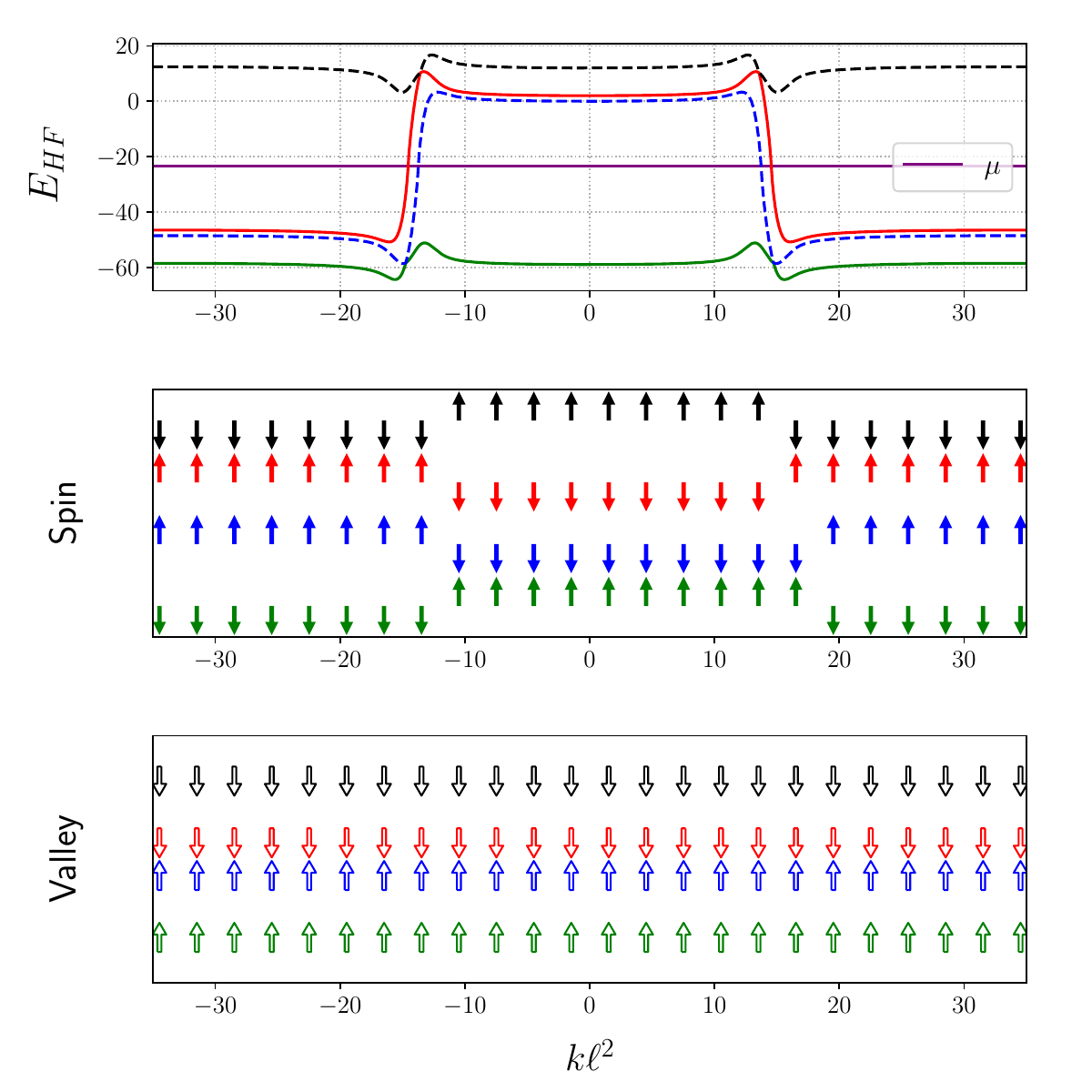}
    \end{subfigure}
    \hspace{-0.2cm}
    \begin{subfigure}{0.33\textwidth}
        \caption{$\Bp=4B_\perp^0$}
        \includegraphics[width=\textwidth,height=7cm]{HF_B_perb_4.0_gxy_-12.0_gz_8.0_Ez_2.0_Ev_5.0.pdf}
    \end{subfigure}
    \caption{Hartree-Fock energies (first row), spin structure (second row), and valley structure (third row) of each energy level for three different values of the perpendicular magnetic field $B_\perp^0,2B_\perp^0,4B_\perp^0$. At $B_\perp^0$ the Hamiltonian parameters are $\Wt^0=20, E_C^0=30, u_z^0=2.0, u_{xy}^0=-3.0, E_Z^0=0.5, E_V=5.0$, identical to  Fig.~(\ref{fig:bulk_ordering_uxy_g_uz_Ev_5}). For $\Bp=B_\perp^0,2B_\perp^0$(first and second column) the occupied states at $\nu=1$ ordered as $K\da < K\ua < K'\ua$. While the spin of each HF level flipped across the interface, the valley indices of each level continue across the interface to make the middle $\nu=-1$ region ordered as $K\ua < K\da < K'\da < K'\ua$. At $\Bp=4B_\perp^0$(last column), the ordering of the occupied $\nu=1$ states changes to $K\da < K'\ua < K\ua $. In this case, the system prefers a valley rotated ground state while the spin of each level remained to be a good quantum number.}
    \label{fig:HF_uxy_g_uz_Ev_5}
\end{figure*}

\begin{figure*}
    \centering
    \begin{subfigure}{0.33\textwidth}
        \caption{$\Bp=1B_\perp^0$}
        \includegraphics[width=\textwidth,height=4cm]{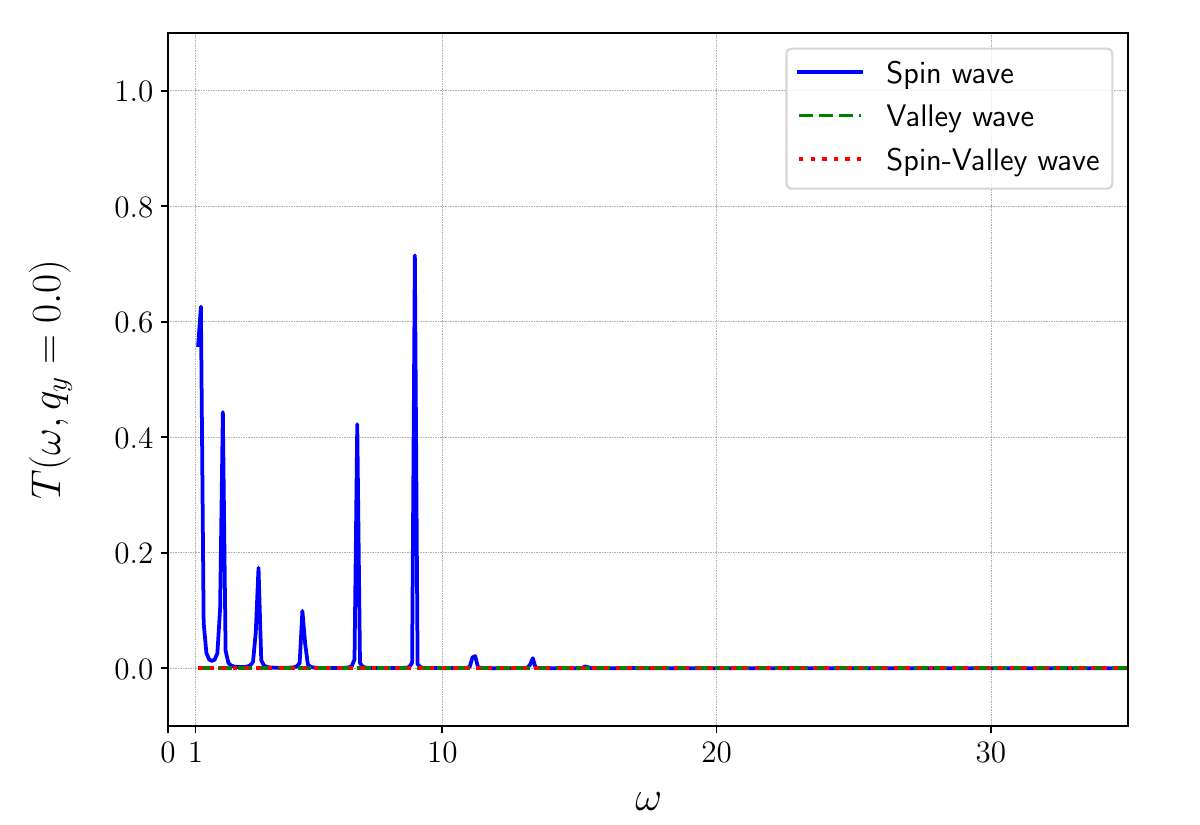}
    \end{subfigure}
    \hspace{-0.2cm}
    \begin{subfigure}{0.33\textwidth}
        \caption{$\Bp=2B_\perp^0$}
        \includegraphics[width=\textwidth,height=4cm]{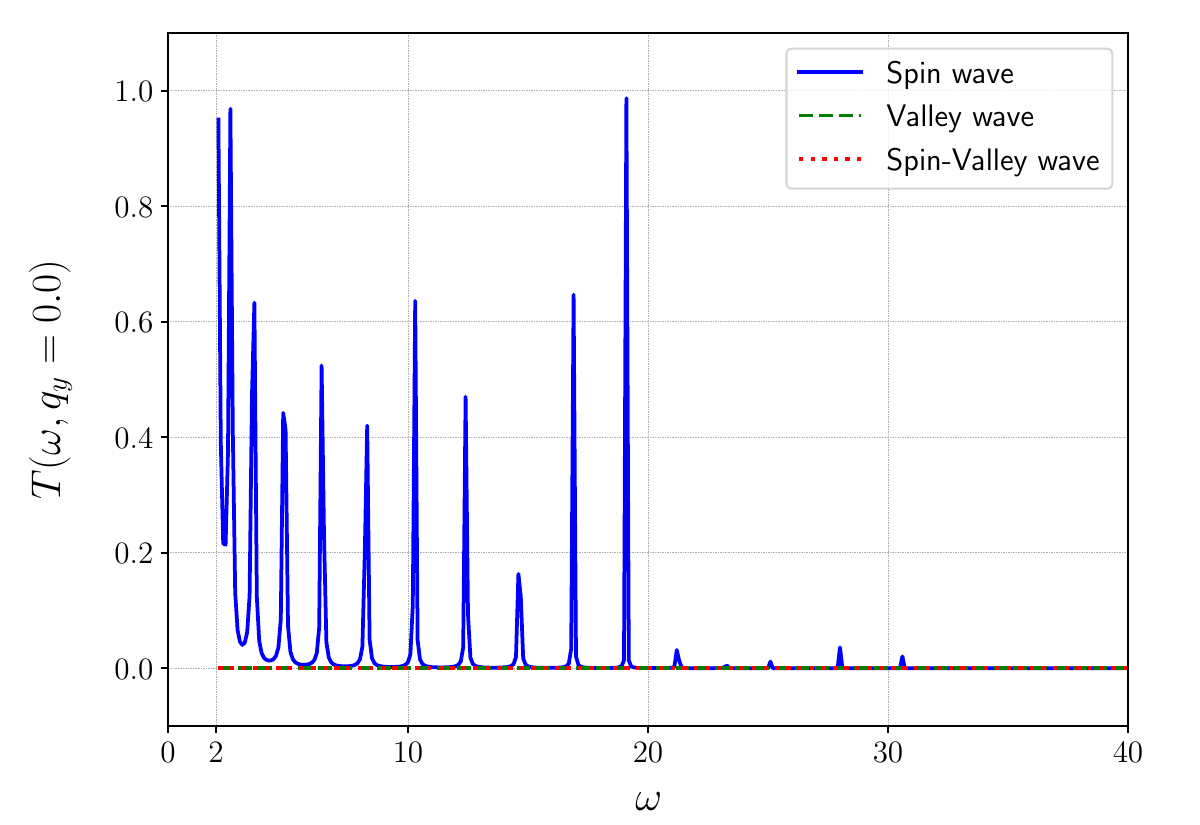}
    \end{subfigure}
    \hspace{-0.2cm}
    \begin{subfigure}{0.33\textwidth}
        \caption{$\Bp=4B_\perp^0$}
        \includegraphics[width=\textwidth,height=4cm]{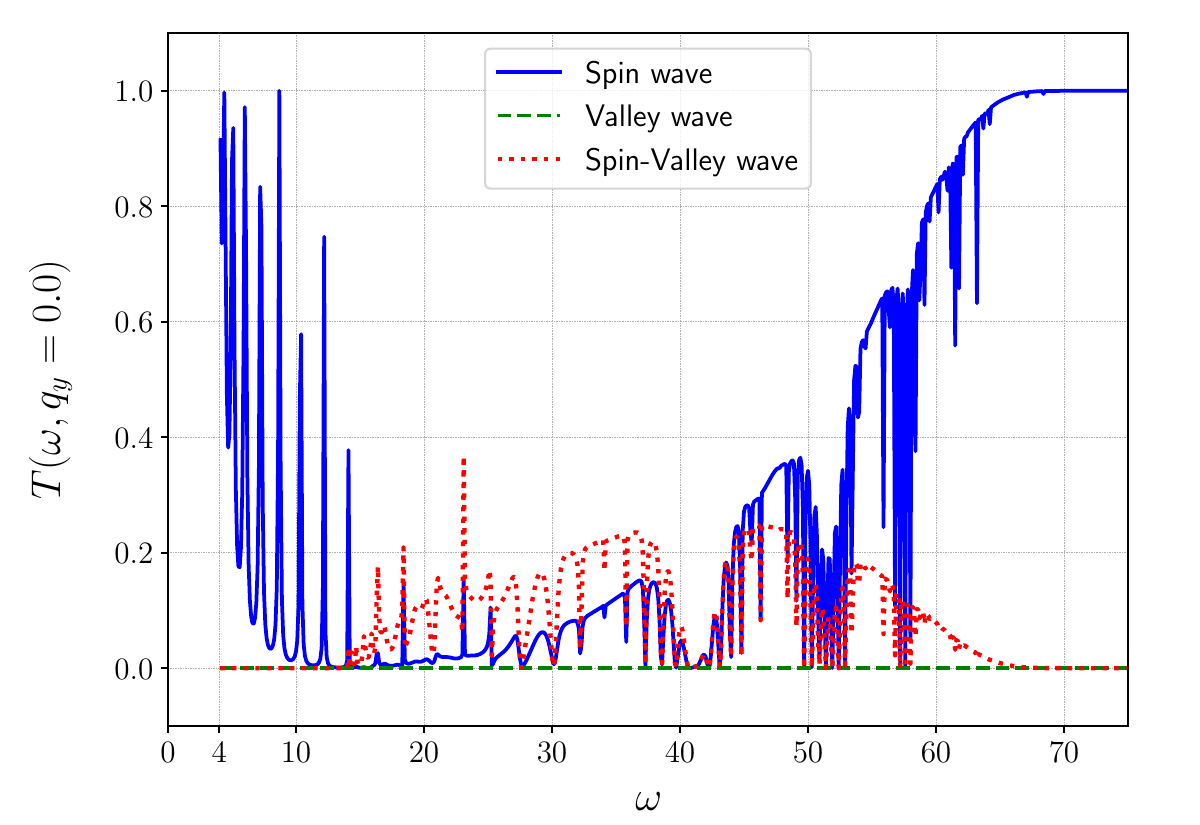}
    \end{subfigure}
    \caption{Transmission amplitudes as a function of incoming magnon energy at $q_y=0$ (normal incidence) corresponding to the HF states in Fig.~(\ref{fig:HF_uxy_g_uz_Ev_5}). At $B_\perp^0$ the Hamiltonian parameters are $\Wt^0=20, E_C^0=30, u_z^0=2.0, u_{xy}^0=-3.0, E_Z^0=0.5, E_V=5.0$. For $\Bp=B_\perp^0,2B_\perp^0$ (first and second column)  the transmission amplitude of the spin magnon is strongly suppressed at higher energies, while both the valley and spin-valley waves are inactive. This is consistent with the previous cases in Fig.~(\ref{fig:TDHF_uz_g_uxy_Ev_5}) where the valley of each HF level remains polarized. For $\Bp=4B_\perp^0$(last column) we find completely different behavior. At high energies the spin magnon is completely transmitted across the junction. We also find that for $\Bp=4B_\perp^0$ at $\omega > 11=2E_Z+2E_V$ the spin-valley wave is excited.}
    \label{fig:TDHF_uxy_g_uz_Ev_5}
\end{figure*}

\section{Collective excitations}
\label{sec:TDHF}
It has been long known that when one uses the HF approximation for 1-body properties, the collective particle-hole excitations should be treated in the time-dependent Hartree-Fock (TDHF) approximation~\cite{TDHF_Fukuyama_1979,TDHF_Bychkov_1981,TDHF_Halperin_1984,TDHF_review_Negele}. Together, they constitute a conserving approximation~\cite{Baym_Kadanoff} in which the approximate correlation functions respect gauge invariance. In the TDHF approach we start with the equations of motion of an arbitrary 1-body operator

\begin{align}
    i\partial_t \hat{P}(t)=[\mathcal{H},\hat{P}(t)]_{HF},
    \label{eq:density_matrix}
\end{align}
where ${\hat P}$ stands for any $c^\dagger_ic_j$. After the commutator is taken, all four-fermi terms are reduced to two-fermi terms using the usual HF rule of considering all possible pairings of operators that have nonzero expectation values in the given SSD state. This results in a closed set of equations for 1-body operators. In translation-invariant problems, such as bulk states, one can further use the conservation of the momentum of the particle-hole pair to reduce the problem to that of diagonalizing a finite-dimensional matrix. We will do this to obtain the $\nu=1$  bulk modes that are the ``in" and ``out" states in our scattering problem. In the inhomogeneous problem we consider, we will assume translation invariance in the $y$ direction, leading to a conserved $y$-momentum $q_y$ for the collective excitations.

To make future manipulations convenient, we first write all one-body operators in the basis that diagonalizes the HF Hamiltonian. Using the index $k$ to label the $y$-momentum (the guiding center location is $X_k=k\ell^2$) and the index $m=0,1,2,3$ to label the HF levels with increasing single-particle energy, we obtain, in this basis,
\begin{align}
   \langle HF| c^\dagger_{k,m}c_{k',n}|HF\rangle= \delta_{kk'}\delta_{m,n} N_F(m,k).
   \label{eq:P0}
\end{align}
Here $N_F(m,k)$ is the occupation of the $m$th HF level at the momentum index $k$. Since we are at $T=0$ this number can only be zero or unity. Let us define the particle-hole operator with momentum labels  $k,q_y$  as 
\begin{align}
    {\hat O}_{k,q_y,m,n}=c^{\dagger}_{k-q_y,m}c_{k,n}
    \label{eq:dP}
\end{align}

Operationally, this means the interaction matrix elements have to be rotated into this basis as follows
\begin{align}
    \Tilde{V}_{ijlm}(k_1,k_2,q_y)=\sum_{\alpha} u_\alpha \big[ (U^\dagger(k_1-q_y) \tau_\alpha U(k_1))_{ij}
    \nonumber \\ (U^\dagger(k_2+q_y) \tau_\alpha U(k_2))_{lm} \big] , 
    \nonumber \\
    V_{ijlm}^{C}(k_1,k_2,q_y)=E_c \big[ (U^\dagger(k_1-q_y) U(k_1))_{ij} 
    \nonumber \\ (U^\dagger(k_2+q_y) U(k_2))_{lm} \big],
\end{align}
where $U(k)$ is the unitary $4\times4$ matrix that rotates the states from the original basis to the basis that diagonalizes the HF Hamiltonian.
The TDHF equations now reduce to 
\begin{align}
    i\partial_t{\hat O}_{k,q_y,m,n}=\sum_{k',m',n'} \mathbb{K}_{k,m,n}^{k',m',n'}(q_y) {\hat O}_{k',q_y,m',n'}
\end{align}
We look for eigenmodes to these equations. Assuming that a particular eigenmode is expressed as 
\begin{align}
    {\hat O}_{\alpha,q_y}=\sum\limits_{k,m,n}\Psi^{\alpha}_{k,m,n} {\hat O}_{k,q_y,m,n}
\end{align}
we obtain the TDHF equation or generalized RPA equation\cite{Wei_Huang_MacDonald2021} in frequency space as
\begin{align}
    \sum_{k',m',n'} \mathbb{K}_{k,m,n}^{k',m',n'}(q_y) \Psi^{\alpha}_{k',m',n'}(q_y,\omega) = \omega_{\alpha} \Psi^\alpha_{k,m,n}(q_y,\omega)
    \label{eq:TDHF}
\end{align}
with the kernel 
\begin{align}
    \mathbb{K}_{k,m,n}^{k',m',n'}(q_y)=[\epsilon_{m'}(k'-q_y)-\epsilon_{n'}(k')]\delta_{k k'}\delta_{m m'}\delta_{n n'} + 
    \nonumber \\
    \frac{\ell}{L_y}[N_F(n',k')-N_F(m',k'-q_y)] \times
    \nonumber \\
    [F_{sr}(k'-k,q_y)\Tilde{V}_{n'm'mn}(k'-q_y,k,-q_y) + 
    \nonumber \\
    F_{lr}(k'-k,q_y)V^{C}_{n'm'mn}(k'-q_y,k,-q_y) - 
    \nonumber \\
    F_{sr}(q_y,k-k')\Tilde{V}_{mm'n'n}(k'-q_y,k,k'-k) - 
    \nonumber \\
    F_{lr}(q_y,k-k')V^{C}_{mm'n'n}(k'-q_y,k,k'-k)]
    \label{eq:Kernel}
\end{align}
where the short-range kernel $F_{sr}$ and the long-range kernel $F_{lr}$ are defined by

\begin{align}
    F_{sr}(k_1,k_2)= \sqrt{2 \pi} ~e^{-\frac{[k_1^2+k_2^2]\ell^2}{2}}, 
    \nonumber \\
    F_{lr}(k_1,k_2)= \int_{-\infty}^{\infty} \ell dq_x  \frac{e^{-[\frac{(q_x^2+k_2^2)\ell^2}{2}]} \cos(q_x k_1 \ell^2)}{\sqrt{q_x^2 \ell^2 + k_2^2 \ell^2 + q_0^2}} .
\end{align}
and $\epsilon_m(k)$ is the HF energy of the self-consistent single particle state  $|k,m\rangle$.

In the next subsection we present the bulk collective excitation for $\nu=1$, and in the following subsection we turn to the main topic of the paper, the transmission of magnons through the $\nu=1|-1|1$ junction.

\subsection{Bulk collective excitations of $\nu=1$}
\label{sec:bulk_collective_modes}
Bulk systems are translation invariant in both directions, hence the HF states $|k,m \rangle \equiv |m \rangle$, their energies $\epsilon_m(k) \equiv \epsilon_m$, and the occupation of each HF level $N_F(m,k) \equiv N_F(m)$ are all independent of momentum index $k$. In this case, one can define an additional conserved momentum $q_x$ for the eigenoperators. The RPA kernel defined in Eq.~(\ref{eq:Kernel}) is made block diagonal in momentum space $\vec{q}=(q_x,q_y)$ by the following Fourier transformation
\begin{align}
    \mathbb{K}_{m,n}^{m',n'}(\vec{q})=\sum_{\Delta k} \mathbb{K}_{k,m,n}^{k',m',n'}(q_y) e^{-iq_x \Delta k \ell^2},~ \Delta k=(k'-k) 
    \nonumber \\
    =\big[\big(\epsilon_{m'}-\epsilon_{n'}\big)-\big(N_F(n')-N_F(m')\big)u_c(q)\big]\delta_{m,m'} \delta_{n,n'}
    \nonumber \\
    + \big(N_F(n')-N_F(m')\big) f(q) \big[\Tilde{V}_{n'm'mn}-\Tilde{V}_{mm'n'n}\big]
    \label{eq:bulk_Kernel}
\end{align}
with 
\begin{align}
    u_c(q)=\frac{E_c \ell^2}{2 \pi} \int d^2k~ \frac{e^{-(k^2 \ell^2)/2}~ e^{-i(\vec{k}.\vec{q})\ell^2}}{\sqrt{k^2 \ell^2+q_0^2}} ,
    \label{eq:uc_q}
\end{align} 
\begin{align}
    f(q)=e^{-\frac{q^2 \ell^2}{2}}
    \label{eq:f_q}    
\end{align}
and 
\begin{align}
    \Tilde{V}_{ijlm}=\sum_{\alpha} u_\alpha (U^\dagger \tau_{\alpha} U)_{ij} (U^\dagger \tau_{\alpha} U)_{lm}
\end{align}
$``U"$ is the unitary matrix that diagonalizes the bulk HF hamiltonian \ref{eq:HFHa_bulk}.

Thus, for the translational invariant bulk, the TDHF equation(\ref{eq:TDHF}) simplifies to
\begin{align}
    \sum_{m' n'} \mathbb{K}_{m,n}^{m',n'}(q) \phi^\alpha_{m'n'}(q)=\omega_\alpha(q) \phi^\alpha_{mn}(q),
    \label{eq:bulk_TDHF}
\end{align}
where
\begin{align}
    \phi^\alpha_{mn}(q)=\frac{1}{L_y} \sum_k \Psi^\alpha_{k-q_y,m,n}(q_y,\omega)~ e^{-iq_x k \ell^2}
\end{align}
are the normalized bulk collective modes of frequency $\omega(q)$, having the normalization condition,
\begin{align}
    \sum_{mn}\big(N_F(n)-N_F(m)\big) \Bar{\phi}_{mn}^{\alpha}(q) \phi_{mn}^{\beta}(q) = \delta_{\alpha \beta}.
    \label{eq:normalization}
\end{align}
Here $\Bar{\phi}^\alpha$ represents the complex conjugation of $\phi^\alpha$.
We are interested in the positive frequency $\omega(q)>0$ collective modes $\phi_{mn}(q)$ here, but it can be shown that the collective modes $\Tilde{\phi}_{mn}(q)$ with frequency $-\omega(q)$ are related to the positive frequency modes by $\Tilde{\phi}_{mn}(q)=\phi_{nm}^*(-q)$.

It is easy to see from the bulk TDHF Eq.~(\ref{eq:bulk_TDHF}) that the collective excitations are superpositions of particle-hole excitations which create a hole in a filled HF level and a particle in an unfilled HF level.
For the $\nu=1$ bulk, the filled states are ${K\uparrow},{K\downarrow},{K'\uparrow}$ and the unfilled state is ${K'\downarrow}$, and as these states preserve the spin and valley indices the bulk RPA kernel in Eq.~(\ref{eq:bulk_Kernel}) is diagonal in the particle-hole basis made of one filled and one unfilled $\nu=1$ HF level. We get three orthonormal bulk collective modes\cite{Wei_Huang_MacDonald2021} for $\nu=1$. (1) The spin wave or the spin magnon mode $\phi_{mn}^s \equiv \phi^s_{({K'\downarrow},{K'\uparrow})}$ conserves valley but flips spin. (2) The valley wave $\phi_{mn}^v \equiv \phi^v_{({K'\downarrow},{K\downarrow})}$ conserves spin but flips the valley.  (3) The spin-valley wave $\phi_{mn}^{sv} \equiv \phi^{sv}_{({K'\downarrow},{K\uparrow})}$ which flips both spin and valley.  Their dispersions are~\cite{Wei_Huang_MacDonald2021} 
\begin{align}
    \omega_s(q)=V_{ex}-u_c(q)+2E_Z+u_z[1-f(q)],\hspace{0.7cm} \nonumber \\
    \omega_v(q)=V_{ex}-u_c(q)+2E_V+u_z[f(q)-1]\hspace{0.7cm} \nonumber \\ 
    -2u_{xy}[1-f(q)],\hspace{3.2cm} \nonumber \\
    \omega_{sv}(q)=V_{ex}-u_c(q)+2E_Z+2E_V+u_z[f(q)-1].\hspace{0.5cm}
    \label{eq:bulk_dispersion}
\end{align}
$u_c(q)$,$f(q)$ are defined previously in equations~\ref{eq:uc_q} and \ref{eq:f_q}.

\begin{figure}
    \centering
    \includegraphics[width=0.4\textwidth,height=5cm]{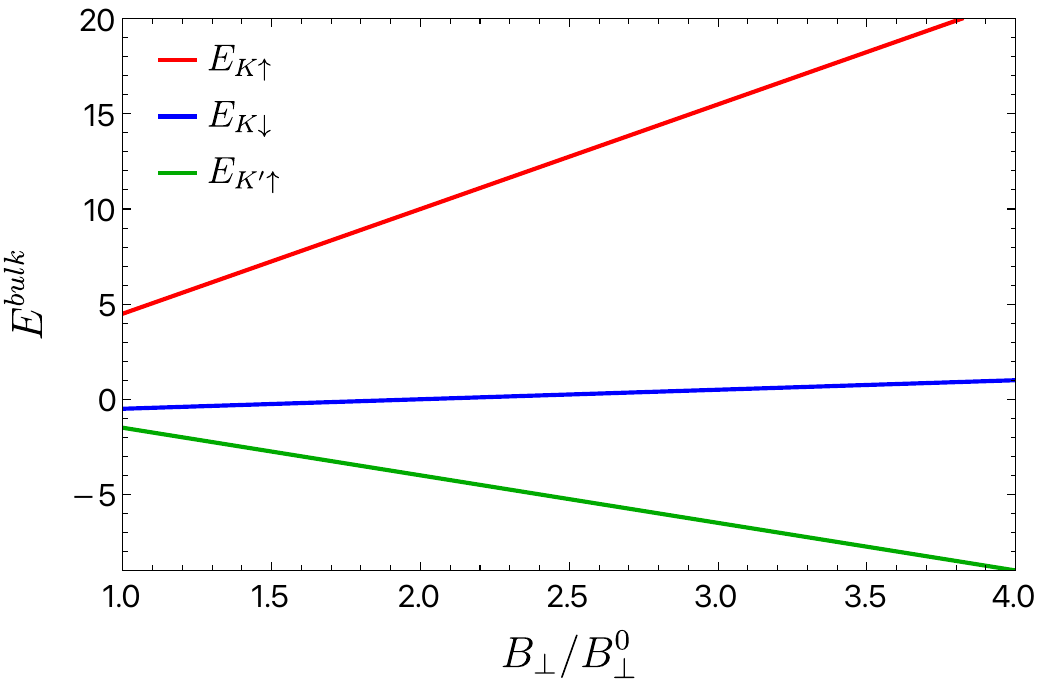}
    \caption{Occupied  single-particle HF energies for bulk $\nu=1$ as a function of $\Bp/B_\perp^0$ with  $V_{ex}=0$, $E_Z^0=0.5$, $E_V=1$, $u_z^0=4$ and $u_{xy}^0=-3$. The parameters are chosen such that it reflects the case when $u_z^0 > |u_{xy}^0| > E_V  >E_Z^0$. As we can see, for this case the ordering of the occupied state is $E_{K'\ua}<E_{K\da}<E_{K\ua}$ which does not change with increasing $\Bp$.}
    \label{fig:bulk_ordering_uz_g_uxy_Ev_1}
\end{figure}

\begin{figure*}
    \centering
    \begin{subfigure}{0.33\textwidth}
        \caption{$\Bp=1B_\perp^0$}
        \includegraphics[width=\textwidth,height=7cm]{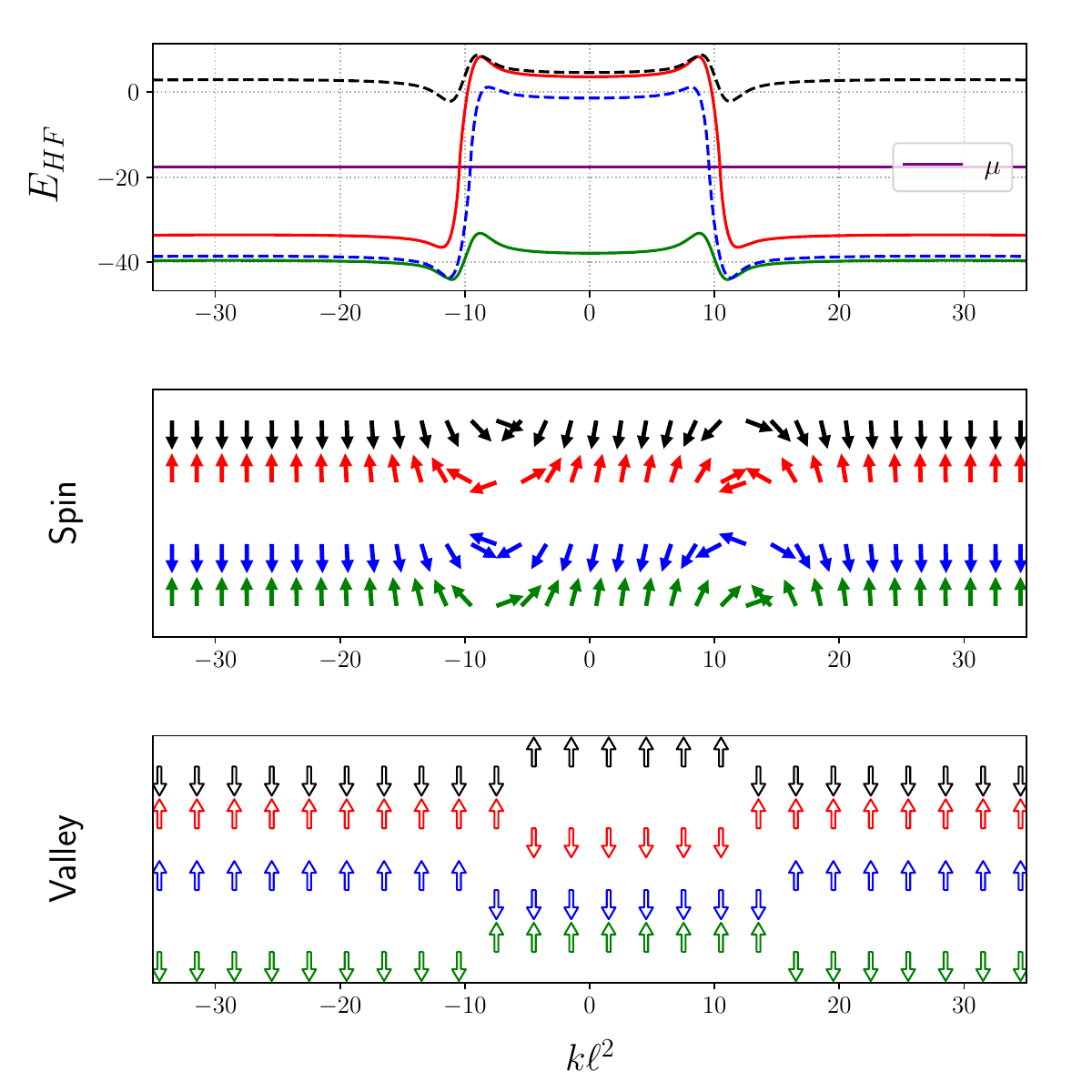}
    \end{subfigure}
    \hspace{-0.2cm}
    \begin{subfigure}{0.33\textwidth}
        \caption{$\Bp=2B_\perp^0$}
        \includegraphics[width=\textwidth,height=7cm]{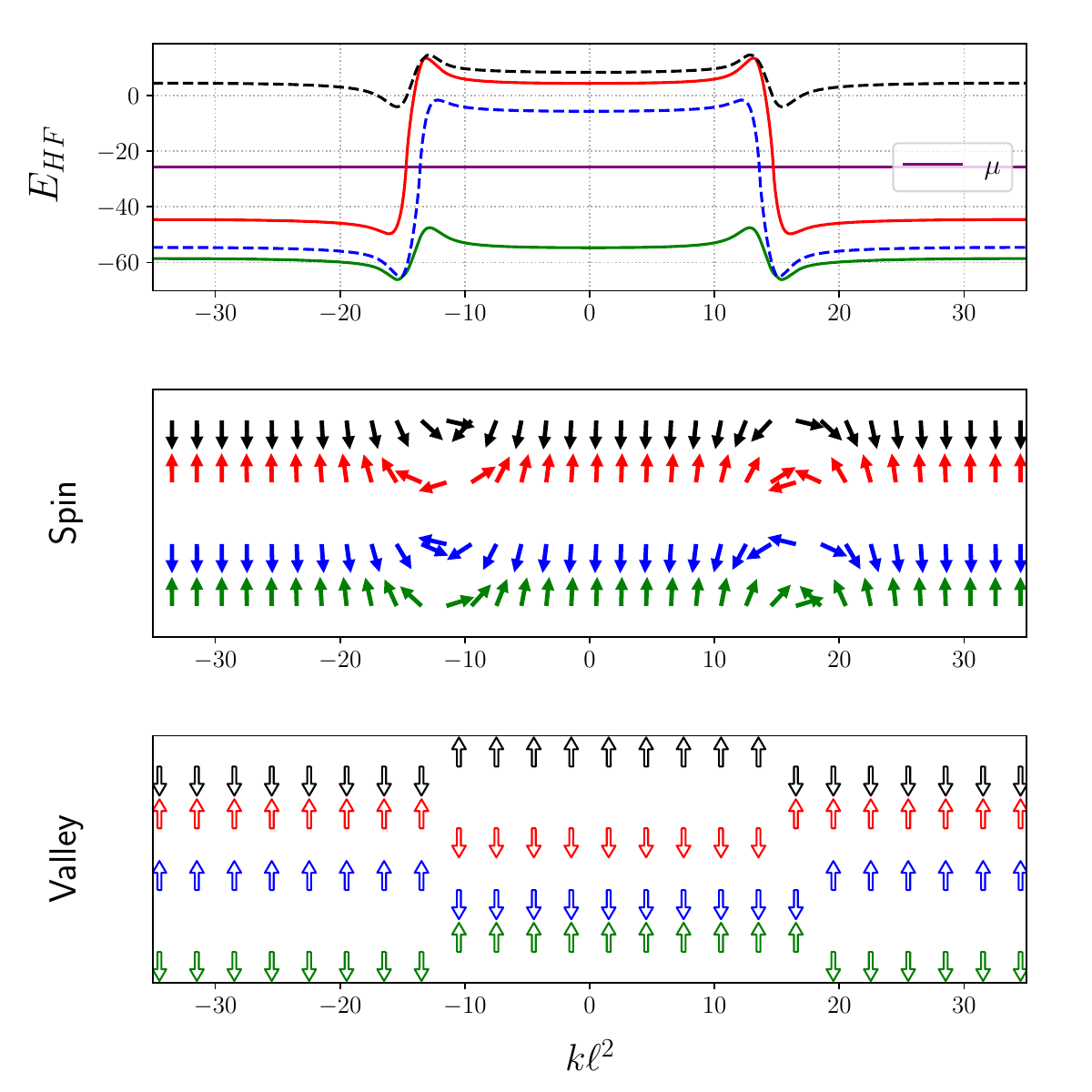}
    \end{subfigure}
    \hspace{-0.2cm}
    \begin{subfigure}{0.33\textwidth}
        \caption{$\Bp=4B_\perp^0$}
        \includegraphics[width=\textwidth,height=7cm]{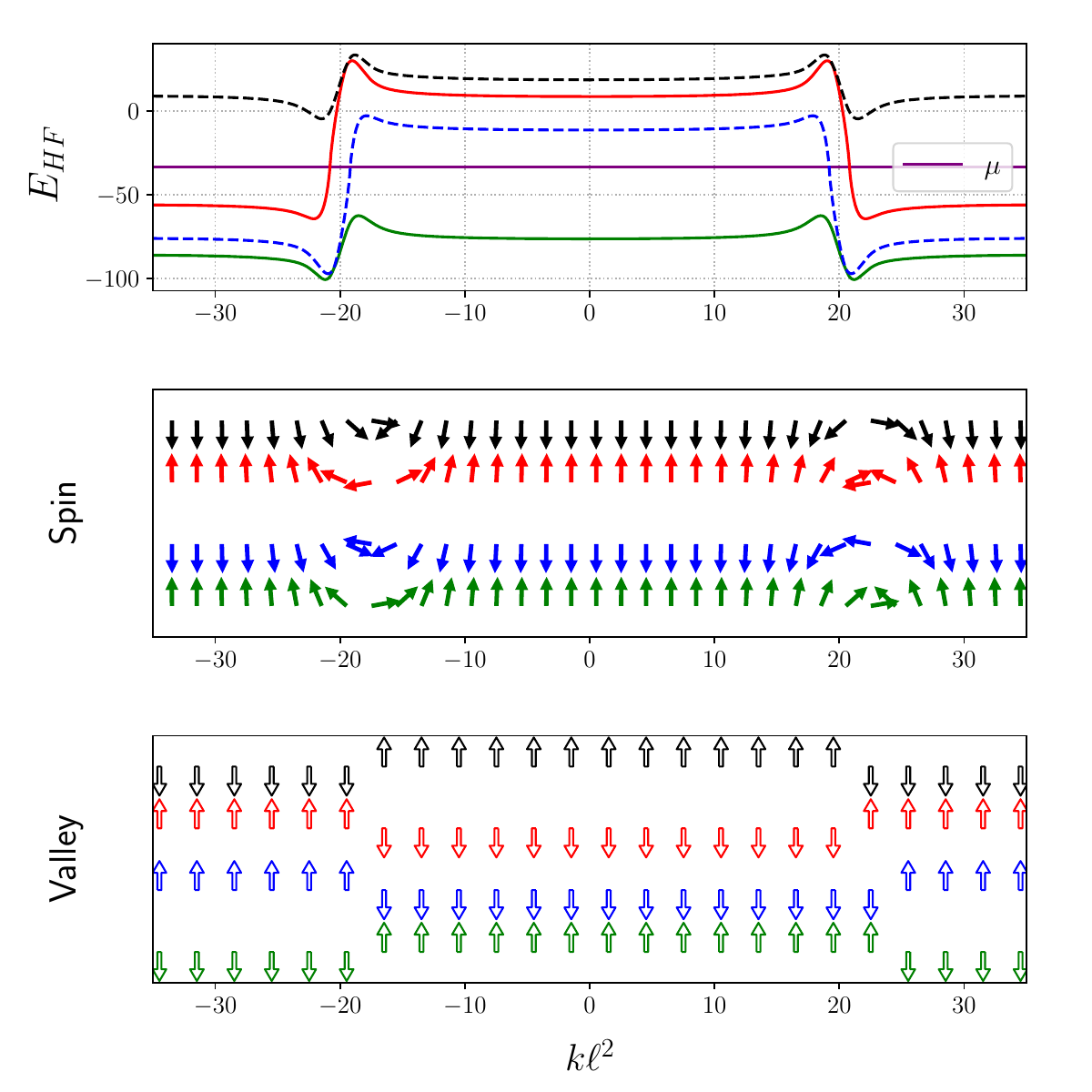}
    \end{subfigure}
    \caption{Hartree-Fock energies (first row), spin structure (second row), and valley structure (third row) of each energy level for three different values of the perpendicular magnetic field $B_\perp^0, 2B_\perp^0, 4B_\perp^0$. At $B_\perp^0$ the Hamiltonian parameters are $\Wt^0=20, E_C^0=30, u_z^0=4.0, u_{xy}^0=-3.0, E_Z^0=0.5, E_V=1.0$, identical to  Fig.~(\ref{fig:bulk_ordering_uz_g_uxy_Ev_1}). For the parameters chosen  the self-consistent HF states always prefer a spin rotated ground state close to the $\nu=1|-1$ interfaces, while the valley  remains a good quantum number. This is similar to the case of $\Bp=4B_\perp^0$ in Fig.~(\ref{fig:HF_uz_g_uxy_Ev_5}).}
    \label{fig:HF_uz_g_uxy_Ev_1}
\end{figure*}

\begin{figure*}
    \centering
    \begin{subfigure}{0.33\textwidth}
        \caption{$\Bp=1B_\perp^0$}
        \includegraphics[width=\textwidth,height=4cm]{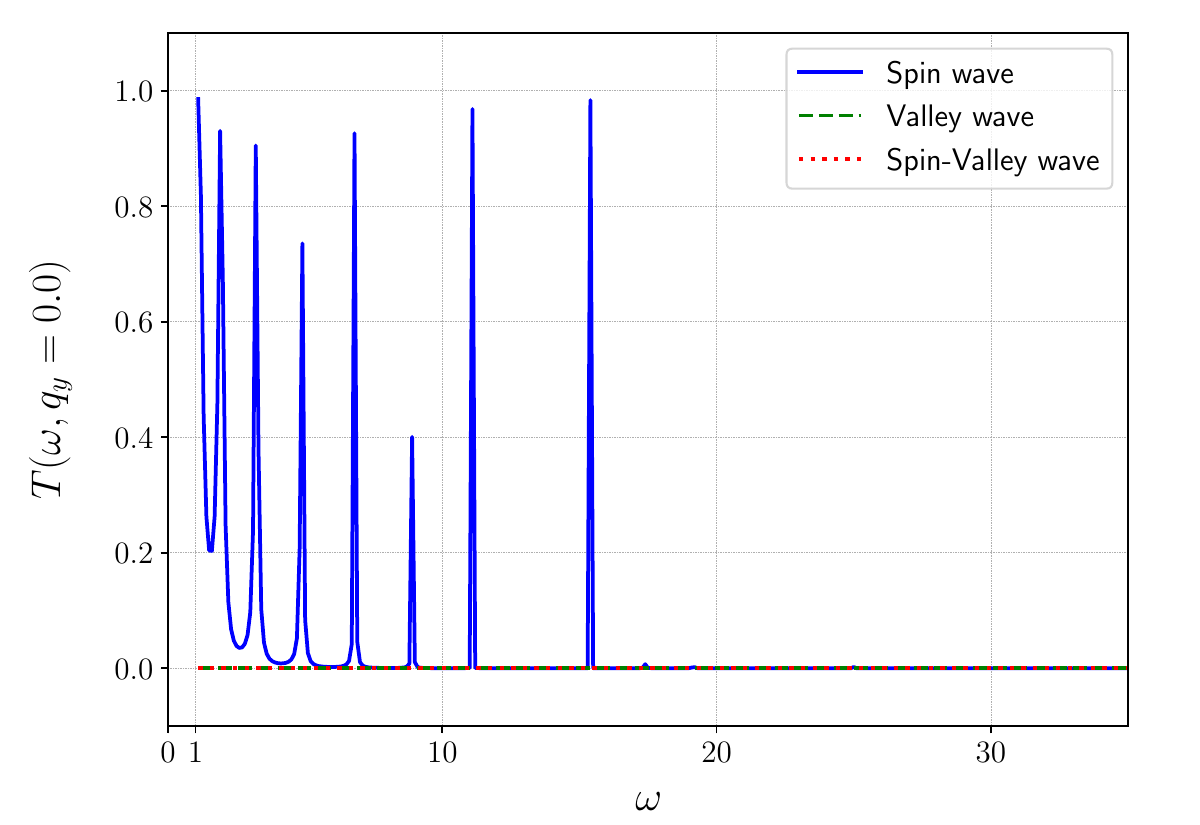}
    \end{subfigure}
    \hspace{-0.2cm}
    \begin{subfigure}{0.33\textwidth}
        \caption{$\Bp=2B_\perp^0$}
        \includegraphics[width=\textwidth,height=4cm]{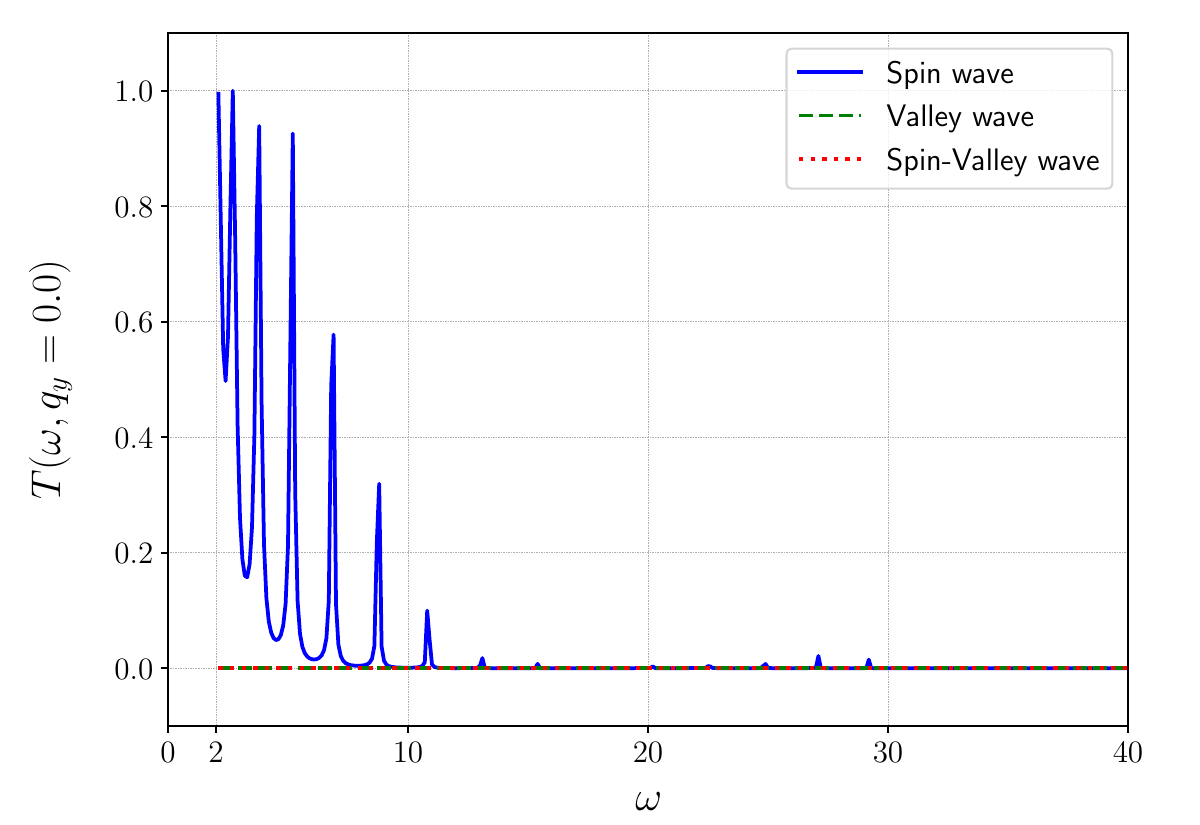}
    \end{subfigure}
    \hspace{-0.2cm}
    \begin{subfigure}{0.33\textwidth}
        \caption{$\Bp=4B_\perp^0$}
        \includegraphics[width=\textwidth,height=4cm]{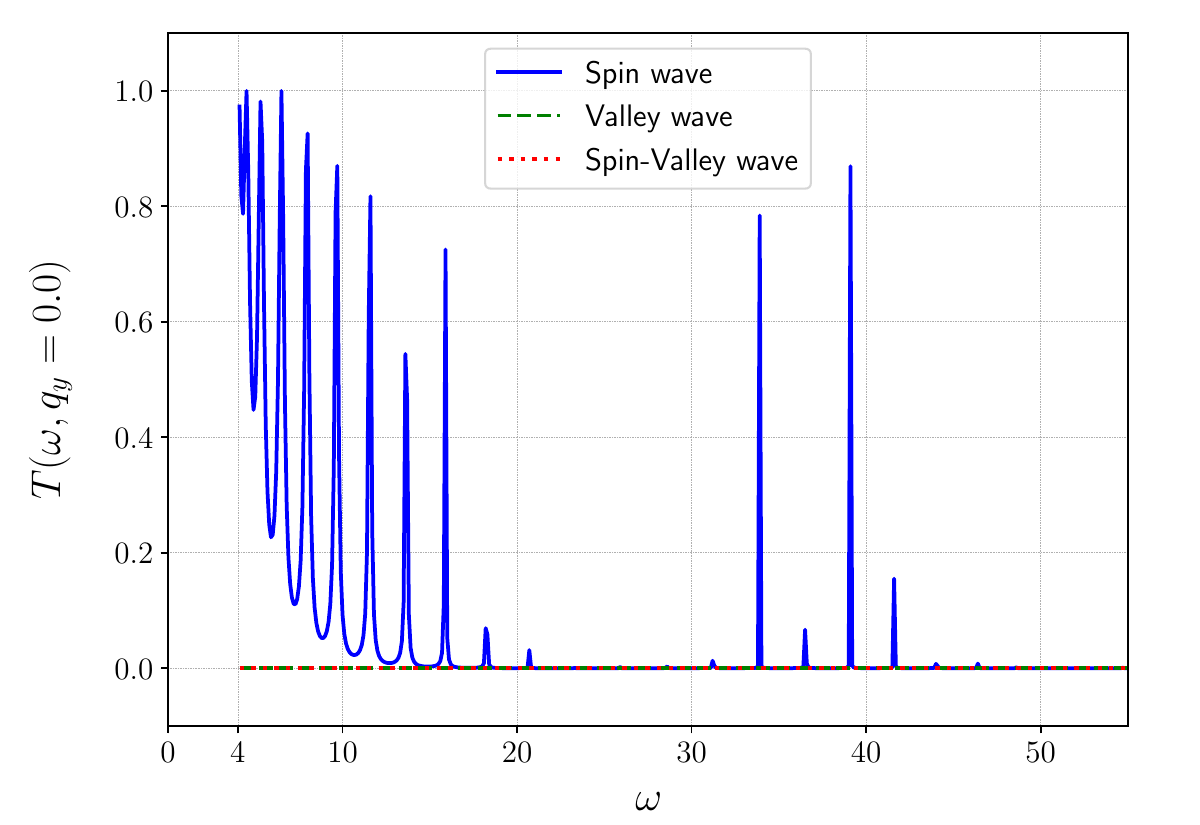}
    \end{subfigure}
    \caption{Transmission amplitudes as a function of incoming magnon energy at $q_y = 0$ corresponding to the HF states in fig(\ref{fig:HF_uz_g_uxy_Ev_1}). At $B_\perp^0$ the Hamiltonian parameters are $\Wt^0=20, E_C^0=30, u_z^0=4.0, u_{xy}^0=-3.0, E_Z^0=0.5, E_V=1.0$. As in Fig.~(\ref{fig:TDHF_uz_g_uxy_Ev_5}), we find that the transmission amplitude of the spin mode is strongly suppressed at higher energies.}
    \label{fig:TDHF_uz_g_uxy_Ev_1}
\end{figure*}

\begin{figure}[h]
    \centering
    \includegraphics[width=0.4\textwidth,height=5cm]{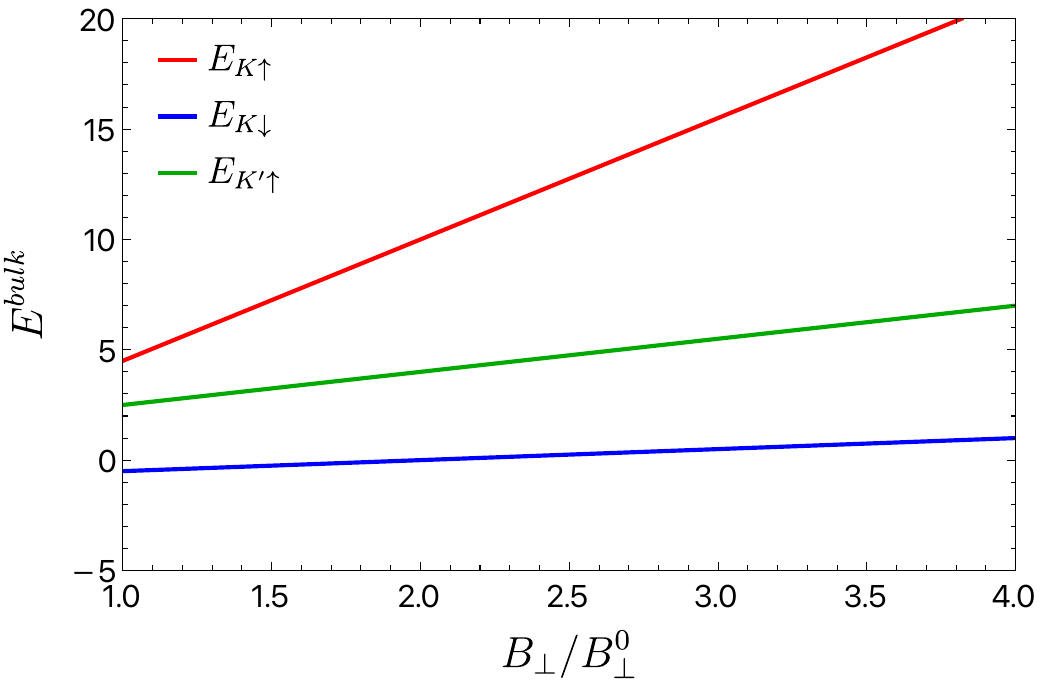}
    \caption{Occupied single-particle HF energies for the bulk $\nu=1$ state as a function of $\Bp/B_\perp^0$ with  $V_{ex}=0$, $E_Z^0=0.5$, $E_V=1$, $u_z^0=2$ and $u_{xy}^0=-3$. We have chosen the parameters such that it reflects the case when $|u_{xy}^0|> u_z^0 > E_V  >E_Z^0$. In this case we find the bulk ordering is $E_{K\da} < E_{K'\ua} < E_{K\ua}$, which does not change with $\Bp$.}
    \label{fig:bulk_ordering_uxy_g_uz_Ev_1}
\end{figure}

\begin{figure*}
    \centering
    \begin{subfigure}{0.33\textwidth}
        \caption{$\Bp=1B_\perp^0$}
        \includegraphics[width=\textwidth,height=7cm]{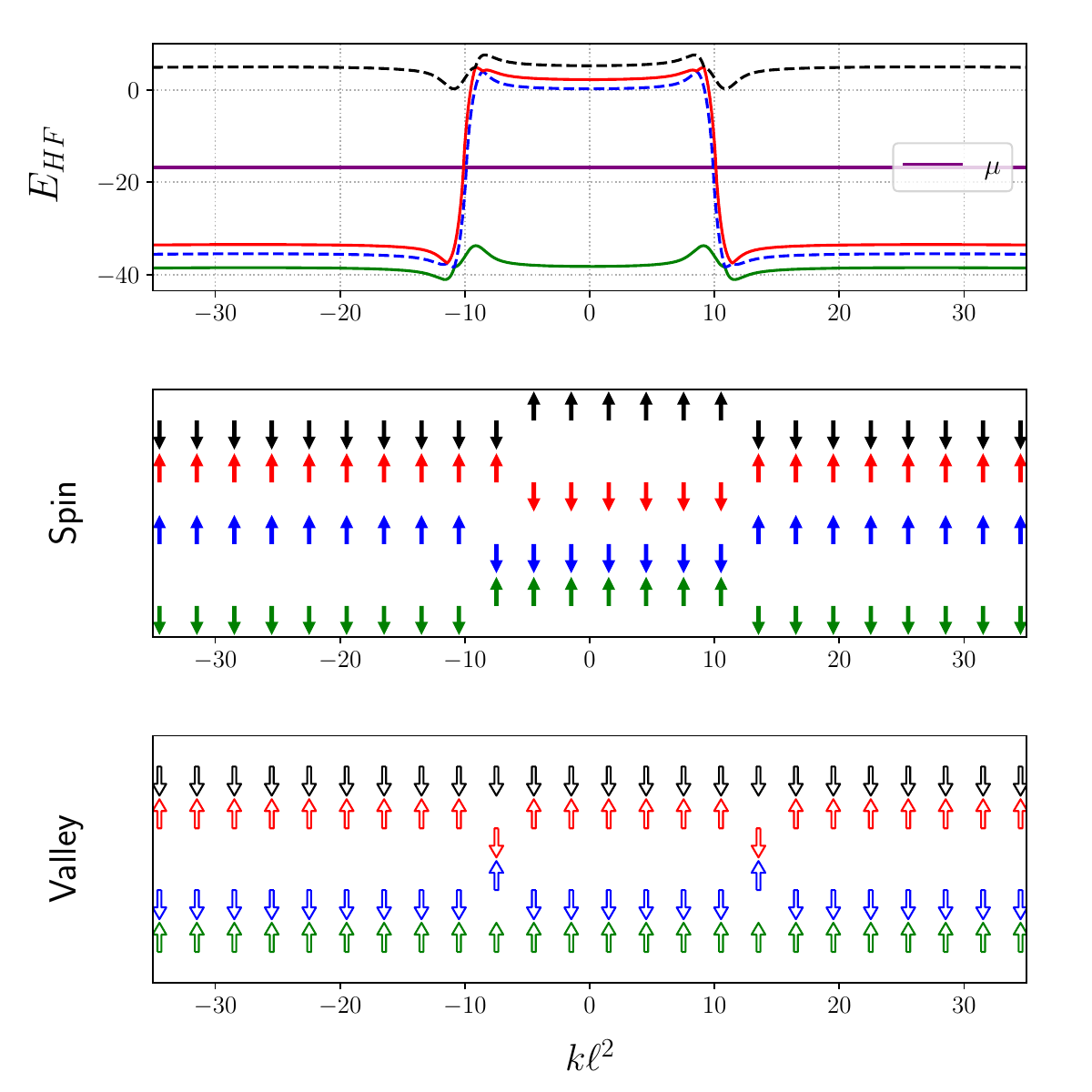}
    \end{subfigure}
    \hspace{-0.2cm}
    \begin{subfigure}{0.33\textwidth}
        \caption{$\Bp=2B_\perp^0$}
        \includegraphics[width=\textwidth,height=7cm]{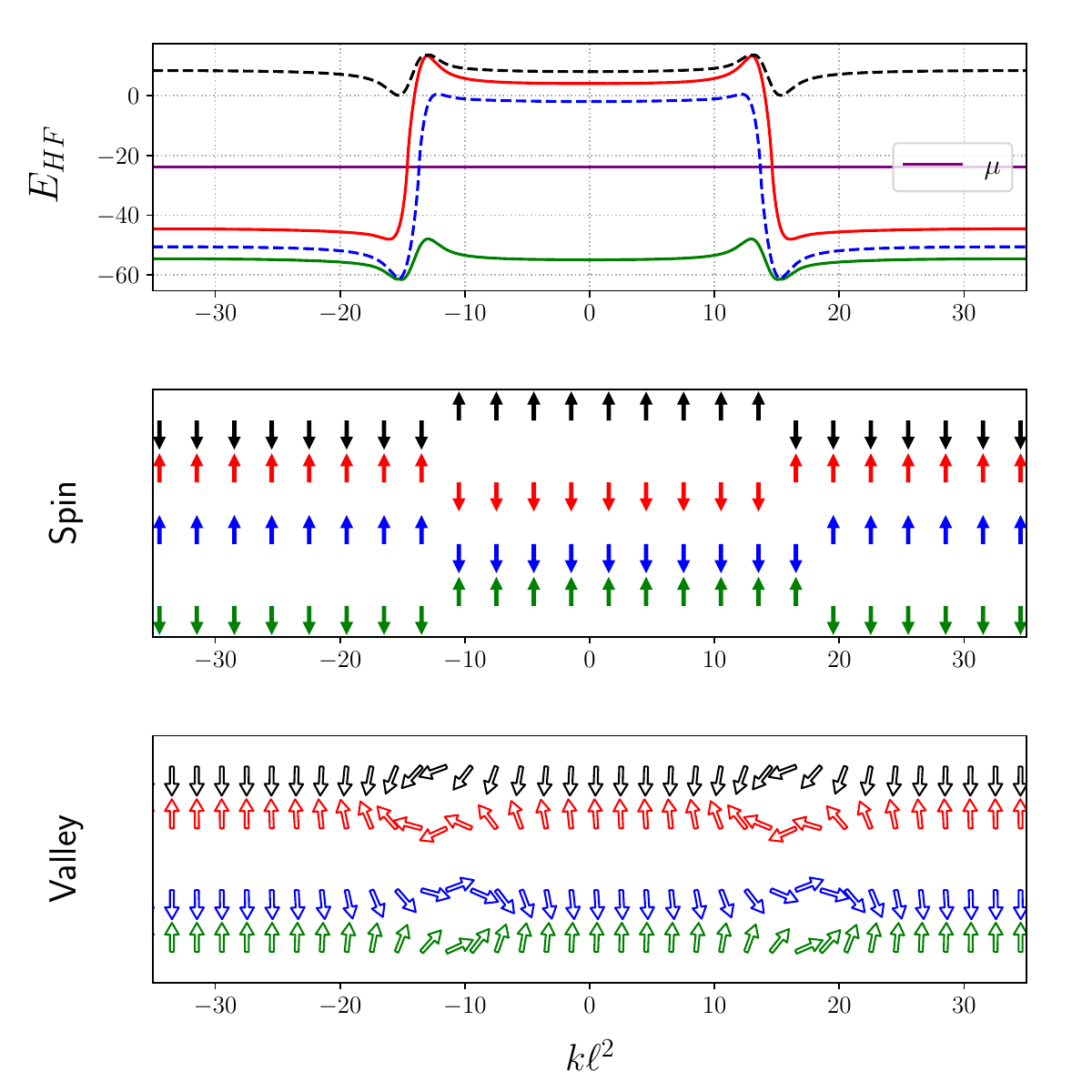}
    \end{subfigure}
    \hspace{-0.2cm}
    \begin{subfigure}{0.33\textwidth}
        \caption{$\Bp=4B_\perp^0$}
        \includegraphics[width=\textwidth,height=7cm]{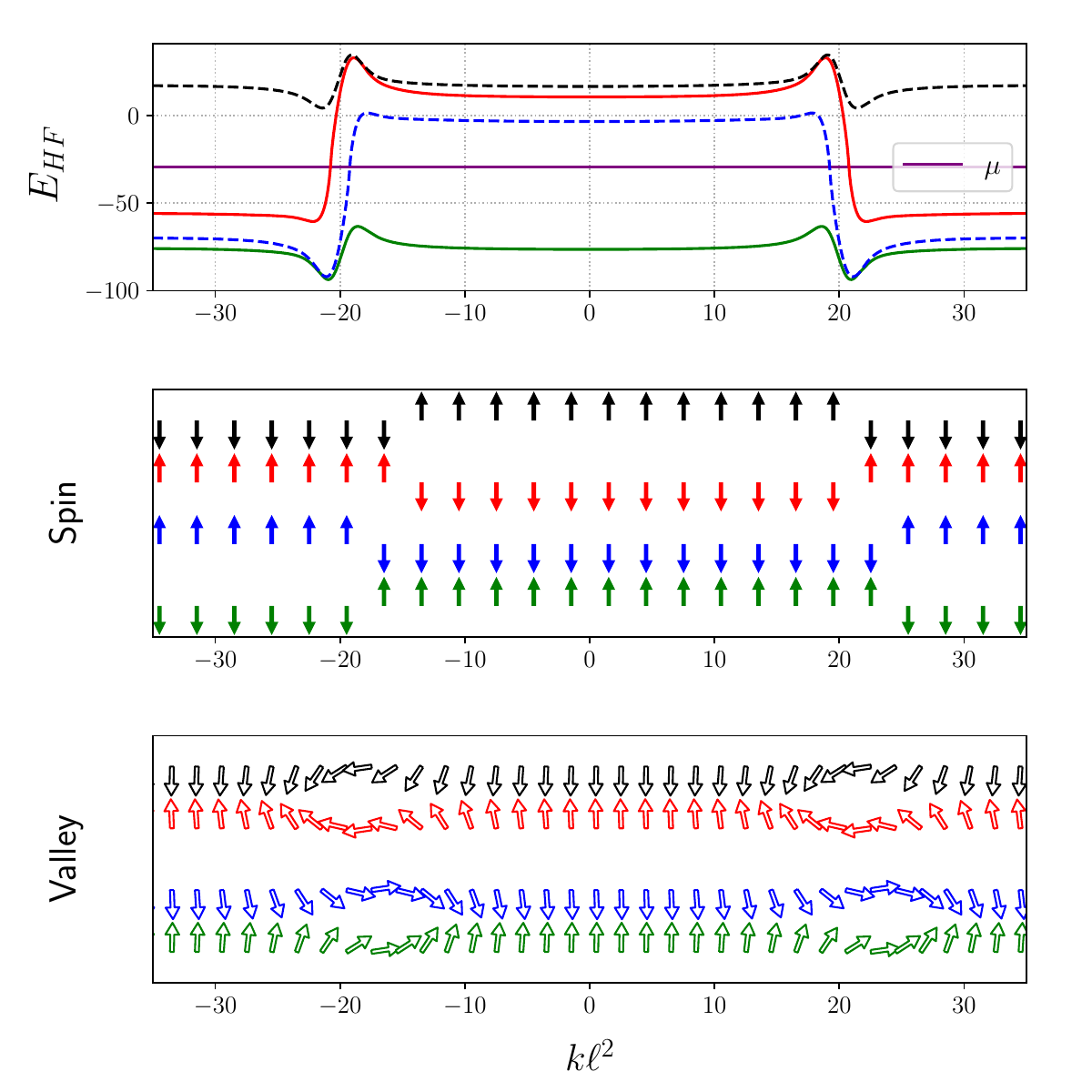}
    \end{subfigure}
    \caption{Hartree-Fock energies (top panel), spin structure (middle panel), and valley structure (bottom panel) of each energy level for three different values of the perpendicular magnetic field $B_\perp^0,2B_\perp^0,4B_\perp^0$. At $B_\perp^0$ the Hamiltonian parameters are $\Wt^0=20,~ E_C^0=30,~ u_z^0=2.0,~ u_{xy}^0=-3.0,~ E_Z^0=0.5,~ E_V=1.0$, which are the same parameters we chosen in Fig.~(\ref{fig:bulk_ordering_uxy_g_uz_Ev_1}). Here we find initially for $B_\perp^0$(a), both the spin and valley of each HF level remain as a good quantum number. With increasing $\Bp$ in (b) and (c), although the spin of each self-consistent level continues to behave as a good quantum number, the valley rotates across the interfaces.}
    \label{fig:HF_uxy_g_uz_Ev_1}
\end{figure*}

\begin{figure*}
    \centering
    \begin{subfigure}{0.33\textwidth}
        \caption{$\Bp=1B_\perp^0$}
        \includegraphics[width=\textwidth,height=4cm]{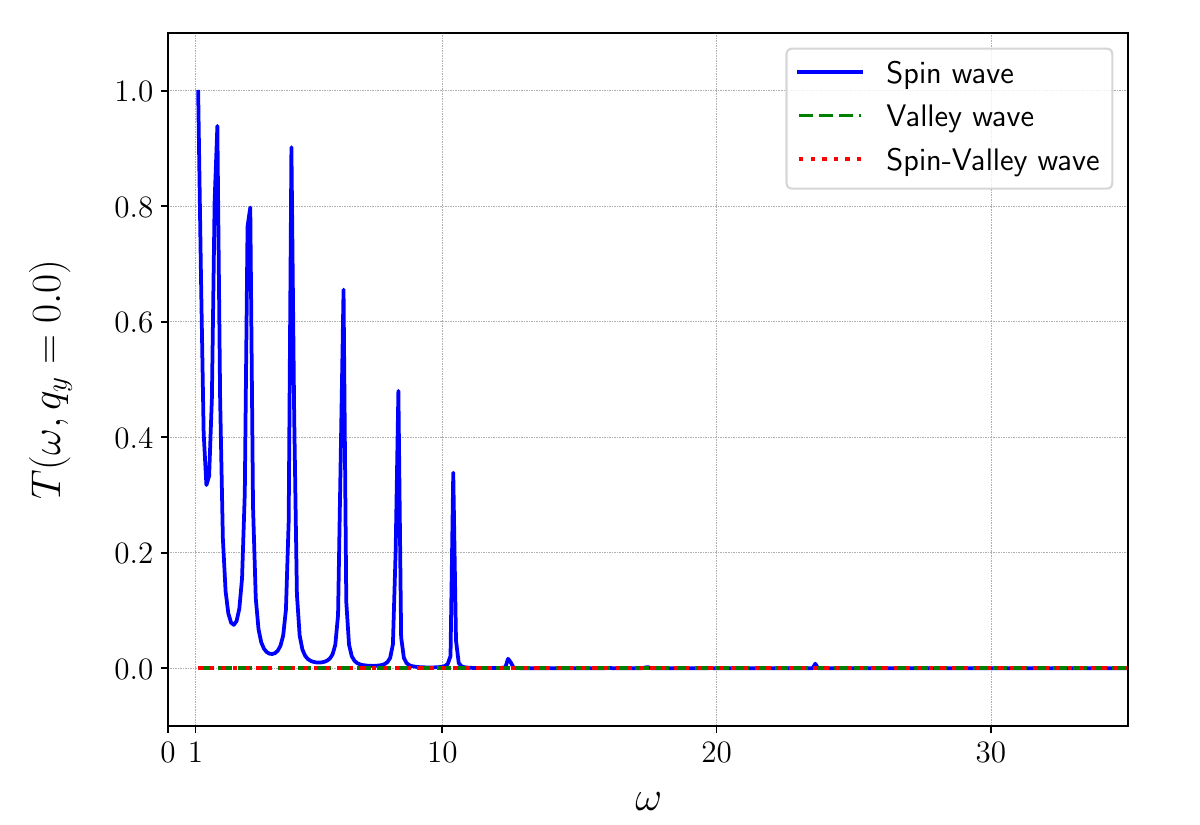}
    \end{subfigure}
    \hspace{-0.2cm}
    \begin{subfigure}{0.33\textwidth}
        \caption{$\Bp=2B_\perp^0$}
        \includegraphics[width=\textwidth,height=4cm]{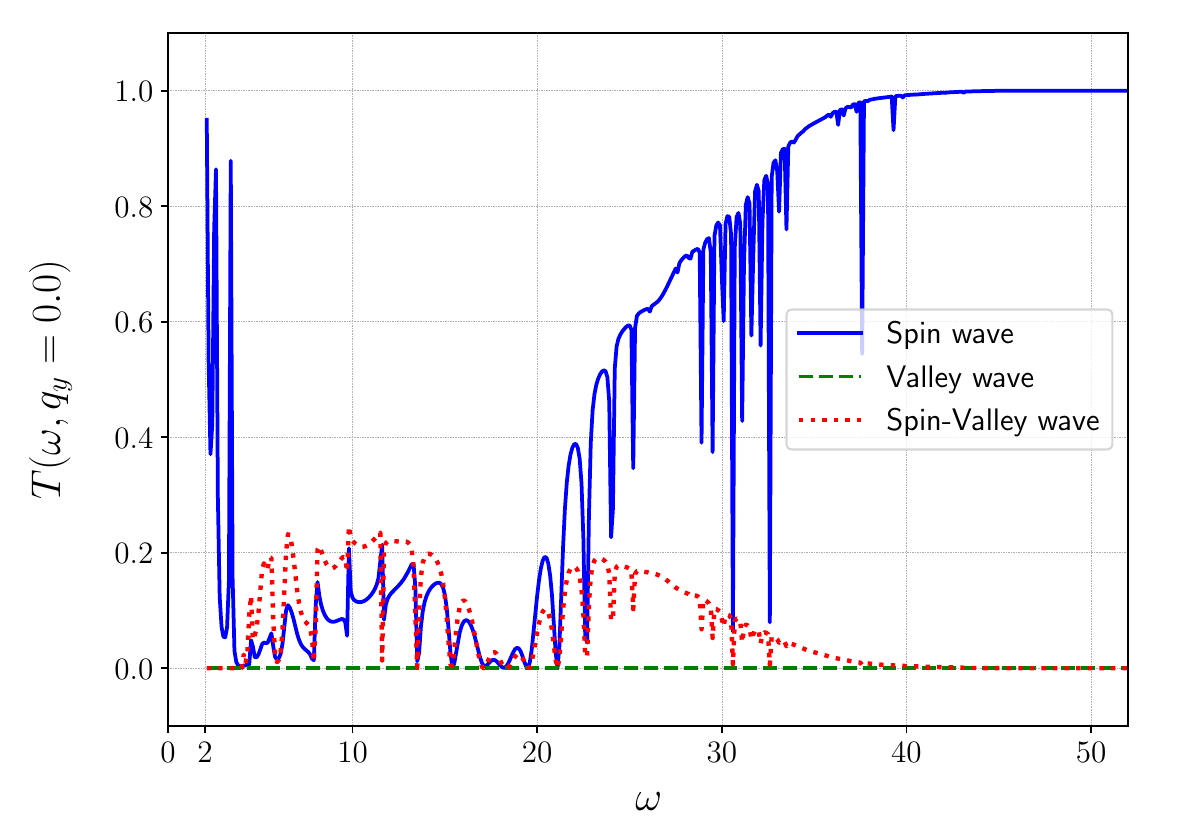}
    \end{subfigure}
    \hspace{-0.2cm}
    \begin{subfigure}{0.33\textwidth}
        \caption{$\Bp=4B_\perp^0$}
        \includegraphics[width=\textwidth,height=4cm]{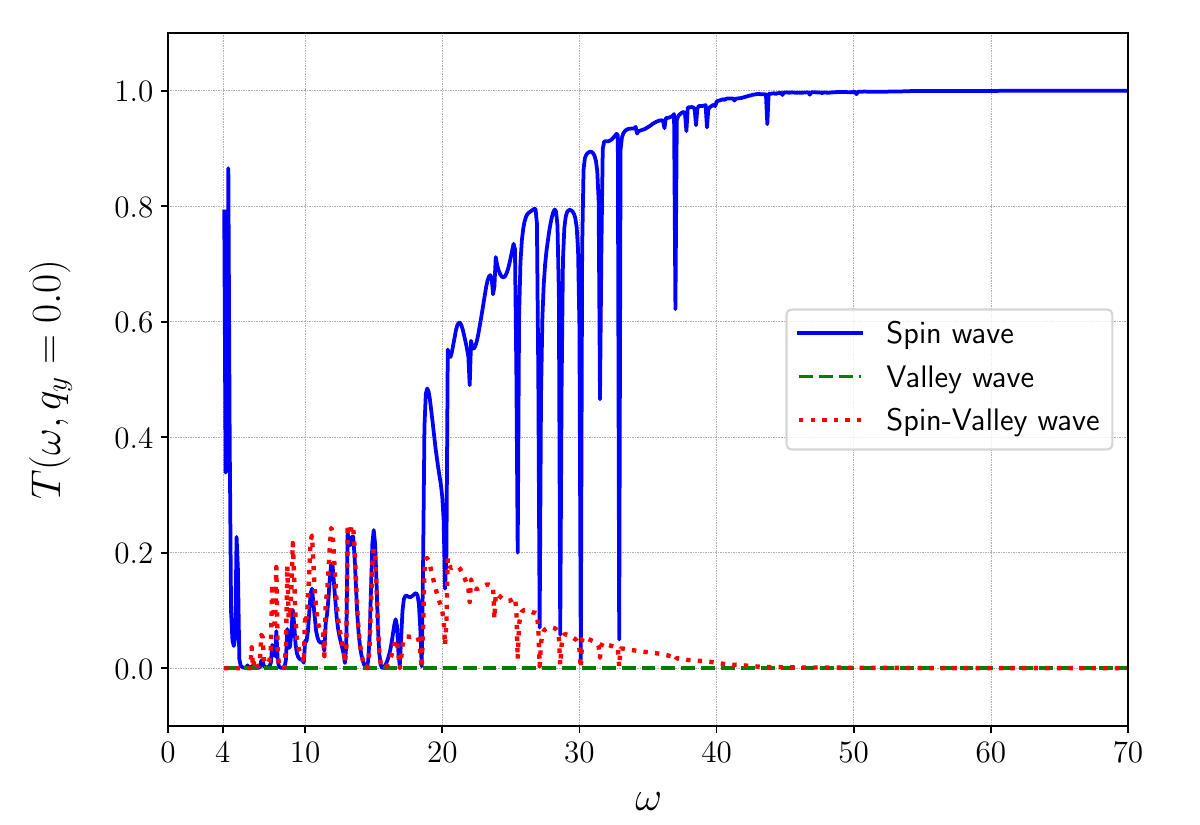}
    \end{subfigure}
    \caption{Transmission amplitudes  as a function of incoming magnon energy at $q_y=0$ for the HF state in Fig.~(\ref{fig:HF_uxy_g_uz_Ev_1}). At $B_\perp^0$ the Hamiltonian parameters are $\Wt^0=20,~ E_C^0=30,~ u_z^0=2.0,~ u_{xy}^0=-3.0,~ E_Z^0=0.5,~ E_V=1.0$. The transmission amplitude of the spin wave shows behavior consistent with Fig.~(\ref{fig:TDHF_uxy_g_uz_Ev_5}), i.e.,  when the valley remains a good quantum number, as in (a), the transmission goes to zero at larger energies and when the valley rotates, as in (b) and (c), the transmission goes to unity at high energies. For (b) and (c) at intermediate energies the spin-valley wave is also excited.}
    \label{fig:TDHF_uxy_g_uz_Ev_1}
\end{figure*}

\subsection{Magnon Scattering setup for $\nu=1|-1|1$ junction}
We will follow the method of Ref.~\cite{Wei_Huang_MacDonald2021}, and solve the following elastic scattering problem. A bulk $\nu=1$ spin magnon is sent in from the asymptotic $x\to-\infty$ $\nu=1$ region towards the junction. Given the geometry of our system, we assume that the $y$-momentum of the magnon $q_y$ is conserved. Note that the Hamiltonian of the system conserves total $S_z$. The spin magnon and the spin-valley magnon both have $S_z=-1$, because they both involve one spin-$\frac{1}{2}$ electron flipping its spin from $\uparrow$ to $\downarrow$. The valley magnon does not carry spin. This means that at the junction, the spin magnon can mix with other spin magnons and spin-valley magnons with the same $q_y$, but not with a valley magnon, because the total $S_z$ has to be conserved. The outgoing waves are either reflected or transmitted, and by the above logic, have to be either spin magnons or spin-valley magnons with the same $q_y$ and the same energy (elastic scattering conserves energy). The outgoing waves are also assumed to be detected in the asymptotic bulk regions $x\to\pm\infty$. The system is divided into three regions: In the two regions $|x|>x_0$ the HF ground state is the bulk state of $\nu=1$, while for $|x|<x_0$ the HF state for each guiding center is obtained by the method described in the previous sections. Note that the middle region is typically quite a bit bigger than the size of $\nu=-1$ region as defined by the background potential, Eq.~\ref{eq:bg}, because the one-body density matrix takes several magnetic lengths to relax to its bulk value, as seen in Figs.~\ref{fig:HF_uz_g_uxy},~\ref{fig:HF_uxy_g_uz}. We will label the guiding centers forming the region $|x|<x_0$ by the  guiding centers $[X_i$,$i=1,...,N]$. We study the transmission probability of an incoming collective mode in $X<X_1$ to an outgoing collective mode in $X>X_N$. For  $X \leq X_1$ we have, very generally, 
\begin{align}
    \Psi_{X,m,n}(q_y,\omega)=\frac{\phi_{mn}^s}{\sqrt{v_s}}\big[e^{iq_x^sX}+r_s e^{-iq_x^sX}\big]
    \nonumber \\
    +\frac{r_v \phi_{mn}^v}{\sqrt{v_v}}~e^{-iq_x^vX}+\frac{r_{sv} \phi_{mn}^{sv}}{\sqrt{v_{sv}}}~e^{-iq_x^{sv}X}
    \label{eq:incoming_wavefunc}
\end{align}
where $r_{s},r_{v},$ and $r_{sv}$ are the reflection coefficients for the spin wave, valley wave and spin-valley wave respectively.  The momentum vector $q_x$ for each bulk collective mode is determined as the positive solution of the equation expressing the conservation of energy 
\begin{equation}
    \omega_\alpha(q_x^\alpha,q_y)=\omega.
    \label{eq:kinematic}
\end{equation} 
Here, $v_\alpha(q_x^\alpha,q_y)=(\frac{dq_x^\alpha}{d\omega})^{-1},~\alpha=s,v,sv$ are the group velocities of the spin wave,valley wave and spin-valley waves respectively. Recall also  that by $S_z$ conservation valley magnons cannot be generated, and thus $r_{v}=0$. Similarly for $X \geq X_N$ the most general solution is 
\begin{align}
   \Psi_{X,m,n}(q_y,\omega)=\frac{t_s \phi_{mn}^s}{\sqrt{v_s}}~e^{iq_x^sX}+\frac{t_v \phi_{mn}^v}{\sqrt{v_v}}~e^{iq_x^vX}
   \nonumber \\
   +\frac{t_{sv} \phi_{mn}^{sv}}{\sqrt{v_{sv}}}~e^{iq_x^{sv}X} \hspace{1.7cm}
   \label{eq:outgoing_wavefunc}
\end{align}
where $t_{s},t_{v},$ and $t_{sv}$ are the transmission coefficients for the spin wave, valley wave and spin-valley wave respectively. The same logic that dictates $r_v=0$ also forces $t_v=0$. It is also possible that Eq.~\ref{eq:kinematic} has no real solutions for a particular mode in a certain range of $\omega$, in which case that mode will be kinematically forbidden. 

If the full wave function of the collective excitations is known,  the reflection coefficients can be obtained at $X=X_1$ as~\cite{Wei_Huang_MacDonald2021}
\begin{align}
    r_s(q_x,q_y)=\big[\sqrt{v_s}\Bar{\phi}_{mn}^s(q)\Psi_{X_1,m,n}(q_y,\omega)-e^{iq_x^sX_1}]e^{iq_x^sX_1}
    \nonumber \\
    r_v(q_x,q_y)=\sqrt{v_v}\Bar{\phi}_{mn}^{v}(q)\Psi_{X_1,m,n}(q_y,\omega)e^{iq_x^vX_1} \hspace{1.6cm}
    \nonumber \\
    r_{sv}(q_x,q_y)=\sqrt{v_{sv}}\Bar{\phi}_{mn}^{sv}(q)\Psi_{X_1,m,n}(q_y,\omega)e^{iq_x^{sv}X_1} .\hspace{1.35cm}
    \label{eq:reflection_coeff}
\end{align}
Similarly, the transmission coefficient can be found from the wave function at $X=X_N$ as~\cite{Wei_Huang_MacDonald2021}
\begin{align}
    t_s(q_x,q_y)=\sqrt{v_s}\Bar{\phi}_{mn}^{s}(q)\Psi_{X_N,m,n}(q_y,\omega)e^{-iq_x^sX_N} 
    \nonumber \\
    t_v(q_x,q_y)=\sqrt{v_v}\Bar{\phi}_{mn}^{v}(q)\Psi_{X_N,m,n}(q_y,\omega)e^{-iq_x^vX_N} 
    \nonumber \\
    t_{sv}(q_x,q_y)=\sqrt{v_{sv}}\Bar{\phi}_{mn}^{sv}(q)\Psi_{X_N,m,n}(q_y,\omega)e^{-iq_x^{sv}X_N}.
    \label{eq:transmission_coeff}
\end{align}
In writing the above expressions for reflection and transmission coefficients, we have assumed sums over repeated indices.

The next task is to find $\Psi_{X_1,m,n}(q_y,\omega)$ and $\Psi_{X_N,m,n}(q_y,\omega)$ for the $\nu=1|-1|1$ junction. We follow the procedure outlined in Ref.~\cite{Wei_Huang_MacDonald2021}. In this approach, we  take the full TDHF equation, Eq.~\ref{eq:TDHF}, and use the fact that the form of the solution is known in the asymptotic regions $X<X_1,~X>X_N$. We then integrate out these asymptotic regions to obtain an inhomogeneous set of equations only for the region $X_1\leq X\leq X_N$. 
\begin{align}
    \sum_{i'=1}^N \sum_{m'n'}\big[\mathbb{K}_{X_i,m,n}^{X_{i'},m',n'}(q_y) + (\Sigma^L)_{X_i,m,n}^{X_{i'},m',n'}(q_y) + \hspace{1cm}
    \nonumber \\
    (\Sigma^R)_{X_i,m,n}^{X_{i'},m',n'}(q_y) - \omega \delta_{ii'}\delta_{mm'}\delta_{nn'}\big]\Psi_{X_{i'},m',n'}(q_y,\omega)
    \nonumber \\
    =V_{X_i,m,n}^s(q_y,\omega),~ 1\leq i\leq N \hspace{2cm}
    \label{eq:effective_TDHF}
\end{align}
Here 
\begin{align}
    (\Sigma^L)_{X_i,m,n}^{X_{i'},m',n'}(q_y)=\delta_{i',1}\sum_{j<1} \sum_\beta \mathbb{K}_{X_i,m,n}^{X_j,m',n'}(q_y) \delta_{\beta,(m',n')} \nonumber \\ 
    \times \hspace{0.1cm} e^{-iq_x^\beta(X_j-X_1)},
    \nonumber \\
     (\Sigma^R)_{X_i,m,n}^{X_{i'},m',n'}(q_y)=\delta_{i',N}\sum_{j>N} \sum_\beta \mathbb{K}_{X_i,m,n}^{X_j,m',n'}(q_y) \delta_{\beta,(m',n')} \nonumber \\ 
    \times \hspace{0.1cm} e^{iq_x^\beta(X_j-X_N)}
\end{align} 
are the self-energy contributions which arise from integrating out the asymptotic regions. With $\beta$ running over all the indices $(m,n)$ of the bulk collective modes. $\mathbb{K}_{X_i,m,n}^{X_{i'},m',n'}(q_y)$ is given in Eq.~(\ref{eq:Kernel}) with $X_i=k\ell^2$ and $X_{i'}=k'\ell^2$.  The inhomogeneous source term is given by
\begin{align}
    V_{X_i,m,n}^s(q_y,\omega)=\sum_{j<1}\sum_{m'n'} \frac{1}{\sqrt{v_s}} \mathbb{K}_{X_i,m,n}^{X_j,m',n'}(q_y)\phi_{m'n'}^s(q)
    \nonumber \\
    \times \big[e^{-iq_x^s(X_j-2X_1)}-e^{iq_x^sX_j}].
\end{align}
Finally, we solve the inhomogeneous system of matrix equations (\ref{eq:effective_TDHF}) to find $\Psi_{X_1,m,n}$ and $\Psi_{X_N,m,n}$ for general $q_y$ and $\omega$~\cite{Wei_Huang_MacDonald2021} from which one can read out the reflection and transmission coefficients using Eqs.~(\ref{eq:reflection_coeff}),~(\ref{eq:transmission_coeff}).

Let us now turn to the results.

\begin{figure}
    \centering
    \includegraphics[width=0.4\textwidth,height=5cm]{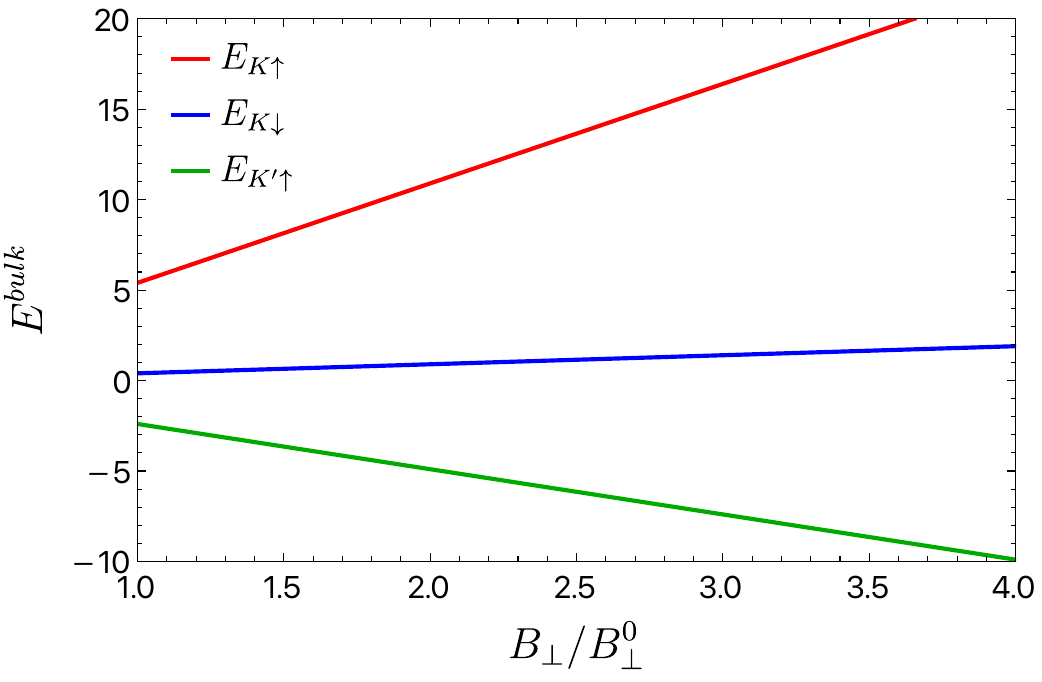}
    \caption{Occupied Hartree-Fock single-particle energies for the bulk $\nu=1$ as a function of $\Bp/B_\perp^0$ with  $V_{ex}=0$, $E_Z^0=0.5$, $E_V=0.1$, $u_z^0=4$ and $u_{xy}^0=-3$. The parameters satisfy $u_z^0 > |u_{xy}^0| > E_Z^0 > E_V $. In this case the ordering of the occupied levels is $E_{K'\ua}<E_{K\da}<E_{K\ua}$ which remains unchanged with increasing $\Bp$.}
    \label{fig:bulk_ordering_uz_g_uxy_Ev_0.1}
\end{figure}

\begin{figure*}
    \centering
    \begin{subfigure}{0.4\textwidth}
      \caption{$\Bp=B_\perp^0$}
      \includegraphics[width=\textwidth,height=7cm]{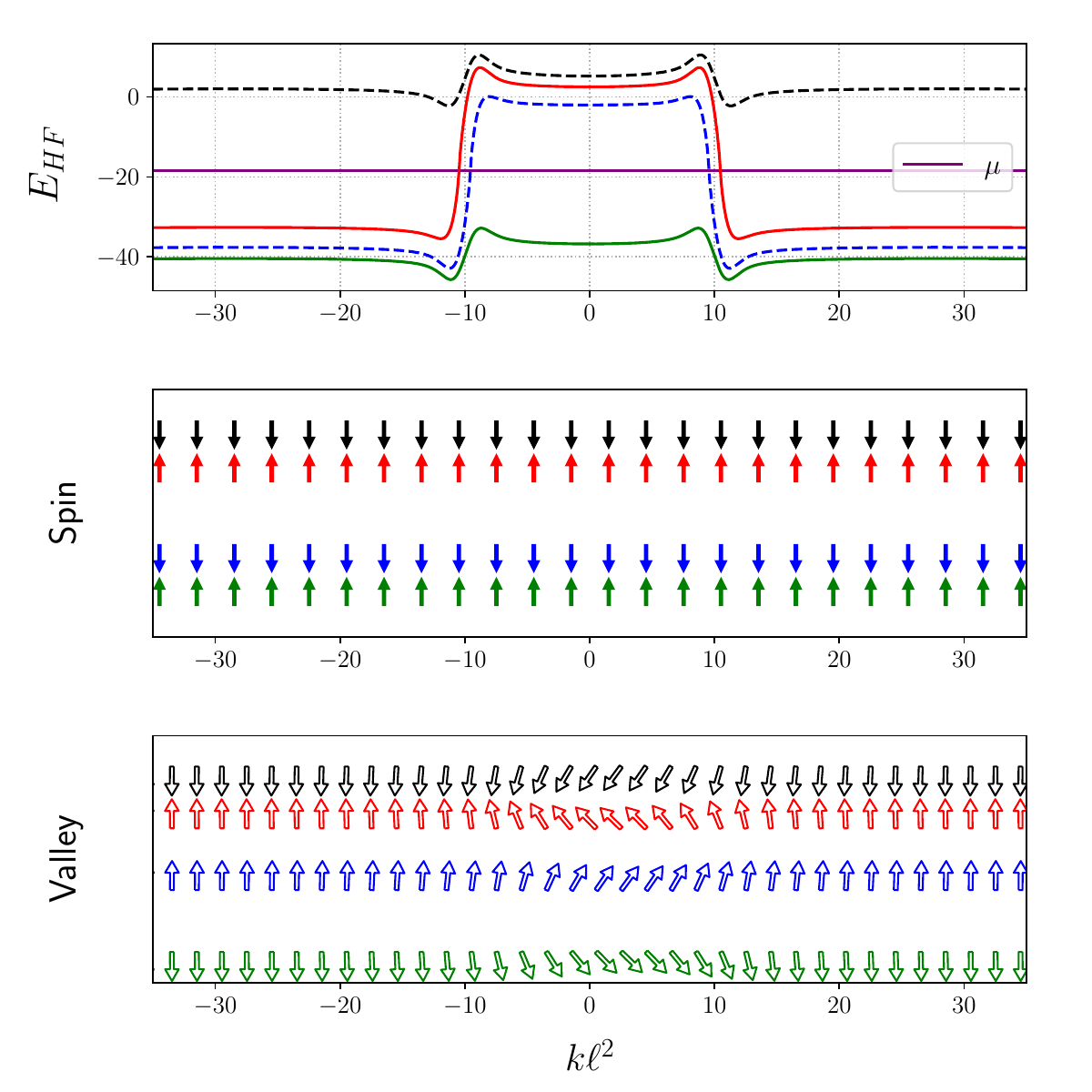}
    \end{subfigure}
    \hspace{0.3cm}
    \begin{subfigure}{0.4\textwidth}
      \caption{$\Bp=2B_\perp^0$}
      \includegraphics[width=\textwidth,height=7cm]{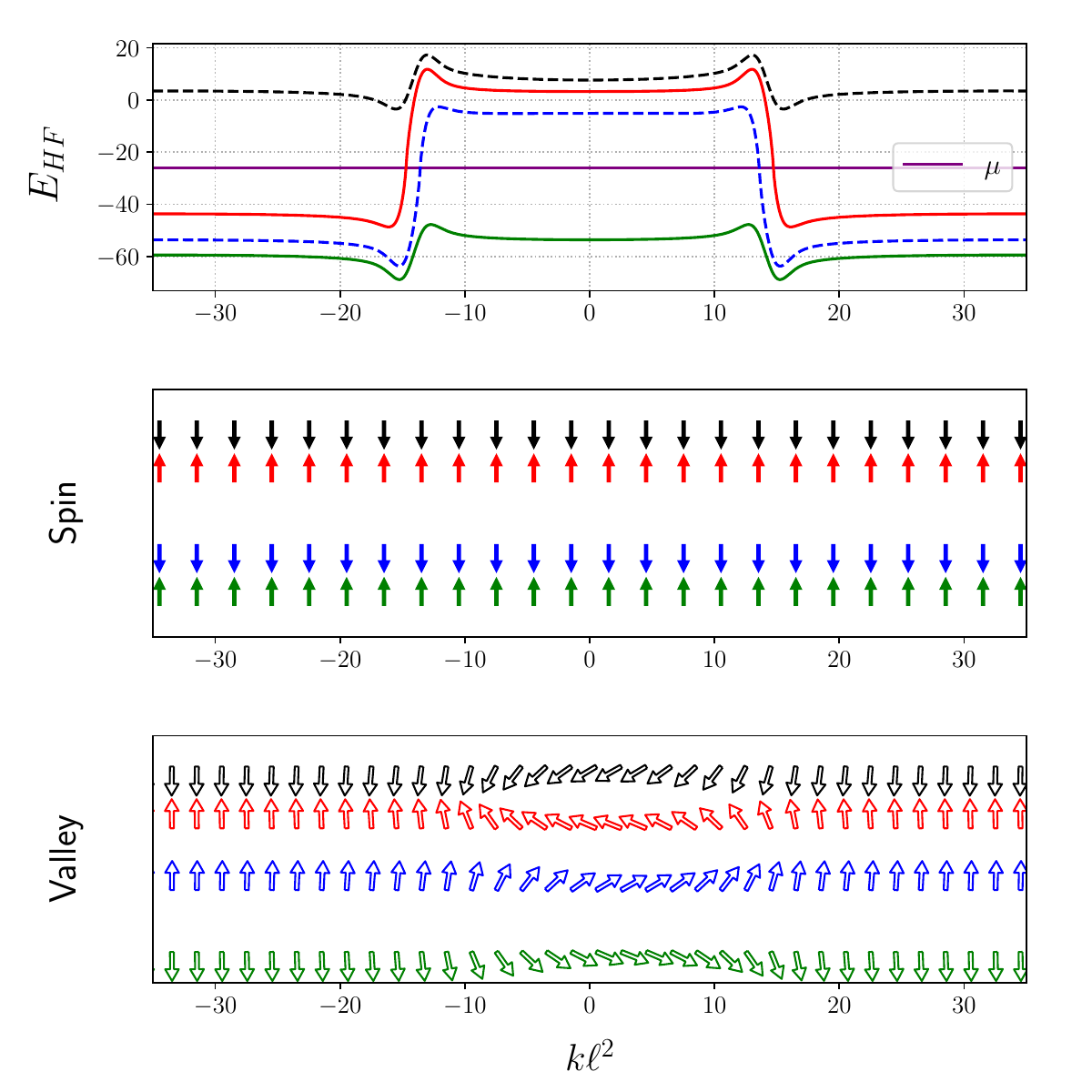}
    \end{subfigure}
    \caption{Hartree-Fock single-particle energies (top panel), spin structure (middle panel), and valley structure (bottom panel) of each energy level for two different values of the perpendicular magnetic field $B_\perp^0, 2B_\perp^0$. At $B_\perp^0$ the Hamiltonian parameters are $\Wt^0=20, E_C^0=30, u_z^0=4.0, u_{xy}^0=-3.0, E_Z^0=0.5, E_V=0.1$ which is same as that of Fig.~(\ref{fig:bulk_ordering_uz_g_uxy_Ev_0.1}). In this case the self-consistent states preserve the spin quantum number while the valley of each level rotates in the middle $\nu=-1$ region. Because of very small but nonzero $E_V$ the occupied state in the middle region has both valley components.}
    \label{fig:HF_uz_g_uxy_Ev_0.1}
\end{figure*}

\begin{figure*}
    \centering
    \begin{subfigure}{0.4\textwidth}
      \caption{$\Bp=1B_\perp^0$}
      \includegraphics[width=\textwidth,height=4cm]{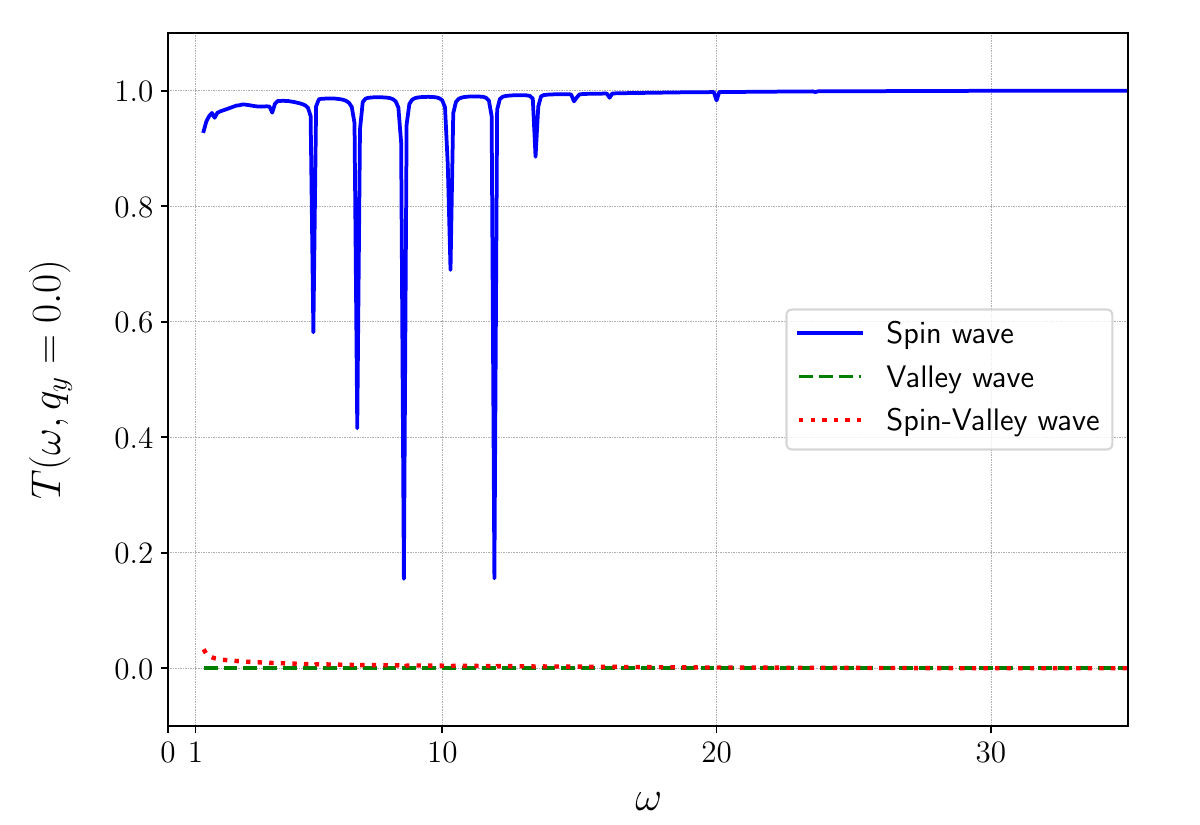}
    \end{subfigure}
    \hspace{0.3cm}
    \begin{subfigure}{0.4\textwidth}
      \caption{$\Bp=2B_\perp^0$}
      \includegraphics[width=\textwidth,height=4cm]{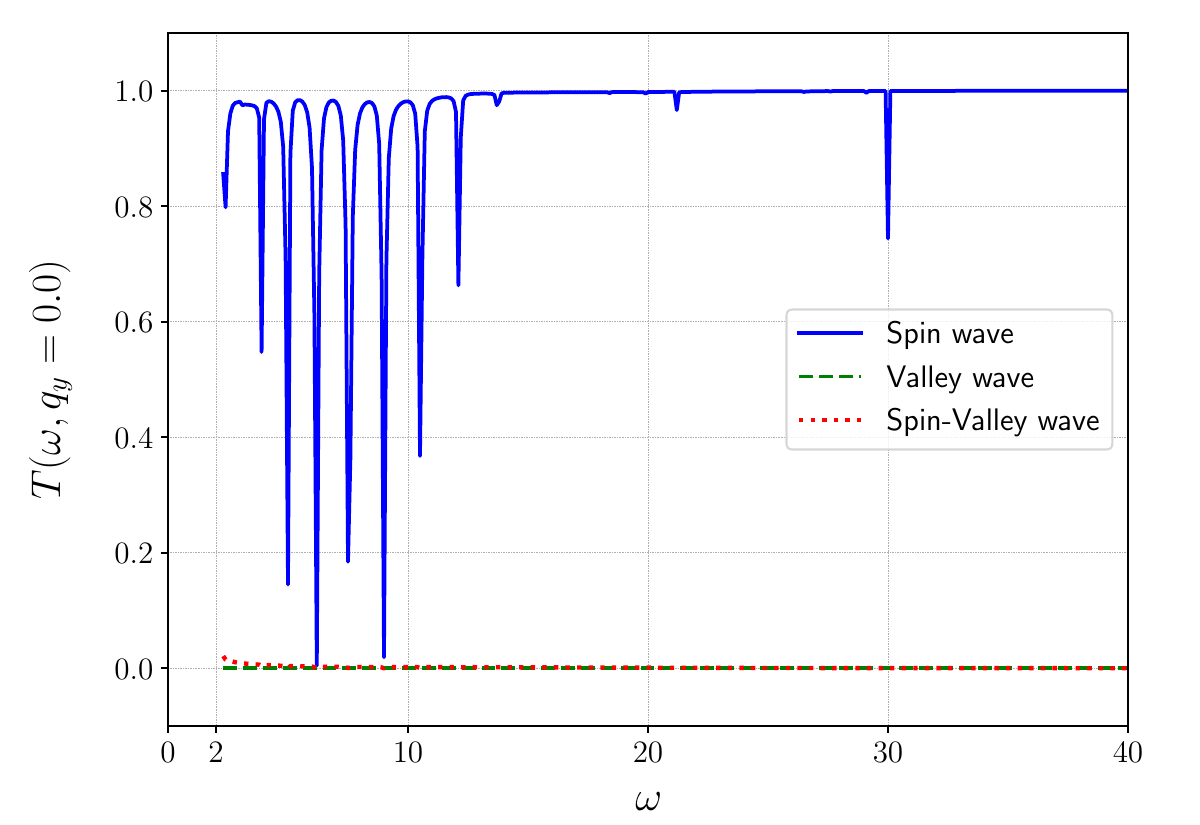}
    \end{subfigure}
    \caption{Transmission amplitudes as a function of incoming magnon energy at $q_y=0$. The parameters correspond to the HF state in Fig.~(\ref{fig:HF_uz_g_uxy_Ev_0.1}). At $B_\perp^0$ the Hamiltonian parameters are $\Wt^0=20, E_C^0=30, u_z^0=4.0, u_{xy}^0=-3.0, E_Z^0=0.5, E_V=0.1$. The spin wave is fully transmitted across the junction with some transmission dips at smaller energies. The qualitative behavior remains the same as $\Bp$ increases.}
    \label{fig:TDHF_uz_g_uxy_Ev_0.1}
\end{figure*}

\begin{figure}
    \centering
    \includegraphics[width=0.4\textwidth,height=5cm]{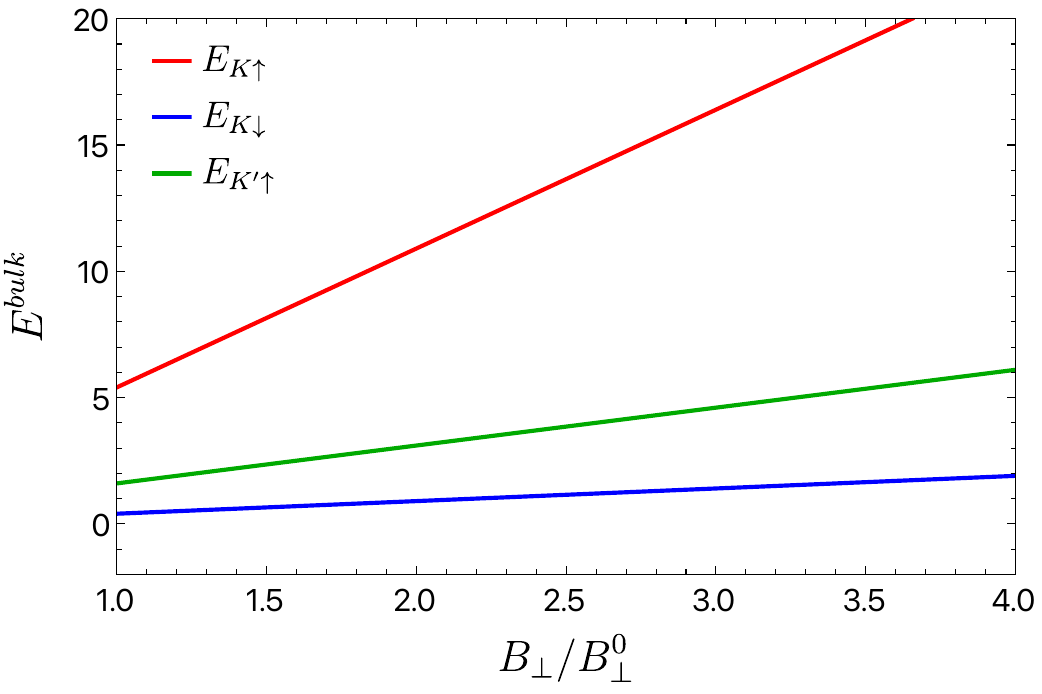}
    \caption{Occupied Hartree-Fock single-particle energies for the bulk $\nu=1$ as a function of $\Bp/B_\perp^0$ with $V_{ex}=0$, $E_Z^0=0.5$, $E_V=0.1$, $u_z^0=2$ and $u_{xy}^0=-3$. The parameters satisfy $|u_{xy}^0|> u_z^0 > E_Z^0 > E_V $. For this case the ordering of the occupied states is $E_{K\downarrow} < E_{K'\uparrow} < E_{K\uparrow}$ which does not change with increasing $\Bp$.}
    \label{fig:bulk_ordering_uxy_g_uz_Ev_0.1}
\end{figure}

\begin{figure*}
    \centering
    \begin{subfigure}{0.4\textwidth}
      \caption{$\Bp=1B_\perp^0$}
      \includegraphics[width=\textwidth,height=7cm]{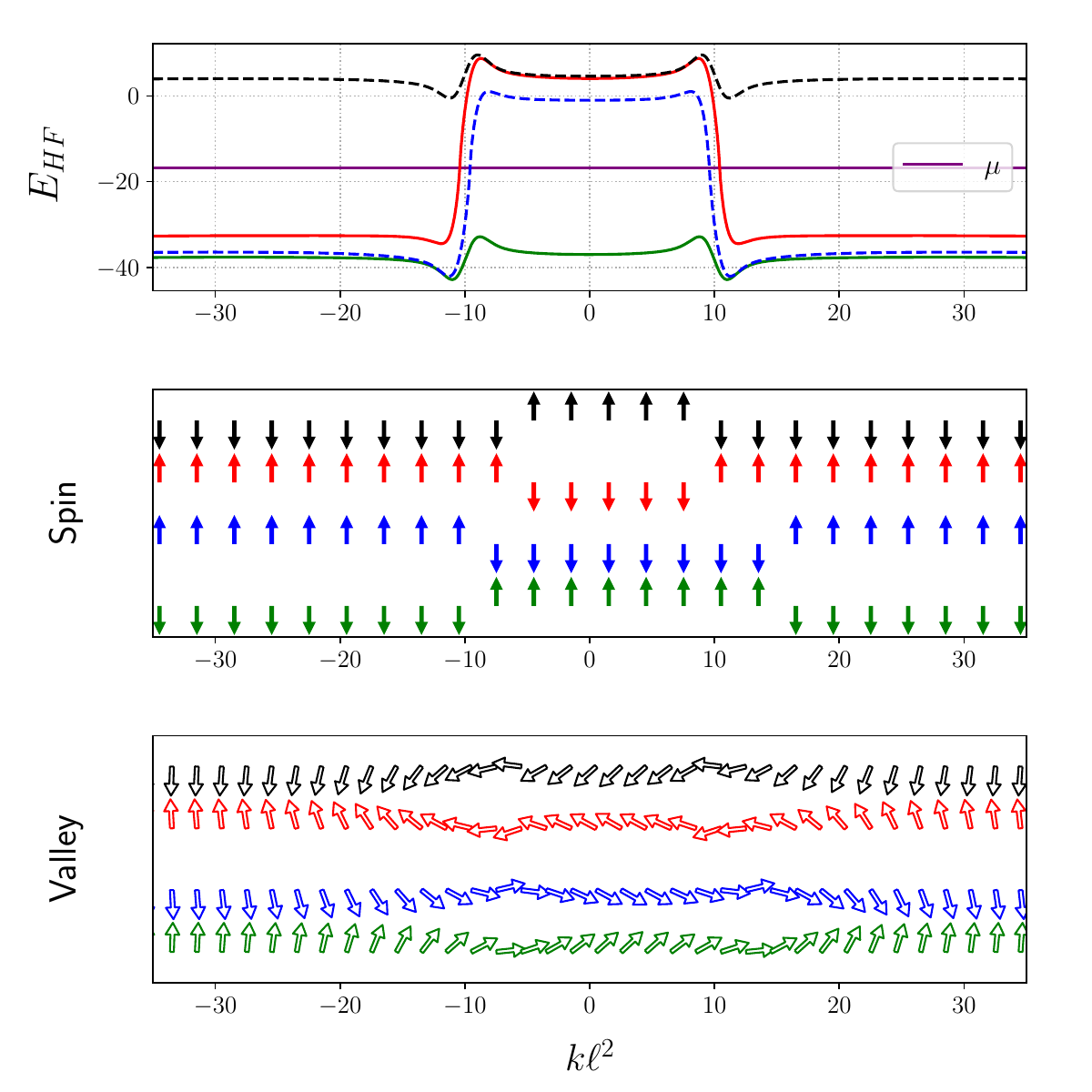}
    \end{subfigure}
    \hspace{0.3cm}
    \begin{subfigure}{0.4\textwidth}
      \caption{$\Bp=2B_\perp^0$}
      \includegraphics[width=\textwidth,height=7cm]{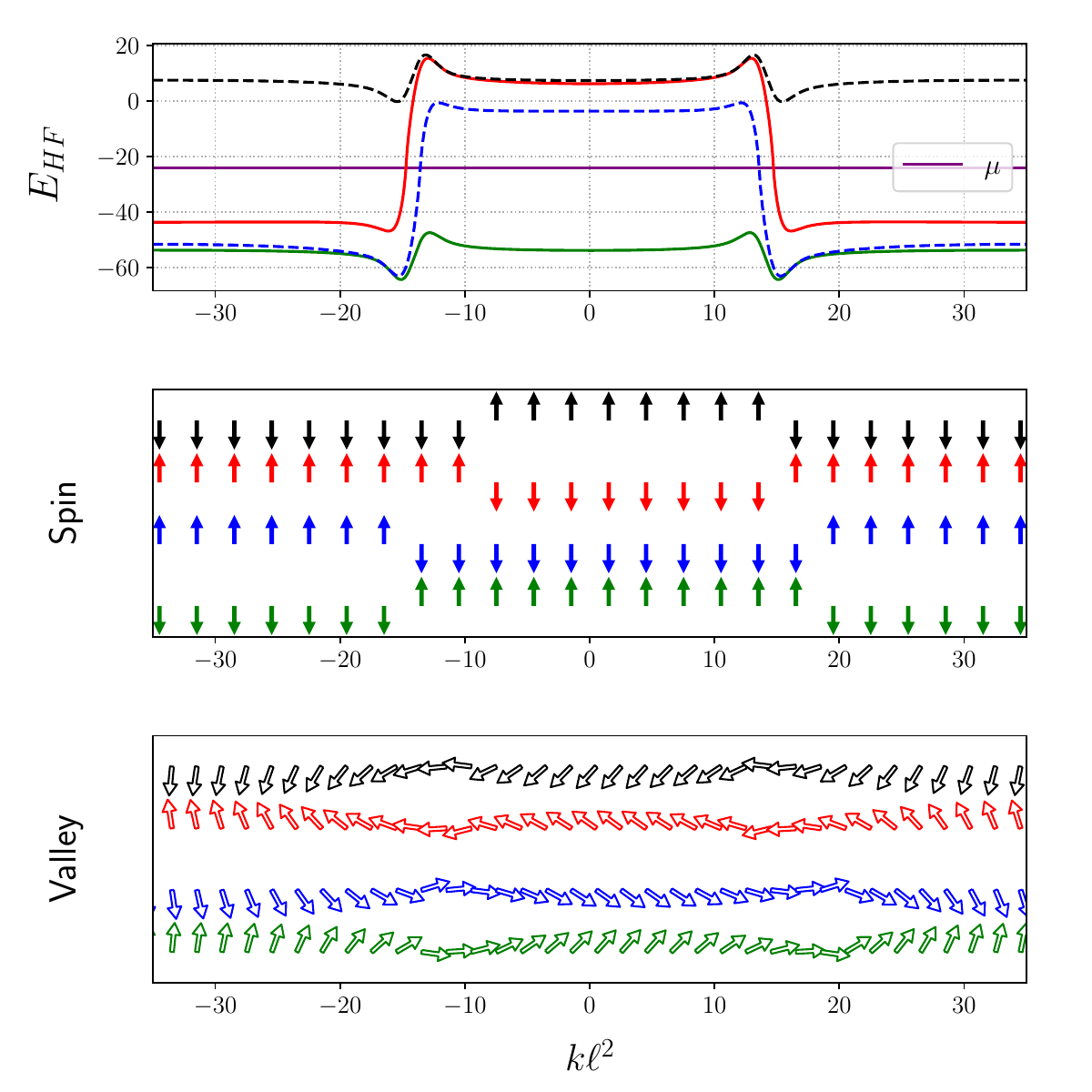}
    \end{subfigure}
    \caption{Hartree-Fock single-particle energies (top panel), spin structure (middle panel), and valley structure (bottom panel) of each energy level for two different values of the perpendicular magnetic field $B_\perp^0, 2B_\perp^0$. At $B_\perp^0$ the Hamiltonian parameters are $\Wt^0=20, E_C^0=30, u_z^0=2.0, u_{xy}^0=-3.0, E_Z^0=0.5, E_V=0.1$, same as that in Fig.~(\ref{fig:bulk_ordering_uxy_g_uz_Ev_0.1}). For this case we find the spin of each HF level flips across the interface and the valley rotates continuously across the junction. The occupied state in the $\nu=-1$ region has both valley components.}
    \label{fig:HF_uxy_g_uz_Ev_0.1}
\end{figure*}

\begin{figure*}
    \centering
    \begin{subfigure}{0.4\textwidth}
      \caption{$\Bp=1B_\perp^0$}
      \includegraphics[width=\textwidth,height=4cm]{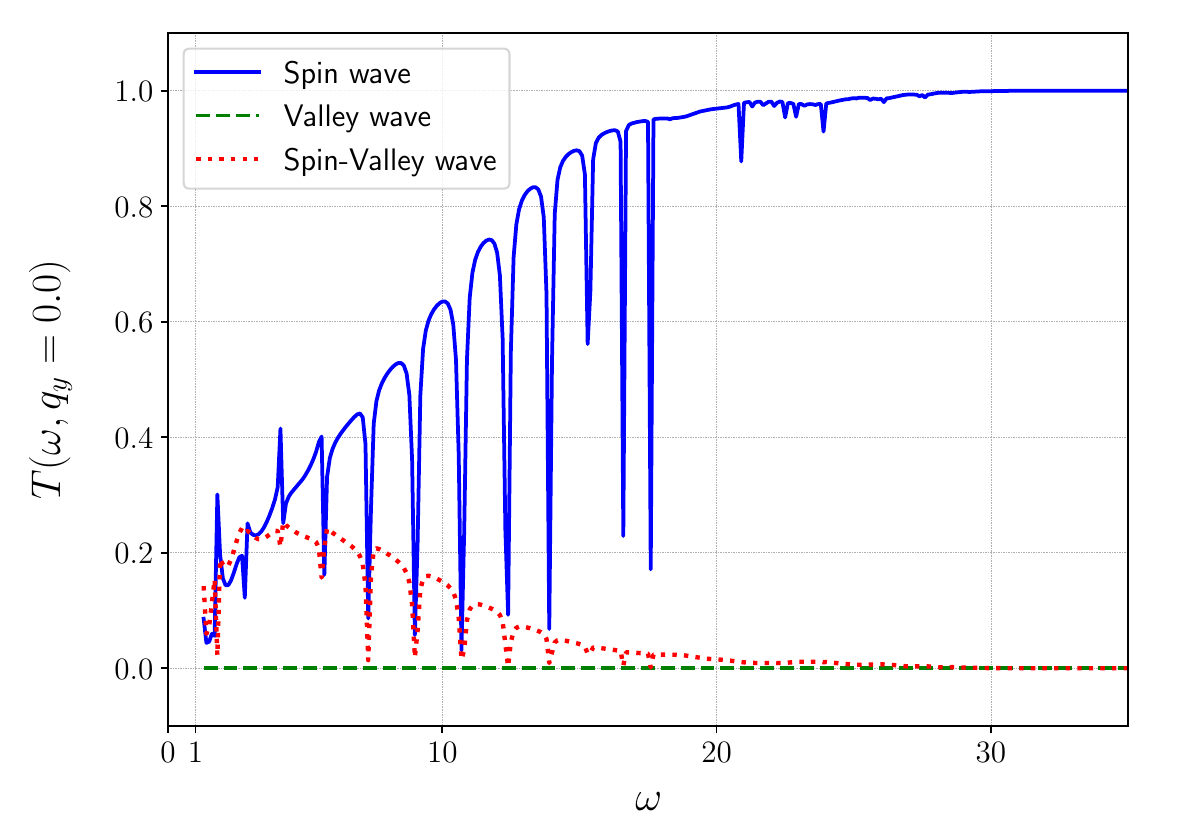}
    \end{subfigure}
    \hspace{0.3cm}
    \begin{subfigure}{0.4\textwidth}
      \caption{$\Bp=2B_\perp^0$}
      \includegraphics[width=\textwidth,height=4cm]{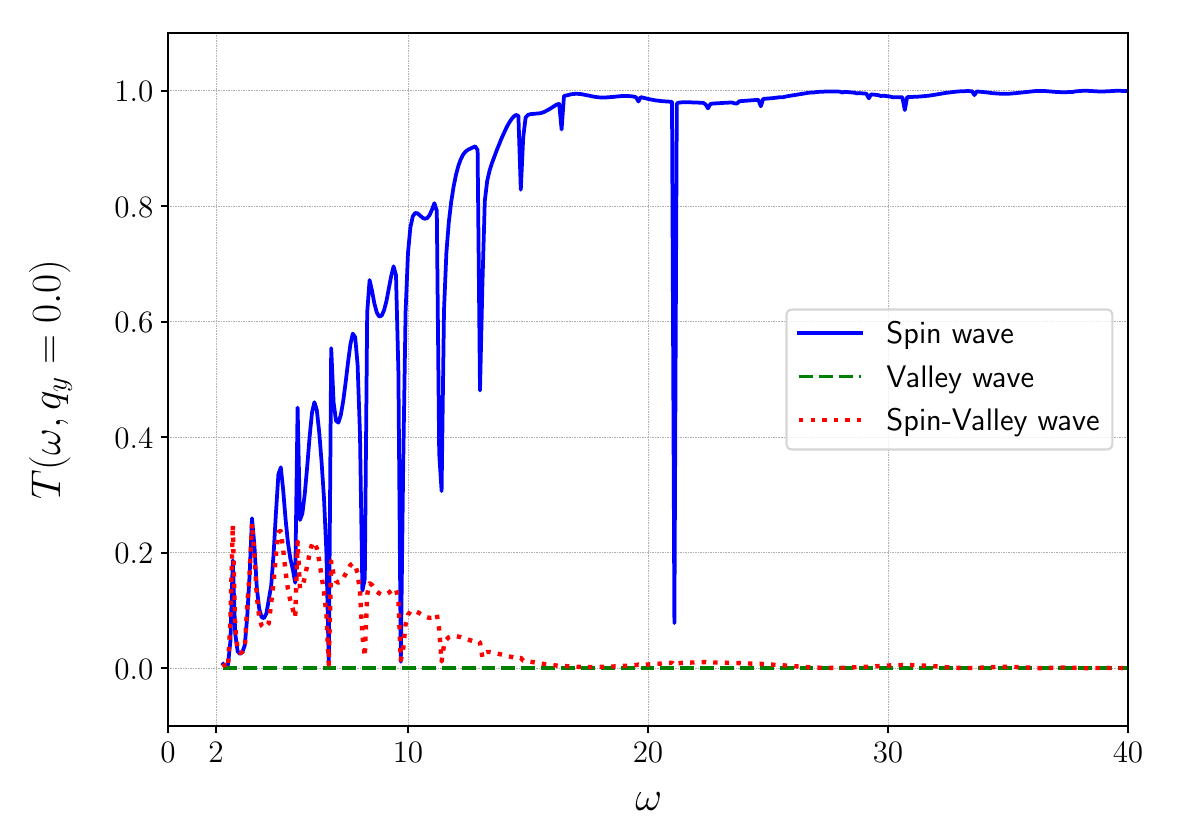}
    \end{subfigure}
    \caption{Transmission amplitudes  as a function of incoming magnon energy at $q_y=0$ for the HF states in Fig.~(\ref{fig:HF_uxy_g_uz_Ev_0.1}). At $B_\perp^0$ the Hamiltonian parameters are $\Wt^0=20, E_C^0=30, u_z^0=2.0, u_{xy}^0=-3.0, E_Z^0=0.5, E_V=0.1$. As in previous cases where the valley of each HF level rotates, we find the spin waves are transmitted perfectly at large energies. }
    \label{fig:TDHF_uxy_g_uz_Ev_0.1}
\end{figure*}

\section{Magnon transmission across the $1|-1|1$ junction}
\label{sec:results}

In view of the large number of parameters in the problem, $E_V,~B_\perp,~u^0_z,~u_{xy}^0,\Wt^0$, we need to organize the results. The case $E_V=0$ was already addressed in Ref.~\cite{Wei_Huang_MacDonald2021}, and hence we will always take $E_V>0$ in what follows. We will focus solely on the regime of short-range interactions where $u^0_z>0,~u_{xy}^0<0$, believed to be realized in graphene. Here there are two major cases, (i) $u_z>|u_{xy}|$ and (ii) $u_z<|u_{xy}|$. Our second organizing principle is to start with small $\Bp$ and go towards large $\Bp$. At small $B_\perp$, for samples  $E_V$ will typically be larger than the short-range couplings (due to the dependence of $u_z,~u_{xy}$ on $B_\perp$), whereas at larger $B_\perp$ the short-range couplings may be of the same order or even larger than $E_V$. The relative magnitudes of the short-range couplings vs $E_V$ have profound effects on the self-consistent structure of the interfaces, and thus on the magnon transmission amplitudes across the system. One of the primary motivations of this work is to use such as study to constrain the ratio $u_z/|u_{xy}|$. 

In view of these considerations, this section is organized as follows: In Subsection~\ref{subsec:large_Ev}, we examine the case $E_V> u^0_z,|u_{xy}^0|$ over a range of $B_\perp$. Results for both $u_z>|u_{xy}|$ and $u_z<|u_{xy}|$ will be presented. We find that for $u_z>|u_{xy}|$ there is no qualitative change in the magnon transmission amplitudes as $B_\perp$ increases, while for $u_z<|u_{xy}|$ the magnon transmission amplitude at high energy changes dramatically as $B_\perp$ increases. Interestingly, in this latter case, a spin-valley magnon is also excited. Next, in Subsection~\ref{subsec:intermediate_Ev}, we examine the case of intermediate $E_V$, with the same subcases as in Subsection~\ref{subsec:large_Ev}. Finally, in Subsection~\ref{subsec:small_Ev}, we consider the case when $E_V$ is the smallest energy scale in the problem, with the inequalities $|u_\alpha^0| > E_Z^0 > E_V$.

\subsection{Magnon transmission for large $E_V$}
\label{subsec:large_Ev}

Throughout this section, we assume $E_V > |u_\alpha^0| > E_Z^0$ with $\alpha=z,xy$. 

Let us start with the case $u_z^0 > |u_{xy}^0|$. We choose the parameters $\Wt^0=20, E_C^0=30, u_z^0=4.0, u_{xy}^0=-3.0, E_Z^0=0.5, E_V=5.0$, which are the same parameters that we used to illustrate the $\nu=1$ bulk ordering in Fig.~(\ref{fig:bulk_ordering_uz_g_uxy_Ev_5}). Now, as seen in Fig.~(\ref{fig:bulk_ordering_uz_g_uxy_Ev_5}), there are three possible orderings of the energies of the filled one-body states as a function of $\Bp$. To sample all the orderings we have chosen three values of $\Bp=B_\perp^0,2B_\perp^0,4B_\perp^0$. 

The self-consistent HF results are shown in Fig.~\ref{fig:HF_uz_g_uxy_Ev_5}, where it is seen that the system prefers to rotate the spins through the interface regions, while the valley degree of freedom remains polarized. This is to be expected because of the large value of $E_V$. Thus there is no qualitative change in the nature of the interface as $\Bp$ increases. The transmission amplitudes for the bulk collective modes for $\Bp=B_\perp^0,2B_\perp^0,4B_\perp^0$ are shown in Fig.~\ref{fig:TDHF_uz_g_uxy_Ev_5}.

As expected from the adiabatic continuity seen in the HF configurations, there is no qualitative change in the magnon transmission amplitudes as $\Bp$ increases. Furthermore, the spin magnon coming in from the $\nu=1$ bulk to the left is excited in the $K'$ valley, because the only unoccupied state is $K'\downarrow$. However, the $\nu=-1$ bulk has only $K\uparrow$ occupied, hence it's spin magnon is excited in the $K$ valley. This incompatibility, combined with the fact that the valley is a good quantum number, leads to very low transmission of magnons throughout the energy range. There are a few sharp peaks which we attribute to resonances between the cavity magnons inside the $\nu=-1$ region and the incoming magnons, mediated by the collective modes at the interface. It is also worth noting that only the spin-magnon is excited, because the spin-valley magnon has a very high energy due to the large value of $E_V$.

We next turn to the case $u_z^0 < |u_{xy}^0|$. We choose the parameters $\Wt^0=20, E_C^0=30, u_z^0=2.0, u_{xy}^0=-3.0, E_Z^0=0.5, E_V=5.0$, the same as those used in Fig.~\ref{fig:bulk_ordering_uxy_g_uz_Ev_5}. As shown in  Fig.~\ref{fig:bulk_ordering_uxy_g_uz_Ev_5}, the ordering of the one-body levels changes once as $\Bp$ increases. $E_{K\downarrow}$ is always the lowest state. For $\Bp<B_c\approx2.6B_\perp^0$ we have $E_{K'\uparrow}>E_{K\uparrow}$, while for $\Bp>B_c\approx2.6B_\perp^0$, we have $E_{K\uparrow}>E_{K'\uparrow}$. The HF self-consistent results are shown in Fig.~\ref{fig:HF_uxy_g_uz_Ev_5}. 

For $\Bp<B_c$ both spin and valley are good quantum numbers for the one-body levels. The spin of the levels changes via level crossings while the valley quantum number remains unaltered. Once again, we expect the incident magnon from the left to be in the $K'$ valley, while the bulk magnon of $\nu=-1$ is in the $K$ valley. Thus, we can expect very little transmission, except at resonances mediated by the collective modes at the two edges of the $\nu=-1$ region.

This is indeed the case, as seen in Fig.~\ref{fig:TDHF_uxy_g_uz_Ev_5}a,b.  For $\Bp>B_c$, however, the valley is no longer a good quantum number across the system. As can be seen in Fig.~\ref{fig:HF_uxy_g_uz_Ev_5}c, the spin remains a good quantum number. Thus, we can expect mixing between the bulk $K'$ spin magnon of $\nu=1$ and the bulk $K$ spin magnon of $\nu=-1$. This expectation is indeed borne out in Fig.~\ref{fig:TDHF_uxy_g_uz_Ev_5}c, where we see that at high energies spin magnons are almost fully transmitted across the system. This almost perfect transmission at high energies can be understood as follows: high-energy magnons have a very short wavelength, much smaller than the length scale over which the valley superpositions change in Fig.~\ref{fig:HF_uxy_g_uz_Ev_5}. Since the valley is not a good quantum number, the magnon can ``adiabatically" change its valley components as it traverses the interfaces. Furthermore, due to the valley rotations at the interfaces and at the high energy, it now becomes possible to excite the spin-valley magnon, which are shown in the red traces in Fig.~\ref{fig:HF_uxy_g_uz_Ev_5}c.  Since the $q=0$ energy of the spin-valley magnon is $2E_Z+2E_V$, it occurs only for $\omega>11$ in our units.

\subsection{Magnon transmission at intermediate $E_V$}
\label{subsec:intermediate_Ev}
Throughout this subsection, we assume the inequalities $|u_\alpha^0| > E_V > E_Z^0$ with $\alpha=z,xy$.

First, we consider the case where $u_z^0 > |u_{xy}^0|$. We illustrate this case with the parameters $u_z^0=4,~u_{xy}^0=-3,~E_V=1,~E_Z=0.5$, for which the one-body energies are shown as a function of $B_\perp$ in Fig.~\ref{fig:bulk_ordering_uz_g_uxy_Ev_1}. As can be seen, we always have $E_{K'\ua} < E_{K\da} < E_{K\ua} < E_{K'\da}$ for $\nu=1$. The occupied state for $\nu=-1$ is $\Kua$.

The self-consistent HF solutions for $\Bp=B_\perp^0,2B_\perp^0,4B_\perp^0$ are shown in Fig.~ \ref{fig:HF_uz_g_uxy_Ev_1}.  As in the case of large $E_V$, we find that for $u_z>|u_{xy}|$ the system prefers to undergo spin rotations at the interfaces, leaving the valley quantum number conserved. It is therefore not surprising that the transmission amplitudes, shown in Fig.~ \ref{fig:TDHF_uz_g_uxy_Ev_1}, are very similar to those at large $E_V$ (Fig.~\ref{fig:TDHF_uz_g_uxy_Ev_5}). There are sharp peaks at low energies, which we believe represent coupling to the cavity collective modes in the $\nu=-1$ region mediated by the interface collective modes. At high energies the transmission drops to zero.

Next, we consider the case where $u_z^0 < |u_{xy}^0|$. We have illustrated this case with the following choice of parameters: $u_z^0=2,~u_{xy}^0=-3,~E_V=1,~E_Z=0.5$. As shown in Fig.~\ref{fig:bulk_ordering_uxy_g_uz_Ev_1}, the ordering of the one-body energies does not change as $B_\perp$ increases.

The self-consistent HF solutions for $\Bp=B_\perp^0,2B_\perp^0,4B_\perp^0$ are shown in Fig.~\ref{fig:HF_uxy_g_uz_Ev_1}. For this case, we find that for low $B_\perp$ the HF solution conserves both spin and valley quantum numbers. However, beyond a certain $B_\perp$, the system prefers to rotate the valley degree of freedom leaving the spin as a conserved quantum number. The value of this critical $B_\perp$ at which the valley ceases to be a good quantum number is not universal, but depends on the coupling constants as well as on the width of the $\nu=-1$ region.

The corresponding transmission amplitudes are shown in Fig.~\ref{fig:TDHF_uxy_g_uz_Ev_1}. For small $B_\perp$, we see the characteristic magnon transmission associated with a conserved valley quantum number in Fig.~\ref{fig:TDHF_uxy_g_uz_Ev_1}a, very similar to those of Fig.~\ref{fig:TDHF_uz_g_uxy_Ev_1}. However, when the valley is no longer a good quantum number, we switch to the other type of magnon transmission spectrum, similar to that of Fig.~\ref{fig:TDHF_uxy_g_uz_Ev_5}c, with the transmission becoming perfect at high energy, and with spin-valley magnons being excited at intermediate energies.

\subsection{Magnon transmission at very small $E_V$}
\label{subsec:small_Ev}
In this subsection we will take $E_V$ to be the smallest energy scale in the problem, that is, $|u_\alpha^0|>E_Z>E_V$.

First, we consider the case where $u_z^0 > |u_{xy}^0|$. We illustrate this case with the parameter choices $u_z^0=4,~u_{xy}^0=-3,~E_Z=0.5,~E_V=0.1$. The HF one-body energy in the $\nu=1$ bulk are shown in Fig.~\ref{fig:bulk_ordering_uz_g_uxy_Ev_0.1}. As can be seen, the ordering $E_{K\uparrow}>E_{K\downarrow}>E_{K'\uparrow}$  is preserved for all $B_\perp$.

The self-consistent HF solutions for $\Bp=B_\perp^0,2B_\perp^0$ are shown in Fig.~\ref{fig:HF_uz_g_uxy_Ev_0.1}. In this case, the system always conserves the spin quantum number while spontaneously breaking the valley $U(1)$ symmetry, rotating the valley degree of freedom in the interface regions. In our case, because $E_V\neq0$, the occupied state in the $\nu=-1$ region has both valley components as opposed to the case $E_V=0$ examined by Wei et al\cite{Wei_Huang_MacDonald2021}, where the filled state in the $\nu=-1$ region is $\Kpua$. The magnon transmission amplitudes are shown in Fig.~\ref{fig:TDHF_uz_g_uxy_Ev_0.1}. As can be seen the transmission is nearly perfect at all energy, barring a few resonant reflections, presumably due to couplings of cavity modes in the $\nu=-1$ region with the asymptotic modes. This is very similar to the $E_V=0$ case examined earlier in Ref.~\cite{Wei_Huang_MacDonald2021}.

Next we consider the case $|u_{xy}^0| > u_z^0$. The ordering of the HF one-body levels for all $B_\perp$ is $E_{K\uparrow}>E_{K'\uparrow}>E_{K\downarrow}$, as shown in Fig.~\ref{fig:bulk_ordering_uxy_g_uz_Ev_0.1}. The self-consistent HF solutions across the system are shown in Fig.~\ref{fig:HF_uxy_g_uz_Ev_0.1}.

We see that in this case the system spontaneously breaks the valley $U(1)$ symmetries at the interfaces and in the $\nu=-1$ region, while preserving the spin symmetry. The corresponding magnon transmission spectrum is shown in Fig.~\ref{fig:TDHF_uxy_g_uz_Ev_0.1}. The magnon transmission now vanishes at low energies, increasing, and becoming nearly perfect at high energies. There are the usual dips associated with resonant reflections at discrete energies.

\section{Conclusions, caveats, and open questions}
\label{sec:conclusions}

In this work, we study the transmission of spin magnons across a graphene $1|-1|1$ system. Our main motivation for studying this particular setup is to obtain knowledge about the ratio of lattice-scale, ultra-short-range anisotropic couplings $u_z$ and $u_{xy}$. Such couplings are unable to pick out unique bulk ground states at $\nu=\pm1$ in the absence of spin and valley Zeeman couplings. Hence we assumed in this work that $E_Z,E_V>0$, which is realistic for experimentally relevant situations. In physical graphene samples it is believed that $u_z>0$ while $u_{xy}<0$, and that the two have roughly the same magnitude. Furthermore, their ratio can be altered by Landau-level mixing~\cite{RG_Murthy_Shankar_2002,RG_Bishara_Nayak_2009,RG_Peterson_Nayak_2013,RG_Sodemann_MacD2013} and the screening environment~\cite{RG_Wei_Sodemann_Huang_2024}. Pinning down this ratio and its evolution with screening and other tuning parameters would be invaluable in determining the phases of the $\nu=0$ system, which is yet to be fully understood. 

In contrast to $\nu=0$, where the nature of the phase is not fully understood, the ground states at $\nu=\pm1$ in the physical range of couplings with both spin and valley Zeeman couplings nonzero are known to be valley-polarized spin ferromagnets~\cite{Lian_Rosch_Goerbig_2016,Lian_Goerbig_2017}. This makes it easier to determine the values of the couplings themselves. $E_Z$ is known from the total $B$ field, and in principle, one can determine $E_V$ by measuring the gap at charge neutrality at $B_\perp=0$. A great advantage of this setup is that the {\sl in situ} tuning parameter $B_\perp$ allows us to alter the ratio of $E_V$ with respect to the other couplings. 

We used the Hartree-Fock approximation to find the self-consistent one-body states, and the variant of the time-dependent Hartree-Fock approximation developed by Wei {\sl et al}~\cite{Wei_Huang_MacDonald2021} to examine the transmission of magnons across the system. We find that the magnon transmission is quite sensitive to the structure of the interfaces between $\nu=1$ and $\nu=-1$. This structure in turn is dependent on the precise ordering of the HF energy levels in bulk $\nu=1$, and on the width of the intermediate $\nu=-1$ region. We find that in certain cases we can control the ordering of energy levels by tuning $B_\perp$. Experimentally there is a finite range over which $B_\perp$ can be varied, bounded at the lower end by disorder, which destroys the quantum Hall effect at low $B_\perp$, and bounded by a few tens of Tesla at the upper end. 

The first important fact to bear in mind in understanding our results is that for $E_V>0$ the bulk spin magnons have different valley characters in $\nu=1$ and $\nu=-1$. In $\nu=1$ the spin magnon is entirely in the $K'$ valley, while in $\nu=-1$, it is entirely in the $K$ valley. This mismatch is why the nature of the interface is so critical to the transmission of magnons across the system. As a function of the coupling constants and $B_\perp$, the system may prefer to keep both spin and valley $U(1)$ symmetries intact, or spontaneously break either or both of the $U(1)$ symmetries.  In all cases when the valley symmetry is preserved by the HF ground state, the magnon transmission drops to zero at high energies because of the mismatch stated above. If the valley symmetry is spontaneously broken, the magnon transmission becomes nearly perfect at high energies, because the high-energy, short-wavelength magnons can adiabatically follow the valley rotation across the interfaces.

The second important fact in understanding our results is that in the physical region of parameters, $u_z>0,~u_{xy}<0$, when $u_z>|u_{xy}|$ the system prefers to break the spin $U(1)$ symmetry at the interfaces, while in the opposite case $u_z<|u_{xy}|$ the system prefers to break the valley $U(1)$ symmetry (for some particular $\Bp$).  Since the interfaces between $\nu=1$ and $\nu=-1$ can very roughly be thought of as miniature regions of $\nu=0$, this is consistent with the fact that in the corresponding regions the $\nu=0$ bulk ground state breaks exactly those symmetries.

Keeping these two facts in mind, we can easily understand the cases that we considered in Section~\ref{sec:results} with $u_z>|u_{xy}|$. There is no valley rotation in these cases, and thus the magnon transmission drops to zero at high energy. For instance, Fig.~\ref{fig:HF_uz_g_uxy_Ev_5} shows the HF interface structure for the case $u_z^{0}=4,~u_{xy}^{0}=-3$ and $E_V=5$. One can see that there is no valley rotation for any value of $B_{\perp}$. The corresponding magnon transmission is  shown in Fig.~\ref{fig:TDHF_uz_g_uxy_Ev_5}, with the transmission dropping to zero at high energies. Similarly for $E_V=1$ the HF structure of the interfaces is shown in Fig.~\ref{fig:HF_uz_g_uxy_Ev_1}, with no valley rotation for any $B_{\perp}$. The corresponding magnon transmission is shown in Fig.~\ref{fig:TDHF_uz_g_uxy_Ev_1}, and shows vanishing transmission at high energies.  The more interesting case is $u_z<|u_{xy}|$, believed to occur for unscreened or lightly screened samples~\cite{RG_Wei_Sodemann_Huang_2024}. Here a crucial role is played by the valley Zeeman field $E_V$. For vanishing $E_V \to 0$,  the physics of magnon transmission was analyzed by Wei {\sl et al}\cite{Wei_Huang_MacDonald2021}, and it was found that there is nearly perfect transmission at high energy. Focusing on $E_V$ moderate to large, at small $B_\perp$, $E_V>|u_\alpha|,E_z$, the configurations at the interfaces do not break any symmetries, and the magnon transmission vanishes at high energies. This is illustrated in panels (a) and (b) of Fig.~\ref{fig:HF_uxy_g_uz_Ev_5}.  The corresponding magnon transmission vs energy is shown in panels (a) and (b) of Fig.~\ref{fig:TDHF_uxy_g_uz_Ev_5}. However, there is a threshold $B_\perp$ at which the couplings $u_\alpha$ become dominant over $E_V$, beyond which the interfaces break the valley $U(1)$ symmetry (Fig.~\ref{fig:HF_uxy_g_uz_Ev_5}(c)), restoring nearly perfect transmission at high energies as shown in Fig.~\ref{fig:TDHF_uxy_g_uz_Ev_5}(c). This threshold $B_\perp$ depends not only on the couplings, but also on the width of the intermediate $\nu=-1$ region, and can be estimated in our model given the sample geometry.  This is one of the main results of our work, because this threshold field provides quantitative information about the coupling constants.

In this context, the earlier experiments discussed in the introduction on $1|-1|1$, are not exactly in the regime we study, and so it would be very helpful to have detailed experimental studies at varying $B_{\perp}$ and varying sizes of the intermediate region for the $1|-1|1$ system. So far, the most recent investigation has been restricted to fillings $0<\nu<1$ for the intermediate region~\cite{Magnon_transport_Zhou_2022}, which, while physically very interesting, is much more complicated than the $1|-1|1$ system that we analyze. 

Let us turn to some of the assumptions that underlie our approach. We have assumed ultra-short-range interactions throughout. Consequently, to choose unique bulk ground states at $\nu=\pm1$, we assumed nonzero spin and valley Zeeman couplings. Relaxing the ultra-short-range assumption in the physical region of parameters $u_z>0,~u_{xy}<0$ does not change the phases of $\nu=\pm1$, as has been shown in previous work~\cite{Lian_Rosch_Goerbig_2016,Lian_Goerbig_2017}. However, introducing interactions beyond ultra-short-range does produce new phases at $\nu=0$. Because of our sharp interfaces, the $\nu=0$ regions here are fairly narrow, and we believe interactions beyond USR will not have any qualitative effect on our results. Secondly, we have focused on incident magnons with $q_y=0$. In Appendix \ref{sec:app_finiteq}, we show some results for incident magnons with $q_y\neq0$, which look qualitatively similar to our results in Section~\ref{sec:results}. Thirdly, we have ignored disorder and finite temperature effects. Disorder can induce the magnons to scatter elastically, thereby reducing the transmission. At $T\neq 0$, thermally generated collective modes will be present in the system, and could scatter the electrically generated magnons inelastically.

There are many open questions that could in principle be addressed by a detailed analysis of magnon transmission. The most important is the $\nu=0$ state, which is yet to be completely understood. It is believed that Landau level mixing leads to the interactions acquiring a range of the magnetic length $\ell$~\cite{Das_Kaul_Murthy_2022,Phase_diagram_nu0_Suman2023,RG_Wei_Sodemann_Huang_2024}. Introducing such interactions leads to the appearance of new phases which manifest the coexistence of CAF and bond order, and are separated from the bond-ordered and CAF states by second-order phase transitions~\cite{Das_Kaul_Murthy_2022,Phase_diagram_nu0_Suman2023,Stefanidis_Sodemann2023}. It should be possible to vary the screening to make the system traverse this coexistence phase.  Presumably, the magnon transmission properties of this phase differ from that of the standard CAF phase. Furthermore, fractional quantum Hall phases near $\nu=0$ also display a rich variety of phases in the physical regime of parameters, which could be explored via magnon transmission~\cite{Magnon_transport_Zhou_2022}. 
 Another interesting future direction could be exploring shot noise related measurements~\cite{Shot_noise_Yuval_Mirlin2020,Shot_noise_Sourav_Ankur2024,Edge_transport_FQHE_Fauzi_Hirayama2023,Edge_transport_FQHE_Nakamura_Manfra2023,Edge_transport_FQHE_Lai_Kun2013} for the magnon transmission, which could help to reveal the topological orders of the underlying fractional quantum hall systems. We hope to address these and other such questions in the near future. 

\begin{acknowledgements}
The authors would like to thank Nemin Wei and Chunli Huang for illuminating discussions. SJD would like to acknowledge both Harish-Chandra Research Institute, Prayagraj, and  the cluster computing facilities at the International Center for Theoretical Sciences (ICTS), Bangalore. He also acknowledges ICTS for hospitality and visitor funding.
SR and GM would like to thank the VAJRA scheme of SERB, India for its support under grant number VJR/2017/000114. GM thanks the Aspen Center for Physics (NSF Grant No. PHY-2210452 and PHY-160761) for its hospitality. 
\end{acknowledgements}

\appendix
\section{Magnon transmission at finite $q_y$}
All the results presented in the main text are for $q_y=0$. In this Appendix, we examine magnon transmission when the $y$-momentum is nonzero. Although we show the results for a particular choice of $q_y=0.3$, this illustrates the general behaviour of the transmission results for any finite $q_y$.  We find that the magnon transmission results are qualitatively similar to the results for $q_y=0$. 

We organize the results in the same way as in Sec. \ref{sec:results},  considering three different values of the valley potential $E_V$. 

\begin{figure}[H]
    \centering
    \includegraphics[width=0.4\textwidth,height=5cm]{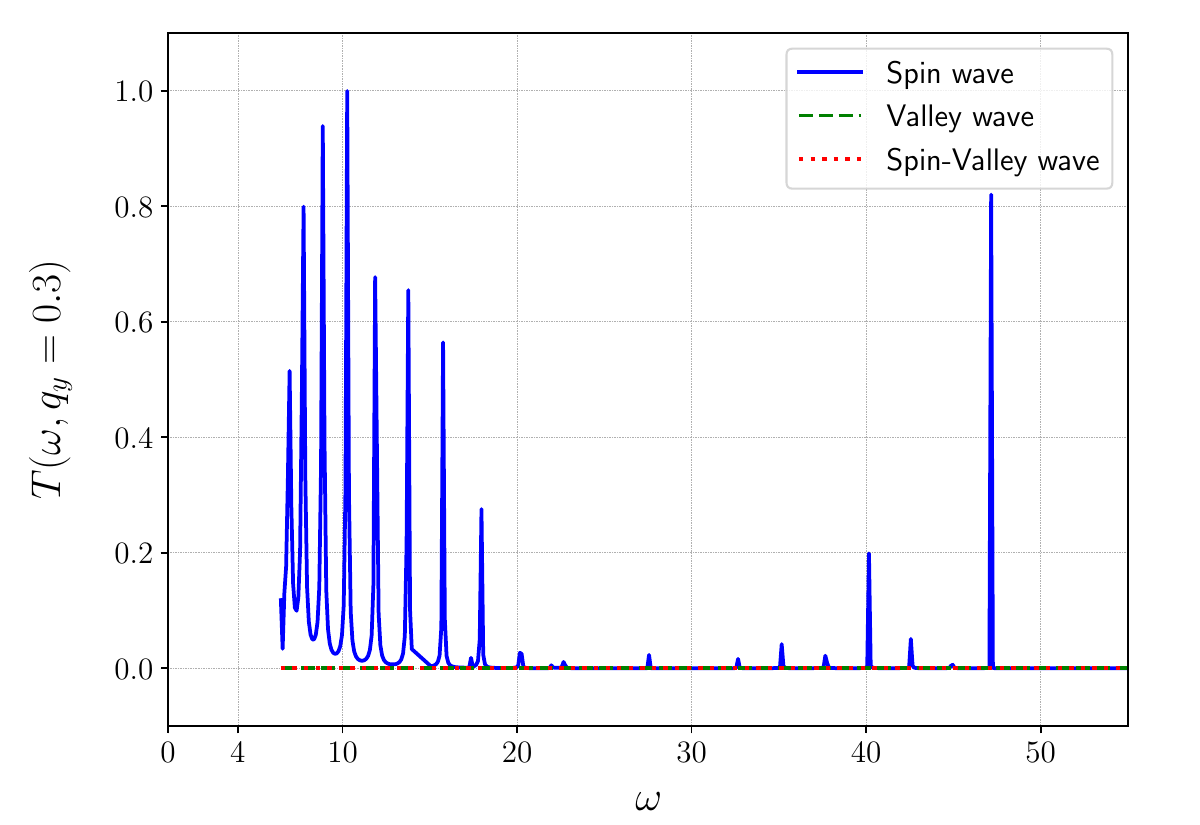}
    \caption{Transmission amplitudes as a function of incoming magnon energy at $q_y=0.3$ and $\Bp=4B_\perp^0$. At $B_\perp^0$ the Hamiltonian parameters are $\Wt^0=20, E_C^0=30, u_z^0=4.0, u_{xy}^0=-3.0, E_Z^0=0.5, E_V=5.0$ and the corresponding HF state is shown in Fig.~\ref{fig:HF_uz_g_uxy_Ev_5}(c). As in the case $q_y=0$, we find that the transmission of the spin-wave is strongly suppressed at most energies, apart from a few isolated resonances.}
    \label{fig:app_uz_g_uxy_Ev_5}
\end{figure}

First, we consider $E_V > |u_\alpha^0| > E_Z^0$ with $\alpha=z,xy$. Magnon transmission for  $u_z > |u_{xy}|$ at $\Bp=4B_\perp^0$ is shown in Fig.~(\ref{fig:app_uz_g_uxy_Ev_5}). We find that the magnon transmission is strongly suppressed, apart from the resonant peaks similar to the $q_y=0$ case. 

\label{sec:app_finiteq}

\begin{figure}[H]
    \centering
    \includegraphics[width=0.4\textwidth,height=5cm]{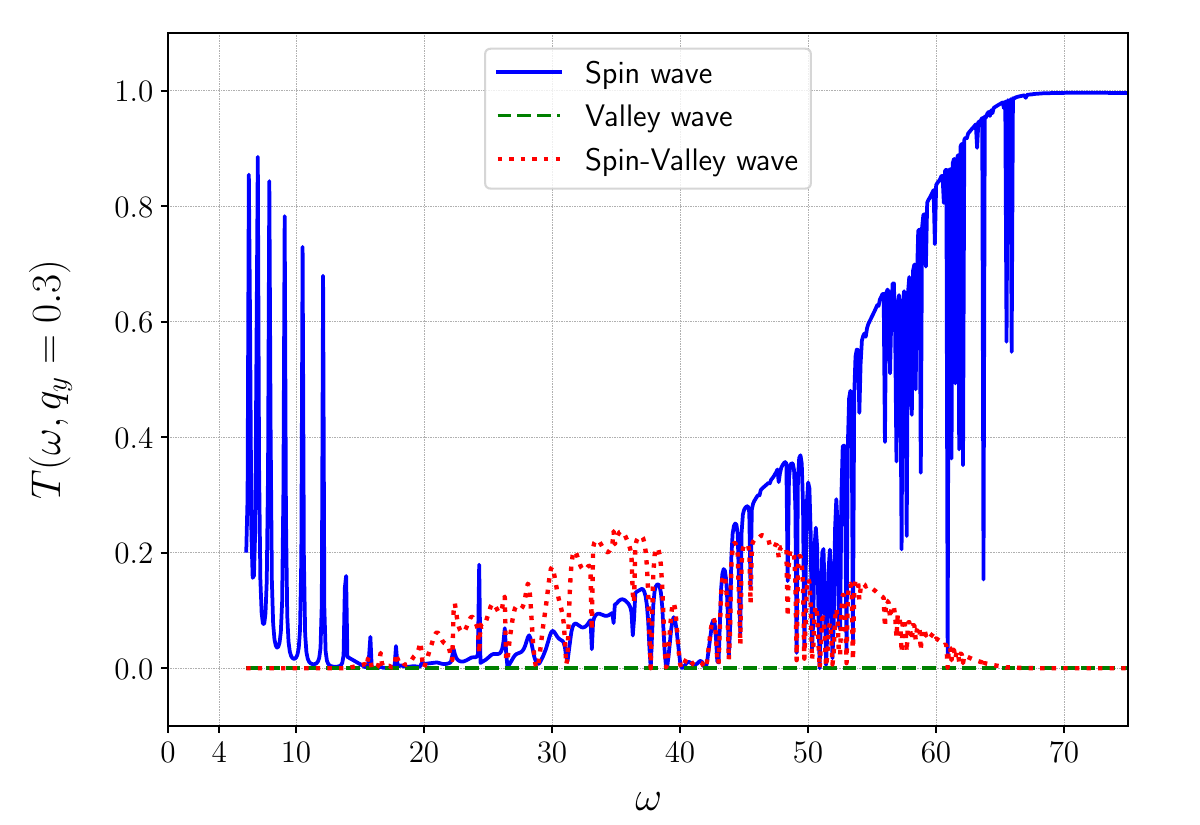}
    \caption{Transmission amplitudes for the collective modes as a function of the incoming magnon energy at $q_y=0.3$ and $\Bp=4B_\perp^0$. At $B_\perp^0$.  The Hamiltonian parameters are $\Wt^0=20, E_C^0=30, u_z^0=2.0, u_{xy}^0=-3.0, E_Z^0=0.5, E_V=5.0$, with the corresponding HF state being shown in Fig.~\ref{fig:HF_uxy_g_uz_Ev_5}(c). As in the case $q_y=0$ (Fig.~\ref{fig:TDHF_uxy_g_uz_Ev_5}(c)), we find that the transmission of the spin wave mode goes to unity at higher energies.}
    \label{fig:app_uxy_g_uz_Ev_5}
\end{figure}

Still staying with  $E_V > |u_\alpha^0| > E_Z^0$, we now consider the case $u_z < |u_{xy}|$ at $\Bp=4B_\perp^0$. The magnon transmission results are shown in Fig.~(\ref{fig:app_uxy_g_uz_Ev_5}).  Here we find that the magnon transmission is small at lower energies and increases with the magnon energy eventually leading to complete transmission at higher energies. The behavior is very similar to that at  $q_y=0$. 

\begin{figure}[H]
    \centering
    \includegraphics[width=0.4\textwidth,height=5cm]{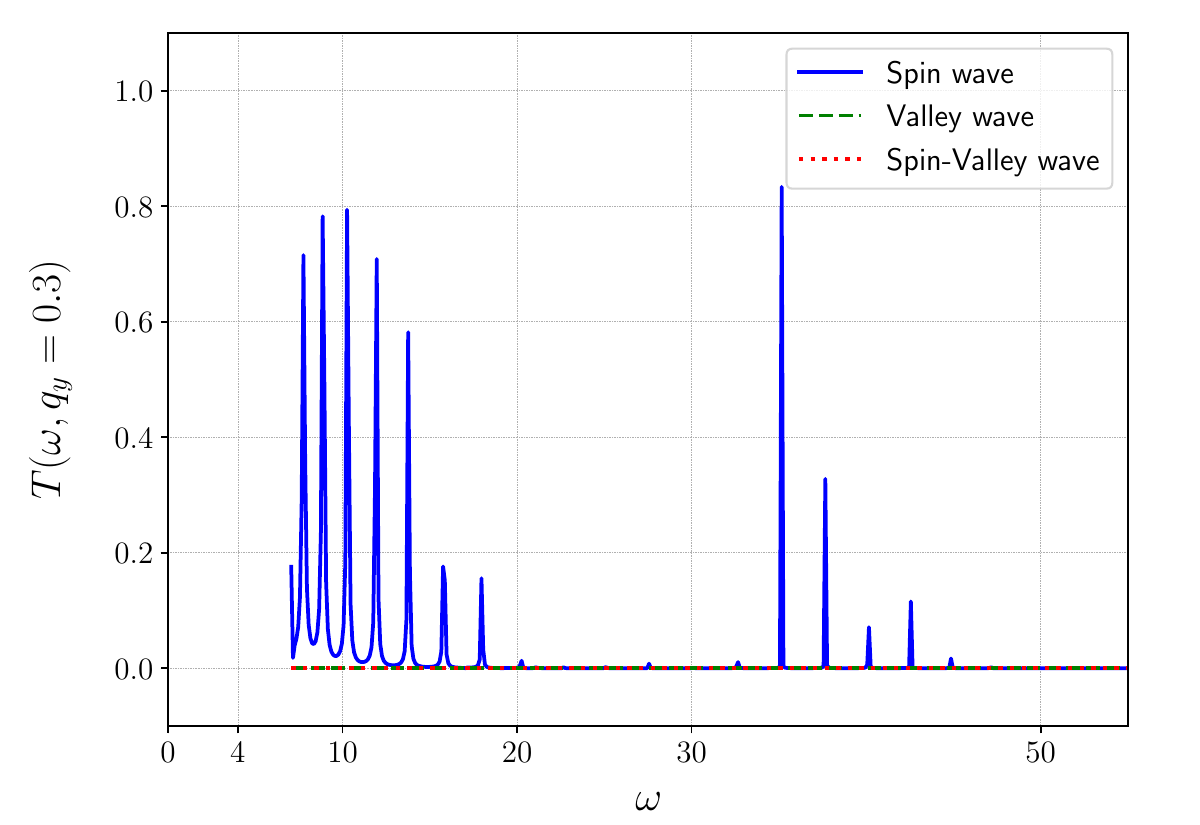}
    \caption{Transmission amplitudes as a function of incoming magnon energy at $q_y=0.3$ and $\Bp=4B_\perp^0$. At $B_\perp^0$ the Hamiltonian parameters are $\Wt^0=20, E_C^0=30, u_z^0=4.0, u_{xy}^0=-3.0, E_Z^0=0.5, E_V=1.0$ and the corresponding HF state is shown in Fig.~\ref{fig:HF_uz_g_uxy_Ev_1}(c). We see qualitatively similar behavior to the $q_y=0$ results shown in Fig.~\ref{fig:TDHF_uz_g_uxy_Ev_1}(c).}
    \label{fig:app_uz_g_uxy_Ev_1}
\end{figure}

\begin{figure}[H]
    \centering
    \includegraphics[width=0.4\textwidth,height=5cm]{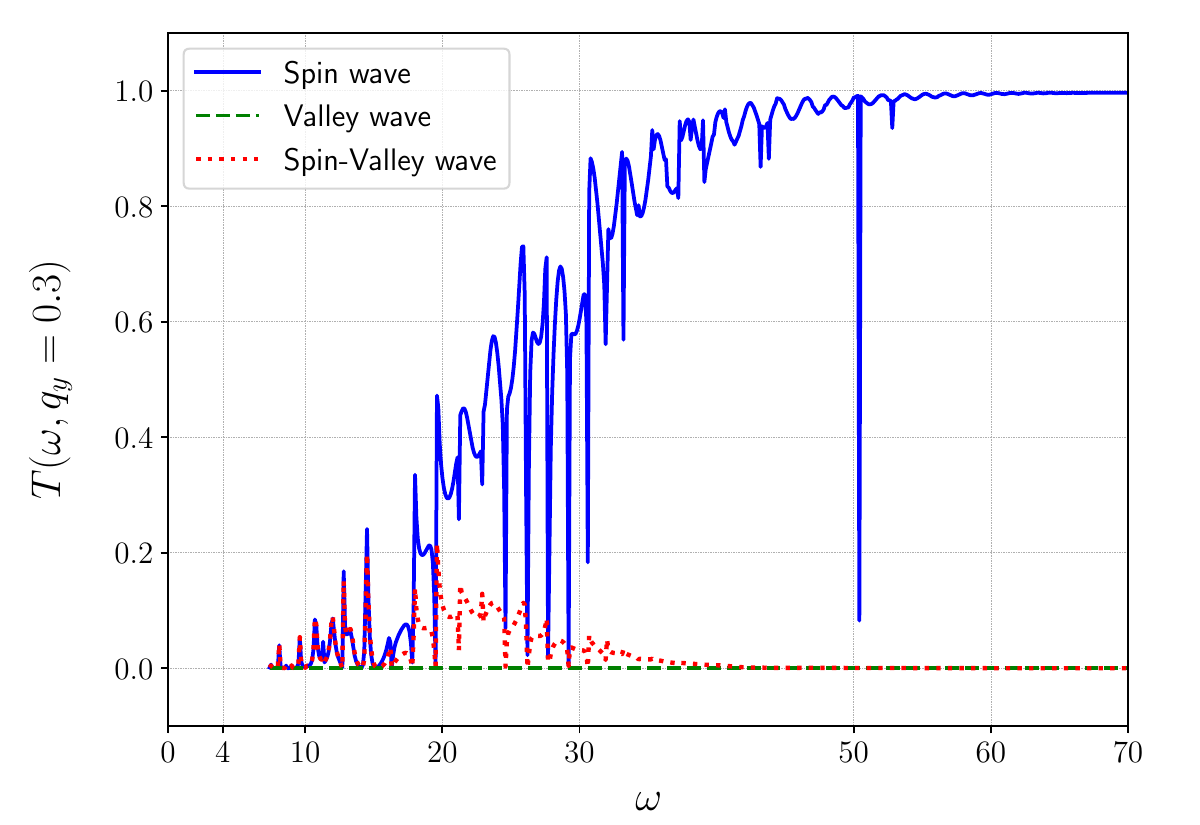}
    \caption{Transmission amplitudes as a function of incoming magnon energy at $q_y=0.3$ and $\Bp=4B_\perp^0$. At $B_\perp^0$ the Hamiltonian parameters are $\Wt^0=20, E_C^0=30, u_z^0=2.0, u_{xy}^0=-3.0, E_Z^0=0.5, E_V=1.0$ and the corresponding HF state is shown in Fig.~\ref{fig:HF_uxy_g_uz_Ev_1}(c). Once again, we see qualitatively similar behavior to the $q_y=0$ results shown in Fig.~\ref{fig:TDHF_uxy_g_uz_Ev_1}(c).}
    \label{fig:app_uxy_g_uz_Ev_1}
\end{figure}

Next, we consider intermediate values of $E_V$ such that $ |u_\alpha^0| > E_V > E_Z^0$ with $\alpha=z,xy$.
The case of  $u_z > |u_{xy}|$ is shown in Fig.~(\ref{fig:app_uz_g_uxy_Ev_1}), while the case with $|u_{xy}| > u_z$ is shown in Fig.~(\ref{fig:app_uxy_g_uz_Ev_1}). Both the results are for $\Bp=4B_\perp^0$.  Here too, the results are qualitatively similar to the $q_y=0$ results.

\begin{figure}[H]
    \centering
    \includegraphics[width=0.4\textwidth,height=5cm]{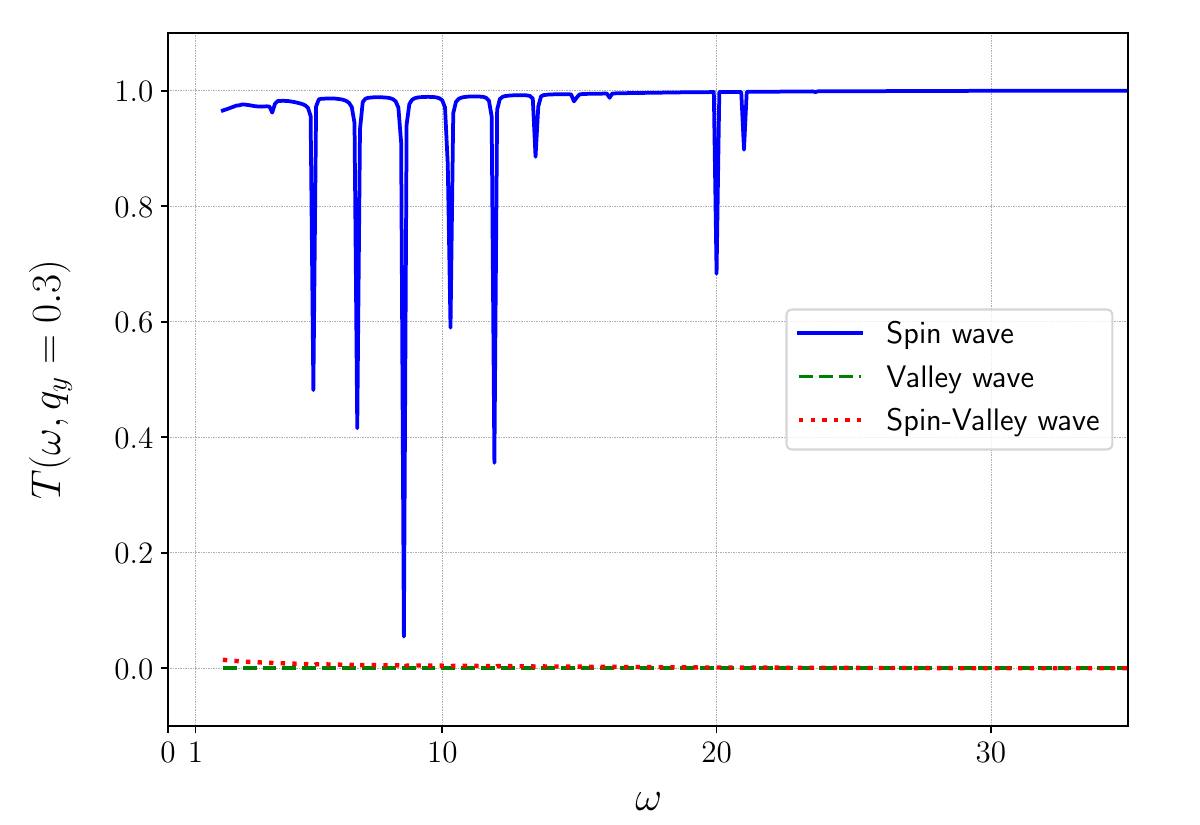}
    \caption{Transmission amplitudes for all the collective modes as a function of incoming magnon energy at $q_y=0.3$ and $\Bp=B_\perp^0$. At $B_\perp^0$ the Hamiltonian parameters are $\Wt^0=20, E_C^0=30, u_z^0=4.0, u_{xy}^0=-3.0, E_Z^0=0.5, E_V=0.1$ and the corresponding HF state is shown in Fig.~\ref{fig:HF_uz_g_uxy_Ev_0.1}(a). We can see qualitatively similar behavior to the $q_y=0$ results shown in Fig.~\ref{fig:TDHF_uz_g_uxy_Ev_0.1}(a).}
    \label{fig:app_uz_g_uxy_Ev_0.1}
\end{figure}

\begin{figure}[H]
    \centering
    \includegraphics[width=0.4\textwidth,height=5cm]{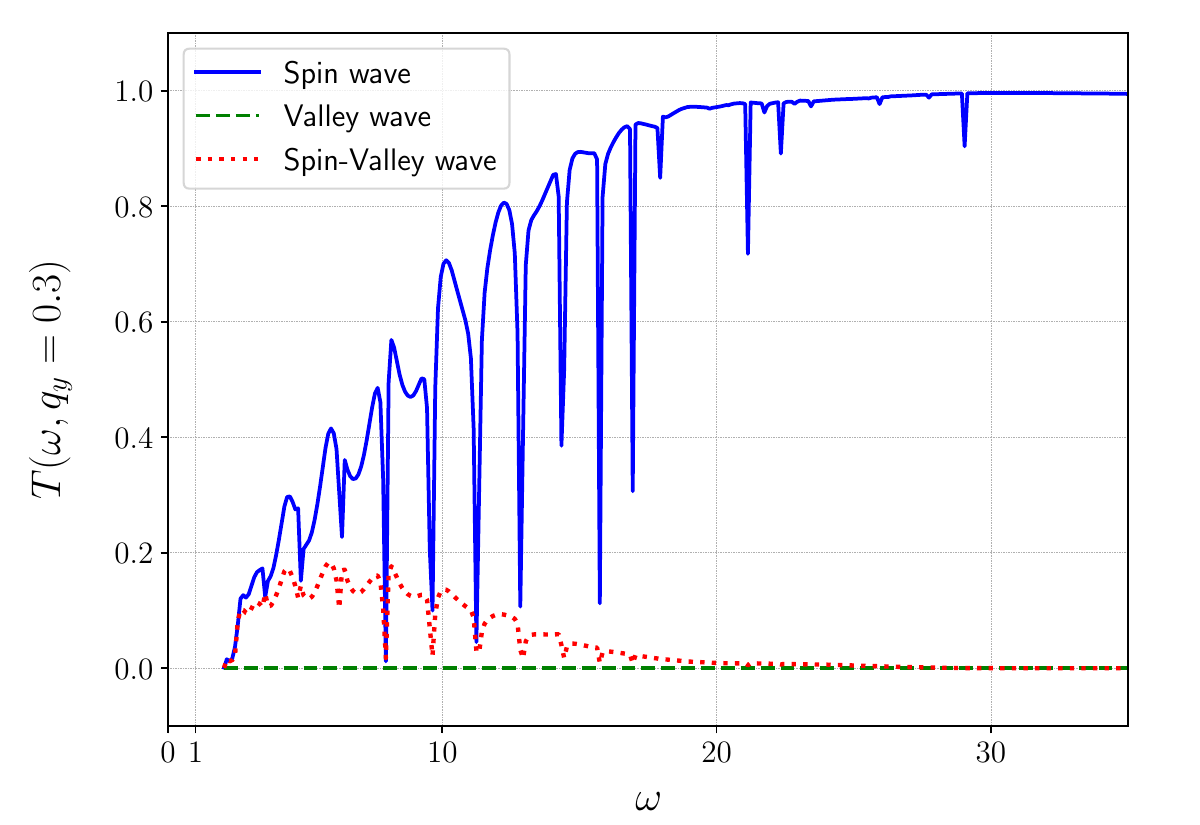}
    \caption{Magnon transmission amplitudes for all the collective modes as a function of incoming magnon energy at $q_y=0.3$ and $\Bp=B_\perp^0$. At $B_\perp^0$ the Hamiltonian parameters are $\Wt^0=20, E_C^0=30, u_z^0=2.0, u_{xy}^0=-3.0, E_Z^0=0.5, E_V=0.1$ and the corresponding HF state is shown in Fig.~\ref{fig:HF_uxy_g_uz_Ev_0.1}(a). Once again, the behavior is qualitatively similar behavior to the $q_y=0$ results shown in Fig.~\ref{fig:TDHF_uxy_g_uz_Ev_0.1}(a).}
    \label{fig:app_uxy_g_uz_Ev_0.1}
\end{figure}

Finally, we consider the case when the valley Zeeman coupling is the smallest scale i.e. $|u_\alpha^0|  > E_Z^0 > E_V$. The case with $u_z > |u_{xy}|$ is shown in Fig.~(\ref{fig:app_uz_g_uxy_Ev_0.1}) and the case with $|u_{xy}| > u_z$ is shown in Fig.~(\ref{fig:app_uxy_g_uz_Ev_0.1}). Both the results are shown at $\Bp=B_\perp^0$. As one can see, for $u_z > |u_{xy}|$,  the magnon transmission remains almost unity apart from the resonant dips similar to the results at $q_y=0$ in Fig.~\ref{fig:TDHF_uz_g_uxy_Ev_0.1}. For $|u_{xy}| > u_z$, on the other hand, as seen before in Fig.~\ref{fig:TDHF_uxy_g_uz_Ev_0.1},  the magnon transmission increases with energy and eventually saturates to unity.

\twocolumngrid
\bibliographystyle{apsrev}
\bibliography{hall}

\end{document}